\documentclass[nofootinbib,floatfix,onecolumn,times]{revtex4-2}
\usepackage{bm}		

\usepackage{graphicx,amssymb}
\usepackage{latexsym,color}
\usepackage{hyperref}
\usepackage[utf8]{inputenc}
\usepackage{natbib}

\usepackage{graphicx}	
\usepackage{amsmath}	

\def\be{\begin{equation}}
\def\ee{\end{equation}}
\def\bea{\begin{eqnarray}}
\newcommand{\cc}{\mathrm{C}}

\def\eea{\end{eqnarray}}

\newcommand{\pp}{\textbf{()}}

\newcommand{\Qa}{\mathcal{Q}}

\newcommand{\il}{~}

\newcommand{\Sa}{\mathrm{S}}
\newcommand{\mnras}{MNRAS}

\newcommand{\aap}{A\&A}

\newcommand{\Em}{\mathcal{E}}
\newcommand{\La}{\mathcal{L}}
\newcommand{\Ta}{{\mbox{\scriptsize  \textbf{\textsf{T}}}}}
\newcommand{\apjs}{APJS}

\count\footins = 1000
\begin{document}

\title{Inter--disks inversion surfaces}


\author{D. Pugliese \&
       Z. Stuchl\'{\i}k 
}

\email{daniela.pugliese@physics.slu.cz}

\affiliation{
Research Centre for Theoretical Physics and Astrophysics, Institute of Physics,
  Silesian University in Opava,
 Bezru\v{c}ovo n\'{a}m\v{e}st\'{i} 13, CZ-74601 Opava, Czech Republic
}

\begin{abstract}
We consider a counter--rotating torus orbiting  a central  Kerr black hole (\textbf{BH})  with dimensionless spin $a$, and   its  accretion flow into the \textbf{BH},   in an agglomerate of an   outer counter--rotating torus and an inner co--rotating torus.
This work focus is the analysis of the inter--disks inversion surfaces.
Inversion surfaces are spacetime surfaces,   defined by the  condition $u^{\phi}=0$ on the flow torodial velocity, located out of the \textbf{BH}  ergoregion, and totally embedding the  \textbf{BH}.
They emerge as   a necessary condition, related  to the  spacetime frame--dragging, for the  counter--rotating flows  into the central Kerr \textbf{BH}.  In our  analysis  we study the inversion surfaces of the Kerr spacetimes    for the counter--rotating flow from the outer torus,  impacting on the inner co--rotating disk. Being   totally or partially embedded in (internal to)  the  inversion surfaces, the inner  co--rotating torus (or jet) could be totally or in part ``shielded",   respectively,  from the impact with  flow   with  $a u^{\phi}<0$.
We prove that, in general, in the spacetimes with $a<0.551$ the  co--rotating  toroids are always external to the accretion flows inversion surfaces.
For  $0.551<a<0.886$,  co--rotating toroids could be  partially internal  (with the disk  inner region, including the inner edge) in the flow  inversion surface.
For \textbf{BHs} with    $a>0.886$,
a co--rotating  torus could be  entirely embedded in the  inversion surface and, for larger spins,  it  is internal to  the inversion surfaces. Tori orbiting in the  \textbf{BH} outer ergoregion are a particular case.
Further constraints   on the \textbf{BHs} spins are discussed in the article.
\end{abstract}

\date{\today}

\maketitle

\section{Introduction}
The  spacetime frame-dragging    influences  the accretion processes    acting, in particular, on the  flows from counter--rotating  accretion  disks, where  the Lense--Thirring effect acts on the accretion  flows with an initial counter--rotating component, reversing its rotation direction, i.e.  the  toroidal component of the velocity  $u^{\phi}$ in the proper frame  along its trajectory.
In this analysis we focus on counter--rotating flows accreting into a central  Kerr Black Hole (\textbf{BH}).
The flow, assumed to be free--falling into the central attractor,  inherits some properties of the counter--rotating accreting   configurations.
Therefore, because of the frame--dragging, the flow  trajectory  is   characterized by the presence of  flow inversion points, defined by the condition  $u^{\phi}=\Omega=0$ on the   toroidal  component of the flow velocity and relativistic angular velocity.
All the   inversion points are located on a surface, the \emph{inversion surface}, placed out of the \textbf{BH} outer ergosurface and dependent  on the flow specific angular momentum only
\cite{retro-inversion,2023EPJC...83..242P,
2023NuPhB99216229P,collect2024}.

 In particular we explore   an aggregate of toroidal  accretion disks orbiting on the equatorial plane of the  Kerr \textbf{BH}, composed by an outer counter--rotating  disk  and  a co--rotating inner disk,  studying the  inversion surfaces of  the counter--rotating flows which are accreting onto the central \textbf{BH}. The inner co--rotating disk  of the  double system, can  be  embedded, external  or partially contained  in the  spacetimes inversion surfaces \emph{(inter--disks inversion surfaces}), according to the \textbf{BHs} spins and the counter--rotating flows specific angular momentum. Our analysis  fixes the more general  conditions where a co--rotating accretion disks   can  impact  with  materials with $a u^\phi<0$ (where $a>0$ is the \textbf{BH} spin).
The disks embedded in an inversion surface will be completely shielded from impact with particles (and photons) with $u^\phi<0$,  accreting into the central Kerr \textbf{BH}.

{It is important to note that the inversion surface (defined as the loci of the vanishing toroidal flow velocity) can be  contained in the region defined by  the surface separating the co-rotating and counter-rotating component,  depending on whether  the co-rotating disks are contained inside the inversion surface or not. In this work we analyse the relation between these surfaces depending on the \textbf{BH} spin and some tori parameters,   leading  to three different scenarios. The toroids can  be  \emph{1.} embedded in an inversion surface; \emph{2.}  located out of  the inversion surface or   \emph{3.} crossing  the inversion surfaces  in particular points being partially contained in  the inversion surface.}
     Our finding therefore ultimately   constrain the case when  the  inner  co--rotating disk can be  replenished with    matter (and photons)  having  $a u^\phi<0$ which is accreting into the central \textbf{BH} or, viceversa  when  the counter--rotating  flow impacts on the co--rotating disk  with   $a u^\phi>0$, influencing the disk dynamics, combining with the co--rotating accreting flows and eventually affecting the co--rotating disk  inner edge (and,  connected with the physics of the inner edge, the launching of materials in jets along the \textbf{BH} axis). This situation will be  constrained according to
  the  flow parameters and the \textbf{BH}  dimensionless spin  $a$.

Counter--rotating accretion flows   are well  known in the  \textbf{BH} Astrophysics\footnote{See for example  \cite{Murray,Kuznetsov,Violette,Ensslin:2002gn,Beckert:2001az,Kim:2016qsr,Barrabes:1995ue,2010ApJ...710..859E,
Christodoulou,Nixonx,Garofalo,2010ASPC..427....3V,Volonteri:2002vz,2012MNRAS.422.2547N,Dyda:2014pia,
2016A&A...591A.114A,Cui,Rao,Reis,Middleton-miller-jones,Morningstar,Cowperthwaite}}.
Counter--rotation can   also  distinguish \textbf{BHs} with or without jets\footnote{Counter--rotating tori and jets were studied   in relation to
radio--loud \textbf{AGN} and   double radio source associated  with galactic nucleus \cite{2010ApJ...710..859E,GarofaloEvans}.}
\cite{Ensslin:2002gn,Beckert:2001az}.
Counter--rotation of the  extragalactic microquasars have been investigated for example also  as
engines for  jet emission, see
\cite{Middleton-miller-jones,Garofalo}, %
connected to
\textbf{SWIFT J1910.2 0546}   \cite{Reis},
and   the faint luminosity of
\textbf{IGR J17091--3624} (a transient X-ray source  believed to be a Galactic \textbf{BH} candidate)  \cite{Rao}. %
(Observational evidence of counter--rotating disks has been  also  discussed in relation to   \textbf{M87}, observed by the Event Horizon Telescope \cite{M87}.)
They can be produced for example during  chaotical, discontinuous   accretion   processes, where aggregates of  co--rotating and counter---rotating  toroids can be mixed \cite{Dyda:2014pia,Aligetal(2013),Carmona-Loaiza:2015fqa,ringed,Multy,long}. Therefore they could also appear in   binary systems with  stellar mass \textbf{BHs} (as  in  \textbf{3C120} \cite{Kataoka,Cowperthwaite}),
in the Galactic binary \textbf{BHs} \cite{Cui,Reis}, and more generally in  \textbf{BH} binary system  (\cite{Morningstar} or \cite{Christodoulou}).
Eventually even misaligned disks  with respect to the central \textbf{BH} spin can form in these situations \cite{Nixon:2013qfa,2015MNRAS.449.1251D,Bonnerot:2015ara,Aly:2015vqa,2022MNRAS.511.3795N,2017MNRAS.469.4483T,Wongetal,2021MNRAS.502.2023P,Zhangetal}.
The Lense--Thirring effect in misaligned disks  can be manifested   in  the Bardeen--Petterson effect,  where the
\textbf{BH} spin  can change  under the action of the tori  torques, with an  originally  misaligned torus  broken due to the  frame--dragging combined to   other factors   as the fluids viscosity, in a double system as the aggregate  considered in this work, composed by  an inner  co--rotating torus  and an counter--rotating outer torus \cite{BP75,Nealon:2015jya,Martin:2014wja,King2005,King:2018mgw,King:2008au,2012ApJ...757L..24N,
2012MNRAS.422.2547N,2006MNRAS.368.1196L,Feiler,King2005}.

\textbf{BHs} could   accrete from disks having  alternately co--rotating and
counter--rotating  senses of rotation
\cite{Murray}. Hence, more complex scenarios envisaging  complicated superimposed sub--structures of counter--rotating and co--rotating materials have been also  studied.
In   \cite{Dyda:2014pia}, for example, a  high-resolution axisymmetric hydrodynamic simulation of viscous counter--rotating
disks was presented for the cases where the two components are vertically separated or  radially separated.
A time-dependent, axisymmetric  hydrodynamic (HD) simulation of complicated composite counter--rotating  accretion disks was studied  in \cite{Kuznetsov}, where the disks
consist of   combined counter--rotating  and corotating components. (It is worth noting that the accretion rates are increased
over that for  co--rotating  disks--see also  \cite{Kuznetsov}.)
Counter--rotating  tori  have  been modelled also as a counter--rotating gas layer on the
surface of a co--rotating disk. %
 Reversals
in the rotation direction of an accretion disk, ``flip--flop" phase, have  been
considered  to explain state transitions\cite{Cui}\footnote{In  X-ray binary, the  \textbf{BH} binaries with no detectable
ultrasoft component above 1--2 keV in their high luminosity state may contain a fast--spinning retrograde \textbf{BH}, and
   the  spectral state transitions can correspond to a temporary ``flip--flop" phase of  disk reversal,
showing the characteristics of both
counter--rotating   and  co--rotating   systems, switching from one state to another
(the hard X--ray luminosity of a co--rotating  system is generally much lower than that of a counter--rotating  system).}.

In this article we  will consider an aggregate of co--rotating and counter--rotating toroids. This system has  been constrained in \cite{pugtot,ringed,open,dsystem} and widely discussed in subsequent papers
\cite{letter,long,Multy,proto-jets,ella-jet,Fi-Ringed,ella-correlation,mnras2,cqg2020}.
Evidences of the presence of a toroids agglomerate   with an inner co--rotating torus and outer counter--rotating  torus has been provided by
\textbf{A}tacama
\textbf{L}arge \textbf{M}illimeter/submillimeter \textbf{A}rray (\textbf{ALMA}).
 It has been assumed that the outer disk could have been  formed in recent times from molecular gas falling. The
inner disks follows the rotation of the galaxy, whereas the outer
disk rotates (in stable orbit) the opposite way.
The interaction between counter--rotating
disks may enhance the accretion rate with a rapid multiple  phases of accretion. The presence of two disks of gas rotating in
opposite directions can be  pointed out from  the observation of gas motion around the \textbf{BH}
inner orbits\footnote{\textbf{ALMA} showed evidence that the molecular torus
consists of counter--rotating and misaligned disks on parsec scales which can explain   the \textbf{BH} rapid growth.
 In \cite{Violette} counter-rotation and high-velocity outflow in the \textbf{NGC1068} galaxy  molecular torus
were studied.
\textbf{NGC 1068}  center hosts a super--massive \textbf{BH}  within a
thick doughnut--shaped  dust and gas  cloud. }.
From methodological view--point, in this work we shall  examine a   free--falling accretion flow constituted by counter--rotating  matter (and photons),  and   axially symmetric equatorial  geometrically thick co--rotating and counter--rotating toroids  orbiting  on the \textbf{BH} equatorial plane.   We use a full  General Relativistic   Hydrodynamical  (GRHD)  model, which include  proto-jets solutions--\cite{abrafra,pugtot}. Proto-jets  are open HD  toroidal configurations, with matter funnels along the \textbf{BH} rotational axis, associated to  thick disks, and emerging under special conditions on the fluid  forces balance.
The focus  of this work is the analysis of the inter--disks inversion surfaces, however, for completeness, we shall  also discuss  the case when one or two of these configurations are proto-jets.

\medskip

In details the plan of this article is  as follows:

The spacetime metric is introduced in Sec.\il(\ref{Sec:quaconsta}). In Sec.\il(\ref{Sec:carter-equa}) there are
the  geodesic equations   of motion   in Kerr  spacetime.
Constraints on the flows and orbiting toroids are  discussed in Sec.\il(\ref{Sec:lblu}).
{Sec.\il(\ref{Sec:thinkc})  details the geometrical thick   accretion disk model in the Kerr spacetime, developing  the analytical conditions  adopted  in the following analysis.    The section   contains also  further notes on  the toroids
extreme points.}
In Sec.\il(\ref{Sec:inversion-spheres}) the
inversion surfaces are introduced.
In Sec.\il(\ref{Sec:center-Tori-crossing}) we  focus on  the co--rotating  toroids orbiting  the  central Kerr \textbf{BH}   in relation to the spacetime  inversion surfaces.
Some preliminary aspects  are explored  in Sec.\il(\ref{Sec:Aspects}), where we  investigate the
inter--disks  inversion surfaces in different Kerr \textbf{BHs} spacetimes,  distinguishing \textbf{BHs} classes according to the \textbf{BH} spins and the inversion surfaces properties in the spacetimes of each class.
In Sec.\il(\ref{Sec:inversion-crossing})
 the  inversion surfaces crossing    the      co--rotating  toroids are examined.   The crossing with the      co--rotating  toroids outer edges is addressed in  Sec.\il(\ref{Sec:outer--edge-crossing}). The surfaces crossing with  the  toroids cusps are studied in  Sec.\il(\ref{Sec:cusp-crossing}), and the case of the inversion surfaces crossing  the toroids  centers is discussed in  Sec.\il(\ref{Sec:torus-center-crossing}).   The  surfaces crossing  the  toroids critical points of pressure and geometrical maxima   are the subject of  Sec.\il(\ref{Sec:geometric--crossing}),  while the   crossing   on    planes different from the equatorial is  examined in  Sec.\il(\ref{Sec:off-equatorial--crossing}).
Conclusions are in Sec.\il(\ref{Conclsusion}).
\section{The spacetime metric}\label{Sec:quaconsta}
In the Boyer-Lindquist (BL)  coordinates
\( \{t,r,\theta ,\phi \}\),
the   Kerr  spacetime line element   reads\footnote{We adopt the
geometrical  units $c=1=G$ and  the $(-,+,+,+)$ signature, Latin indices run in $\{0,1,2,3\}$.  The radius $r$ has unit of
mass $[M]$, and the angular momentum  units of $[M]^2$, the velocities  $[u^t]=[u^r]=1$
and $[u^{\phi}]=[u^{\theta}]=[M]^{-1}$ with $[u^{\phi}/u^{t}]=[M]^{-1}$ and
$[u_{\phi}/u_{t}]=[M]$. For the seek of convenience, we always consider the
dimensionless  energy and effective potential $[V_{eff}]=1$ and an angular momentum per
unit of mass $[L]/[M]=[M]$.}
%
\bea \label{alai}&& ds^2=-\left(1-\frac{2Mr}{\Sigma}\right)dt^2+\frac{\Sigma}{\Delta}dr^2+\Sigma
d\theta^2+\left[(r^2+a^2)+\frac{2M r a^2}{\Sigma}\sin^2\theta\right]\sin^2\theta
d\phi^2
-\frac{4rMa}{\Sigma} \sin^2\theta  dt d\phi,
\\&&\label{Eq:delta}
\mbox{where}\quad
\Delta\equiv a^2+r^2-2 rM\quad\mbox{and}\quad \Sigma\equiv a^2 (1-\sin^2\theta)+r^2.
\eea
 $M$ is the   mass parameter,  $J$ is  the   total angular momentum,  $a=J/M\geq0$ is the metric spin.
  A Kerr \textbf{BH} is defined  by the condition $a\in[0,M]$, the extreme Kerr \textbf{BH}  has dimensionless spin $a/M=1$ and the non-rotating   case $a=0$ is the   Schwarzschild \textbf{BH} solution.
  (The Kerr naked singularities (\textbf{NSs}) have  $a>M$.)
    The \textbf{BH}  inner and outer   horizons are
\bea
   r_-\leq r_+,\quad \mbox{where} \quad r_{\pm}\equiv M\pm\sqrt{M^2-a^2}
\eea
respectively. The spacetime outer    ergosurface   is the radius
\bea\label{Eq:sigma-erg}
r_{\epsilon}^{+}\equiv M+\sqrt{M^2- a^2 (1-\sigma)}\quad\mbox{with}\quad \sigma\equiv \sin^2\theta\in [0,1],
\eea
where
  $r_{\epsilon}^+=2M$  in the equatorial plane $\theta=\pi/2$ ($\sigma=1$).
The spacetime outer ergoregion  is  the region  $]r_+,r_{\epsilon}^+
]$.
In the following, we will  use  dimensionless    units with $M=1$ (where $r\rightarrow r/M$  and $a\rightarrow a/M$).

\subsection{Geodesics equations}\label{Sec:carter-equa}
The  geodesic equations   of motion   in Kerr  spacetime   are:
\bea&&\label{Eq:eqCarter-full}
 \dot{t}=\frac{1}{\Sigma}\left[\frac{P \left(a^2+r^2\right)}{\Delta}-{a \left[a \Em (\sin\theta)^2-\La\right]}\right],\quad
\dot{r}=\pm \frac{\sqrt{R}}{\Sigma};\quad \dot{\theta}=\pm \frac{\sqrt{T}}{\Sigma},\quad\dot{\phi}=\frac{1}{\Sigma}\left[\frac{a P}{\Delta}-\left[{a \Em-\frac{\La}{(\sin\theta)^2}}\right]\right];
\eea
(\cite{Carter}),
where $\dot{q}$ indicates the derivative of any quantity $q$  with respect  the proper time  (or  a properly defined  affine parameter for the light--like orbits), and
\bea\nonumber
& P\equiv \Em \left(a^2+r^2\right)-a \La,\quad & R\equiv P^2-\Delta \left[(\La-a \Em)^2+\mu^2 r^2+\Qa\right],\\
\label{Eq:eich}  &T\equiv \Qa-
(\cos\theta)^2 \left[a^2 \left(\mu^2-\Em^2\right)+\left(\frac{\La}{\sin\theta}\right)^2\right], %
\eea
with   $\mu=0$ for light-like particles.
 Quantity  $ \Qa$ is  the Carter constant of motion, and   $(\Em, \La)$ are constants of geodesic  motions
 defined  as
\bea&&\label{Eq:EmLdef}
\Em=-(g_{t\phi} \dot{\phi}+g_{tt} \dot{t}),\quad \La=g_{\phi\phi} \dot{\phi}+g_{t\phi} \dot{t},\quad  g_{ab}u^a u^b=\kappa \mu^2,
\eea
from   the Kerr geometry   Killing field   $\xi_{\phi}\equiv \partial_{\phi}$,
     and  $\xi_{t}\equiv \partial_{t}$,
where $u^a\equiv\{ \dot{t},\dot{r},\dot{\theta},\dot{\phi}\}$, and  $\kappa=(\pm,0)$ is a normalization constant ($\kappa=-1$ for  test particles).

 The constant $\La$ in Eq.\il(\ref{Eq:EmLdef}) may be interpreted       as the axial component of the angular momentum  of a test    particle following
timelike geodesics and $\Em$  represents the total energy of the test particle
 related to the  radial infinity, as measured  by  a static observer at infinity.
We  introduce also  the specific  angular momentum
 \bea&&\label{Eq:flo-adding}
\ell\equiv\frac{\La}{\Em}=-\frac{g_{\phi\phi}u^\phi  +g_{\phi t} u^t }{g_{tt} u^t +g_{\phi t} u^\phi} =-\frac{g_{t\phi}+g_{\phi\phi} \Omega }{g_{tt}+g_{t\phi} \Omega},
\eea
where  $\Omega \equiv{u^\phi}/{u^{t}}$ is the relativistic angular velocity, related to the static observer at infinity.
%
%
  %
With  $a>0$, the  fluids and particles  counter--rotation  (co--rotation) is defined by $\ell a<0$ ($\ell a>0$).
The equatorial plane, $\sigma=1$, is the  metric symmetry plane  and   the  equatorial  circular geodesics are confined on the equatorial  plane as a consequence of the metric tensor symmetry under reflection through  the plane $\theta=\pi/2$.
 Furthermore, static  observers with  four-velocity   $\dot{\theta}=\dot{r}=\dot{\phi}=0$
cannot exist inside the ergoregion, then trajectories   $\dot{r}\geq0$, including photons  crossing the outer ergosurface  and escaping outside
in the region $r\geq r_{\epsilon}^+$ are possible\footnote{We assume $\Em>0$ (and $\dot{t}>0$). This condition  for co--rotating fluids in the ergoregion has to be discussed further. In the ergoregion  particles can also have $\La=0$ (i.e. $\ell=0$). However this condition characterizing the ergoregion  is not associated to geodesic  circular motion in the \textbf{BH} spacetimes. There are no solutions in general for $
\dot{t}\geq 0$, $\Em<0$ $T\geq0$ and $ \ell<0$ (for $r>r_+,a\in [0,1], \sigma\in [0,1]$). We assume the so--called positive root states $\dot{t}>0$. 
}.

 \section{{Geodesics structures and Toroids}}\label{Sec:lblu}
\begin{figure}
\centering
\includegraphics[width=8cm]{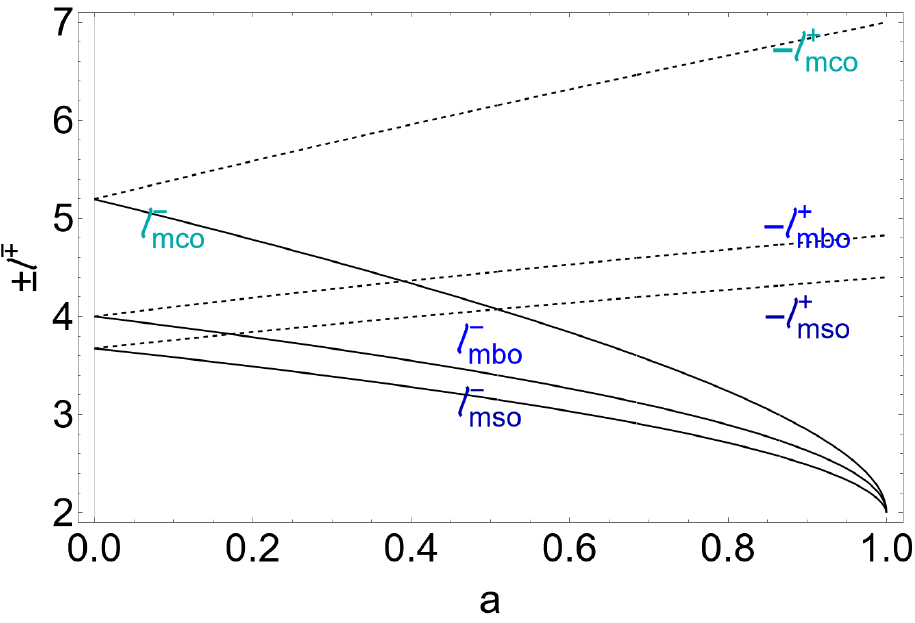}
\includegraphics[width=8cm]{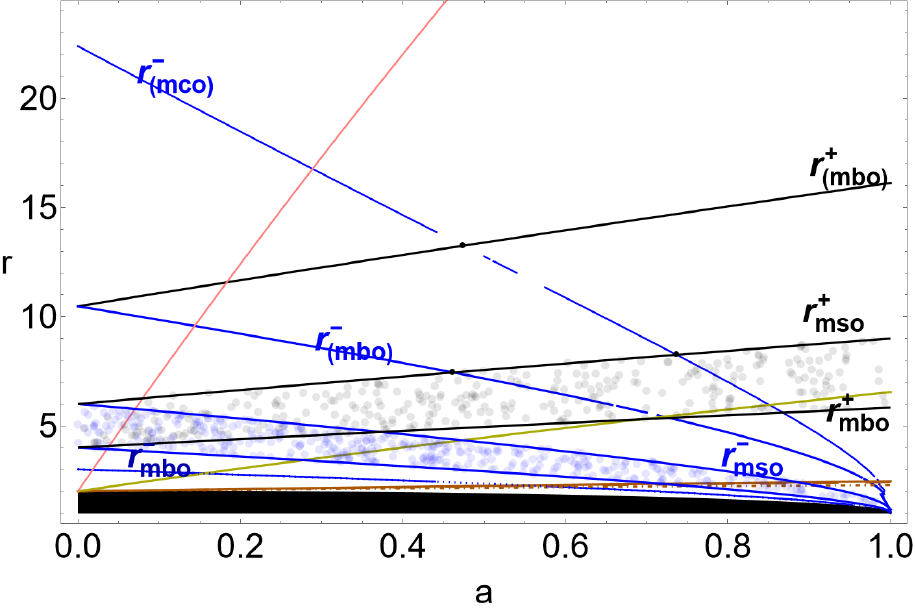}
\caption{Left panel: Parameter $\ell^\pm$ (specific angular momentum) defined in Eq.\il(\ref{Eq:flo-adding}) for co--rotating $(-)$ (solid) and counter--rotating $(+)$ orbits (dashed) as function of the \textbf{BH}  dimensionless spin $a$.
$mbo$ is for marginally bounded circular orbit, $mco$ is for marginally  circular orbit, $mso$ is for marginally stable circular orbit.   Right panel:
Solid curves are $r_{mco}^-<r_{mbo}^-<r_{mso}^-<r_{(mbo)}^-<r_{(mco)}^-$  defined in  Eqs\il(\ref{Eq:def-nota-ell}). Blue dotted (gray dotted) region is the range $[r_{mbo}^-,r_{mso}^-]$  ($[r_{mbo}^+,r_{mso}^+]$) for the location of the  co--rotating (counter--rotating)  disk cusp.   Brown solid, dashed and dotted--dashed curves are the inversion radii $r_{\Ta}(\ell_{mso}^+)>r_{\Ta}(\ell_{mbo}^+)>r_{\Ta}(\ell_{mco}^+)$. Pink (yellow) curves are for  $\ell^+=\ell^+_{mso}/100$ ($\ell^+=\ell^+_{mso}/10$). All quantities are dimensionless.}\label{Fig:Plotfondo}
\end{figure}

 %
 %
 %
 The  spacetime equatorial circular geodesics structures are a main  property  of the background  governing
 the  accretion  disk   physics, and they  are
  constituted by the \emph{marginal  circular orbit} for timelike particles, $r_{mco}^{\pm}$,  which is also the photon circular  orbit, the \emph{marginal  circular bound orbit}, $r_{mbo}^{\pm}$, and the \emph{marginal stable circular orbit}, $r_{mso}^{\pm}$ for co--rotating $(-)$ and counter--rotating $(+)$ motion--see Figs\il\ref{Fig:Plotfondo}--left panel. \cite{1972ApJ...178..347B}.

The constant specific angular momentum $\ell$ of  Eq.\il(\ref{Eq:flo-adding})   on these orbits are  $\ell^\pm\in \{\ell_{mso}^\pm,\ell_{mbo}^\pm,\ell_{mco}^\pm\}$, where
\bea
\mp\ell_{mso}^\pm\leq \mp\ell_{mbo}^\pm\leq \mp\ell_{mco}^\pm
\eea
 --see Figs\il\ref{Fig:Plotfondo}--right panel.
(In general,  we adopt the notation $\mathrm{Q}_\bullet$ for any quantity $\mathrm{Q}$ evaluated at $r_\bullet$.).
%
%

The  inner edge  of an accretion disk orbiting the central \textbf{BH} on its equatorial plane, is   generally assumed to be  in the range $[r_{mbo}^\pm,r_{mso}^\pm]$, for counter--rotating and co--rotating disks respectively,  and its  precise location depends on the central attractor  features and the disks characteristics.
 This  constraint  on the accretion disk inner edge  is widely adopted (and  well grounded)  in \textbf{BH} Astrophysics, and in the following analysis we   will   use this  assumption  {independently} of  other specific details  of the  accretion disk models.

However, to make our arguments more precise, we   discuss  an  example   of  General Relativistic   Hydrodynamical  (GRHD) toroidal  configurations--\cite{abrafra}.
These configurations  are     used to fix up the initial configurations for  numerical integration of a broad
variety  of   models.
 The methodological advantage in the adoption of these models is that the   general relativistic thick  tori {morphological}  features, related to  the equilibrium (quiescent) and accretion phases as cusped  closed toroidal surfaces,  are    determined    by the centrifugal and gravitational components of the force balance in the disks, rather then the dissipative effects.

 In these models, at the cusp the fluid may be considered pressure-free, and the matter outflows as consequence of an   hydro-gravitational instability  mechanism,  following the violation of mechanical equilibrium  in the balance of the gravitational and inertial forces and the pressure gradients in the disks  \cite{Pac-Wii}).
 Showing   a remarkably good fitting with  the more complex  dynamical models, they are   largely adopted  as initial setup for numerical  simulations  in  several general relativistic magnetohydrodynamic  ({GRMHD}) models, constituting also a  comparative model in  many  numerical analysis of  complex situations--\cite{abrafra,Igumenshchev,Shafee,Fragile:2007dk,DeVilliers}.
 \begin{table*}
 \centering
 \resizebox{.91\textwidth}{!}{%
\begin{tabular}{l|ll}
\hline
 $\mathbf{L_1}$& $
\mp \ell^{\pm}\equiv[\mp \ell_{mso}^{\pm},\mp\ell_{mbo}^{\pm}[$:& $\pp_1:$ Quiescent (i.e. not cusped) $\cc_1$ and cusped tori $\cc_\times$.
\\&
  The cusp is   $r^{\pm}_{\times}\in]r^{\pm}_{mbo},r^{\pm}_{mso}]$ ($K_{\times}^{\pm}<1$)).
  \\
  &The  torus  center is  $r^{\pm}_{center}\in]r^{\pm}_{mso},r^{\pm}_{(mbo)}]$.
  \\
  \hline
 $\mathbf{L_2}$ &  $\mp\ell^{\pm}\equiv[\mp \ell_{mbo}^{\pm},\mp\ell_{mco}^{\pm}[ $:& $\pp_2:$ Quiescent  tori $\cc_2$ and proto-jets.
\\&
The   torus  cusp is   $r_{\times}^{\pm}\in]r_{mco}^{\pm},r_{mbo}^{\pm}]$ (proto-jets $K_{\times}>1$).\\
&
 The  torus center  is  $r_{center}^{\pm}\in]r_{(mbo)}^{\pm},r_{(mco)}^{\pm}]$.
\\
\hline
 $\mathbf{L_3}$ &  $\mp \ell^{\pm}\geq\mp\ell_{mco}^{\pm}$:&  $\pp_3:$ Quiescent  tori $\cc_3$.
\\
& The torus center is   $r^{\pm}_{center}>r_{(mco)}^{\pm}$.
\\
\hline
  \end{tabular}}
   \caption{Geometrically thick GRHD tori orbiting Kerr \textbf{BHs}.  $\ell$ is the specific fluid angular momentum.
$mbo$ is for marginally bounded circular orbit, $mco$ is for marginally  circular orbit, $mso$ is for marginally stable circular orbit.  $\pm$ is for counter--rotating and co--rotating fluids respectively. Radii   $r_{(mbo)}^{\pm}$ and $r_{(mco)}^{\pm}$ are defined in Eq.\il(\ref{Eq:def-nota-ell}) and,  together  with the radii of  geodesic structure, form the  Kerr spacetime \emph{extended geodesic structure}--Fig.\il\ref{Fig:Plotfondo}. Fluid effective potential $V_{eff}(\ell,a,r)$ defines  the function $K(a,r)=V_{eff}(\ell(r,a),a,r)$.
 Torus cusp $r_\times$ is the  minimum point of pressure and density in the torus. The torus center $r_{center}$ is the fluid maximum point of pressure and density.  Cusped tori have parameter $K=K_\times\equiv K(r_{\times})\in]K_{center}, 1[\subset ]K_{mso}, 1[$, where $K_{center}\equiv K(r_{center})$. Proto-jets are open cusped configurations. Notation
$\cc^{\pm}$ ($\cc_{\times}^{\pm}$) is for quiescent  (cusped) tori,    $\pp^{\pm}$  for quiescent  or  cusped configurations, $r_{outer}$ is the torus outer edge and the notation $\Qa(L_\bullet)$  or $\Qa_\bullet$ for $\bullet\in\{1,2,3\}$ for configurations
$\Qa$ with momenta in the range $\mathbf{L_\bullet}$.}\label{Table:tori}
    \end{table*}

{In Sec.\il(\ref{Sec:thinkc}) we discuss the model in some detail, developing  the analytical conditions  adopted  in the following analysis.}
 \subsection{{Thick accretion disk model in the Kerr spacetime}}\label{Sec:thinkc}
We focus on  general relativistic barotropic  toroidal configurations  which are geometrically thick and  optical opaque. The toroids,  composed by a  one particle-specie   perfect fluid, are  cooled by advection,  and are     axially symmetric and stationary. 
Exploring the  case of
 a one-species particle perfect  fluid (simple fluid),  each  torus is  described by  the  energy momentum tensor
\be\label{E:Tm}
T_{\alpha \beta}=(\rho +p) u_{\alpha} u_{\beta}+\  p g_{\alpha \beta},
\ee
where $\rho$ and $p$ are  the total energy density and
pressure, respectively, as measured by an observer moving with the fluid whose four-velocity $u^{\alpha}$  is
a timelike flow vector field.  (For the
symmetries of the problem, we always assume $\partial_t \mathbf{Q}=0$ and
$\partial_{\phi} \mathbf{Q}=0$, being $\mathbf{Q}$ a generic spacetime tensor.).

\medskip

From the  GRHD equations we find that the  fluid dynamics  is described by the \emph{continuity  equation} and the \emph{Euler equation} respectively:
\bea\label{E:1a0}
u^\alpha\nabla_\alpha\rho+(p+\rho)\nabla^\alpha u_\alpha=0\, ,\quad
(p+\rho)u^\alpha \nabla_\alpha u^\gamma+ \ h^{\beta\gamma}\nabla_\beta p=0\, ,
\eea
where the projection tensor $h_{\alpha \beta}=g_{\alpha \beta}+ u_\alpha u_\beta$ and $\nabla_\alpha g_{\beta\gamma}=0$.
The toroids  are centered on  the  plane $\theta=\pi/2$, and  defined by the constraint
$u^r=0$. (No
motion is assumed in the $\theta$ angular direction, which means $u^{\theta}=0$.)
We assume moreover a barotropic equation of state $p=p(\rho)$.

\medskip

The  continuity equation  in Eq.\il(\ref{E:1a0}) 
%
 %
is  identically satisfied as consequence of these conditions  and, from the Euler  equation (\ref{E:1a0}), one  finds
\bea\label{Eq:scond-d}&&
\frac{\partial_{\mu}p}{\rho+p}=-{\partial_{\mu }W}+\frac{\Omega \partial_{\mu}\ell}{1-\Omega \ell},\quad\mbox{with}\quad   \ell\equiv \frac{L}{E}
\\
&&\nonumber \mbox{where}\quad W\equiv\ln V_{eff}(\ell),\quad  \mbox{and }\quad  V_{eff}(\ell)=\sqrt{\frac{g_{\phi t}^2-g_{tt} g_{\phi \phi}}{g_{\phi \phi}+2 \ell g_{\phi t} +\ell^2g_{tt}}},
\eea
%
where the function $W$ is Paczynski-Wiita  (P-W) potential and  $V_{eff}(\ell)$ provides an effective potential for the fluid, assumed here  to be  characterized by a  conserved and constant specific angular momentum $\ell$.

\medskip

The procedure adopted  in the present article
borrows from the  Boyer theory on the equi--pressure surfaces applied to a  thick  torus:
  the Boyer surfaces are given by the surfaces of constant pressure  or   $\Sigma_{i}=$constant for \(i\in(p,\rho, \ell, \Omega) \), 
   where the angular frequency  is indeed $\Omega=\Omega(\ell)$ and $\Sigma_i=\Sigma_{j}$ for \({i, j}\in(p,\rho, \ell, \Omega) \). ({More generally $\Sigma_{\mathbf{Q}}$ is the  surface $\mathbf{Q}=$constant for any quantity or set of quantities $\mathbf{Q}$.})

\medskip

  The toroidal surfaces  are the equipotential surfaces of the effective potential  $V_{eff}(\ell) $, considered  as function of $r$,  solutions $ \ln(V_{eff})=\rm{c}=\rm{constant}$ or $V_{eff}=K=$constant.

\medskip

\textbf{{Extreme points of the co--rotating  toroids}}

In the first  and  simplest  realization of this   model each orbiting configuration is therefore  parametrized by  a   constant fluid specific angular momentum  $\ell$ (defined in Eq.\il(\ref{Eq:flo-adding})),
as a torus parameter \cite{abrafra,ringed}.
The model is  therefore regulated by the couple of parameters $\mathbf{p}\equiv (\ell,K)$ which together uniquely identify each equi-pressure (equi-potential) surface.

Therefore, from   Eq.\il(\ref{Eq:scond-d}), it is clear that the maximum  and minimum density points (the pressure gradients) in the disk, which are the toroids centers and cusps respectively,  are fixed 
by  the gradients  of a scalar, the fluid effective potential $V_{eff}$ in Eq.\il(\ref{Eq:scond-d}), depending on the  \textbf{BH} spin $a$, the fluid specific angular momentum  $\ell$, and the radial distance from the central attractor (the function is  evaluated on the toroid equatorial plane which is also the \textbf{BH} equatorial plane).

 Note, in general the curve $\ell^\pm_\sigma=$constant,  defined by
$\ell^\pm_\sigma:\partial_y V_{eff}^2=0$,
where   there is $\{z=r \cos\theta,y=r \sin \theta \sin\phi,x=r \sin\theta \cos\phi\}$, provides
 the  pressure (and density) extreme points on the equatorial plane  and counter--rotating and co--rotating  tori geometrical extremes respectively.

The extremes of  the pressure  are therefore regulated by  the angular momentum distributions, as there is  $\ell=\ell(r,\theta;a):\partial_r V_{eff}=0$  on the equatorial plane $\theta=\pi/2$.
Explicitly the specific angular momentum for the co--rotating and counter--rotating toroids are
\bea
&&
\ell^\mp=\frac{-a^3\pm\sqrt{r^3 \left(a^2+(r-2) r\right)^2}+a r (4-3 r)}{(r-2)^2 r-a^2}
\eea
respectively. (Note $\ell^\mp=\ell_\sigma^\mp$ for $\sigma=1$)

\medskip

\textbf{The $K$ parameter and the toroids}

There are three sets of configurations  defined by the  range of values of the momentum $\ell$,  and
$K=$constant  as described in Table\il\ref{Table:tori}.

The flows   considered in this analysis are defined by the   parameter $\ell$  for ``accretion  driven" and  ``proto-jets driven" flows,  having   $\mp\ell^\pm\in [\mp\ell_{mso}^\pm,\mp\ell_{mbo}^\pm]$  related  to  tori or, having   $\mp\ell^\pm\in [\mp\ell_{mbo}^\pm,\mp\ell_{mco}^\pm]$   as proto-jet configurations,  which are open toroidal surfaces   made by funnels of matter, along the  \textbf{BH} symmetry axis having a cusp on the equatorial plane   in the range $r_{J}^\pm\in[r_{mbo}^\pm,r_{mco}^\pm]$--\cite{proto-jets}.

The  tori and proto--jets cusps, corresponding to the minimum points of the hydrodynamic pressure, can be found from $V_{eff}$  of Eq.\il(\ref{Eq:scond-d}) , as the maximum points of the effective potential function with $K<1$ or $K>1$ respectively--see also Tabel\il\ref{Table:tori}.

The location of the configurations centers, $r_{center}^\pm$ (which can be easy found as minimum points of the effective potential $V_{eff}$ in Eq.\il(\ref{Eq:scond-d}) with $\ell=$constant), is constrained by the fluids specific angular momentum    $\ell$,  according   to    the radii  $r_{(mbo)}^{\pm}$ and $r_{(mco)}^{\pm}$, defined by the  relations:
\bea&&\nonumber
r_{\mathrm{(mbo)}}^{\pm}:\;\ell^{\pm}(r_{\mathrm{mbo}}^{\pm})=
 \ell^{\pm}(r_{\mathrm{(mbo)}}^{\pm})\equiv \mathbf{\ell_{\mathrm{mbo}}^{\pm}},\quad
  r_{(mco)}^{\pm}: \ell^{\pm}(r_{mco}^{\pm})=
  \ell^{\pm}(r_{(mco)}^{\pm})\equiv \ell_{mco}^{\pm} \quad \mbox{where} 
  \\&&\label{Eq:def-nota-ell}
r_{mco}^{\pm}<r_{\mathrm{mbo}}^{\pm}<r_{\mathrm{mso}}^{\pm}<
 r_{\mathrm{(mbo)}}^{\pm}<
  r_{(mco)}^{\pm}
  \eea
  as described in Table\il\ref{Table:tori}.
  Radii $(r_{\mathrm{(mbo)}}^{\pm},
  r_{(mco)}^{\pm})$
  form, with the geodesic radii, the \emph{extended geodesic structure} of the Kerr spacetime--Fig.\il\ref{Fig:Plotfondo}.

In the following it will be useful to specify the case for the co--rotating toroids.
The cusps  and center of co--rotating torus, $r_{\times}^-<r_{mso}^-<r_{center}^-$ or proto-jets $r_{J}^-<r_{mso}^-<r_{center}^-$ can be found   in terms of the parameter $\ell$, solving the equation
$\ell^-(r)=\ell$, and there is
\bea&&
r_{center}^-\equiv
\frac{\xi+\Xi}{12}\quad\mbox{and}\quad
r_{\times}^-\equiv \frac{\xi-\Xi}{12},\quad\mbox{where}
\\\nonumber
&&
\xi\equiv \sqrt{3} \sqrt{B_5}+3 \ell^2,\quad
\Xi\equiv \sqrt{6} \sqrt{\frac{3 \sqrt{3} \left(\ell^6-8 B \ell^2+32 B_1\right)}{\sqrt{B_5}}-16 B-\frac{8 \sqrt[3]{2} B_0}{\sqrt[3]{B_4}}-2^{2/3} \sqrt[3]{B_4}+3 \ell^4},
\eea
with
\bea&&
B\equiv a^2-3 a \ell+2 \ell^2, \quad B_0\equiv \left(2 a^2-3 a \ell+\ell^2\right)^2,\\&&\nonumber
B_1\equiv a^2-2 a \ell+\ell^2;
\\\nonumber
&&
 B_2\equiv-72 B B_1 \left(2 a^2+\ell^2\right)+27 B_1 \left(a^2 \ell^4+16 B_1\right)+16 B^3,
\quad B_3\equiv B_2^2-4 (4 B_0)^3,\quad B_4\equiv B_2+\sqrt{B_3},
\\\nonumber
&& B_5\equiv -16 B+\frac{16 \sqrt[3]{2} B_0}{\sqrt[3]{B_4}}+2\ 2^{2/3} \sqrt[3]{B_4}+3 \ell^4.
\eea

 The  closed  surfaces   cross in general  the equatorial plane in  three points, including the inner and outer edges of the closed configurations. The  equi-potential surfaces identify  therefore  the inner  and outer edges of the quiescent  (not cusped)  tori, as
 $r_{inner}<r_{center}<r_{outer}$.

%
%
%



It will be convenient  to use the notation
$\cc^{\pm}$ ($\cc_{\times}^{\pm}$) for {quiescent } (cusped) tori,    $\pp^{\pm}$  for  quiescent  or  cusped configurations, $r_{outer}$ is the torus outer edge, $r_\times$ for the torus cusp, and the notation $\Qa(\mathbf{L_\bullet})$  or $\Qa_\bullet$ for $\bullet\in\{1,2,3\}$ for quantity
$\Qa$ with momenta in the range $\mathbf{L_\bullet}$, according to Table\il\ref{Table:tori}.  Symbols $\lessgtr$
 for two tori refer to the relative position of the tori, for example, we use short notation $\cc_\times^-<\cc_\times^+$  for  $r_{\times}^-<r_{center}^-<r_{outer}^-<r_{\times}^+<r_{center}^+<r_{outer}^+$.
Since the  toroidal configurations  can be corotating  $\ell a>0 $ or counterrotating   $\ell a<0$, with respect to the black hole  $a>0$, then assuming    several  toroidal configurations,  say the couple $(\cc_a, \cc_b)$, with proper angular momentum $(\ell_a, \ell_b)$  orbiting  in   the equatorial plane of a given Kerr \textbf{BH},    they can be
 \emph{$\ell$corotating} disks,  defined by  the condition $\ell_{a}\ell_{b}>0$, or  \emph{$\ell$counterrotating} disks  by the relations   $\ell_{a}\ell_{b}<0$,  where  the two $\ell$corotating tori  can be both corotating $\ell a>0$ or counterrotating  $\ell a<0$ with respect to the central attractor.
%
%

%
\section{Inversion surfaces}\label{Sec:inversion-spheres}
\begin{figure*}
\centering
\includegraphics[width=18cm]{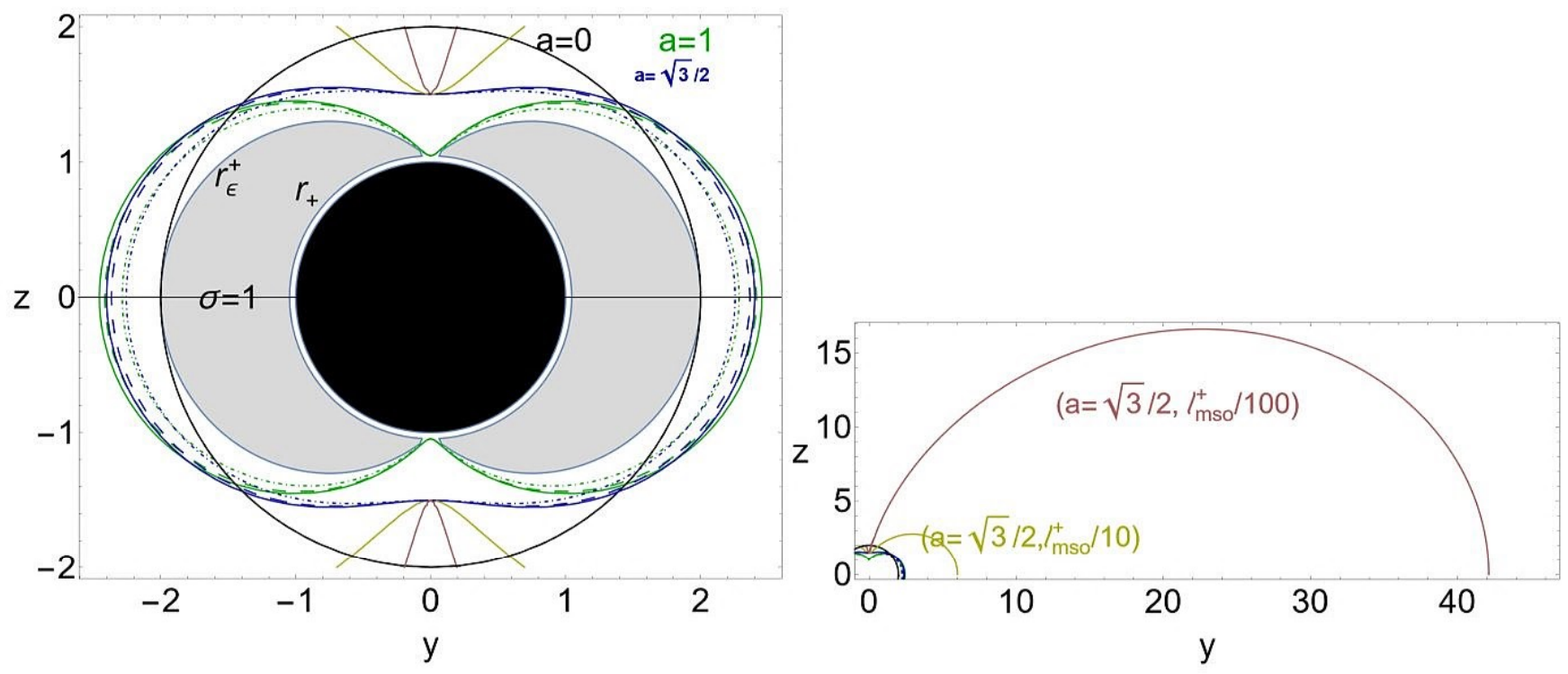}
   \caption{Inversion surfaces in the $z-y$ plane, where  $r= \sqrt{y^2+z^2}$ and $\theta=\arccos({z}/r)$. The counter--rotating flows  inversion points $r_{\Ta}$ are evaluated at fluid specific angular momenta $\ell_{mso}^+$ (plain) and $\ell_{mbo}^+$  (dashed), $\ell_{mco}^+$  (dotted-dashed).   (There is  $\mathrm{Q}_\bullet$ for any quantity  evaluated at $r_\bullet$, and
$mbo$ is for marginally bound circular orbit, $mco$ is for marginally  circular orbit, $mso$ is for marginally stable circular orbit.). All quantities are dimensionless.  Left  panel: green curves represent the case for \textbf{BH} dimensionless spin $a=1$ (extreme Kerr spacetime), black curve is  $a=0$ (Schwarzschild  spacetime), blue curve  is for  $a=\sqrt{3}/2$. Shaded gray region is  the outer ergoregion $]r_{+},r_\epsilon^+]$, for the \textbf{BH} with spin $a=1$, black region is the central \textbf{BH} with spin $a=1$.
   Right  panel:  is  a different view of the left panel for the spin and momenta signed on the panel.}\label{Fig:Plotrte4b}
\end{figure*}

\begin{figure*}
\centering
  \includegraphics[width=18cm]{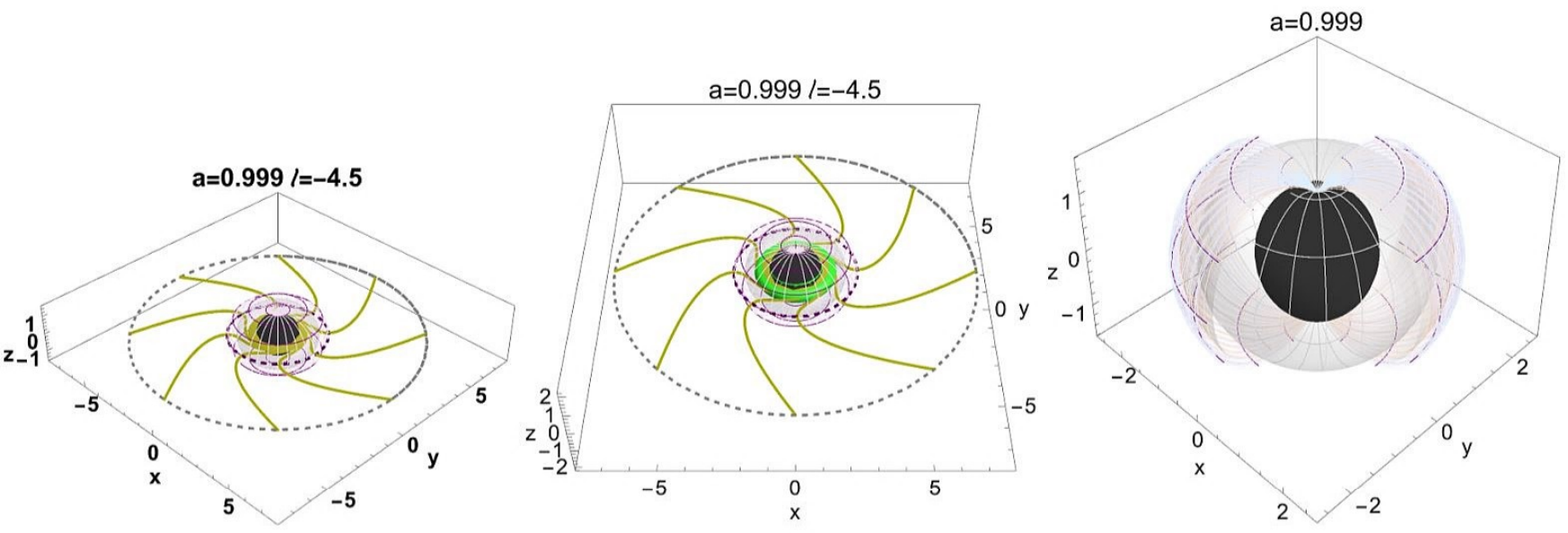}
           \caption{Left panel: tori driven counter--rotating  photons  inversion points in the \textbf{BH} spacetimes with dimensionless spin  $a=0.999$. The impact parameter (specific angular momentum) is    $\ell=-4.5$,  where there is $\{z=r \cos\theta,y=r \sin \theta \sin\phi,x=r \sin\theta \cos\phi\}$ {in dimensionless units}. The parameter $\ell$ is defined  in  Eq.\il(\ref{Eq:flo-adding}).  The panel shows a   view  of the  counter--rotating  flow stream from a torus ({equi-potential surface with $\ell=-4.5$}) inner edge (cusp-dashed gray curve)  to the central \textbf{BH} (black region $r<r_+$, radius $r_+$ is the outer horizon). Flow inversion point $r_{\Ta}$  is plotted as the deep-purple curve.
   Radius $r_{\Ta}$
   lies  in the  inversion corona defined by the
   range $(r_\Ta(\ell_{mso}^+)-r_\Ta(\ell_{mbo}^+))$. (There is  $\mathrm{Q}_\bullet$ for any quantity  evaluated at $r_\bullet$, and
$mbo$ is for marginally bounded circular orbit, $mso$ is for marginally stable circular orbit.). Gray region is the outer ergosurface, light-purple shaded region is the region $r< r_\Ta(\sigma_\Ta)$.  Right panel: Sections of the inversion spheres with radius $r_\Ta(\ell_{mso}^+)>r_\Ta(\ell_{mbo}^+)>
r_\Ta(\ell_{mco}^+)$  for $a=0.999$ (blue, purple and orange surfaces), gray region is the outer ergoregion, black region is the central \textbf{BH}. All quantities are dimensionless.}\label{Fig:Ppoblotrte4a}
\end{figure*}

\begin{figure*}
\centering
\includegraphics[width=18cm]{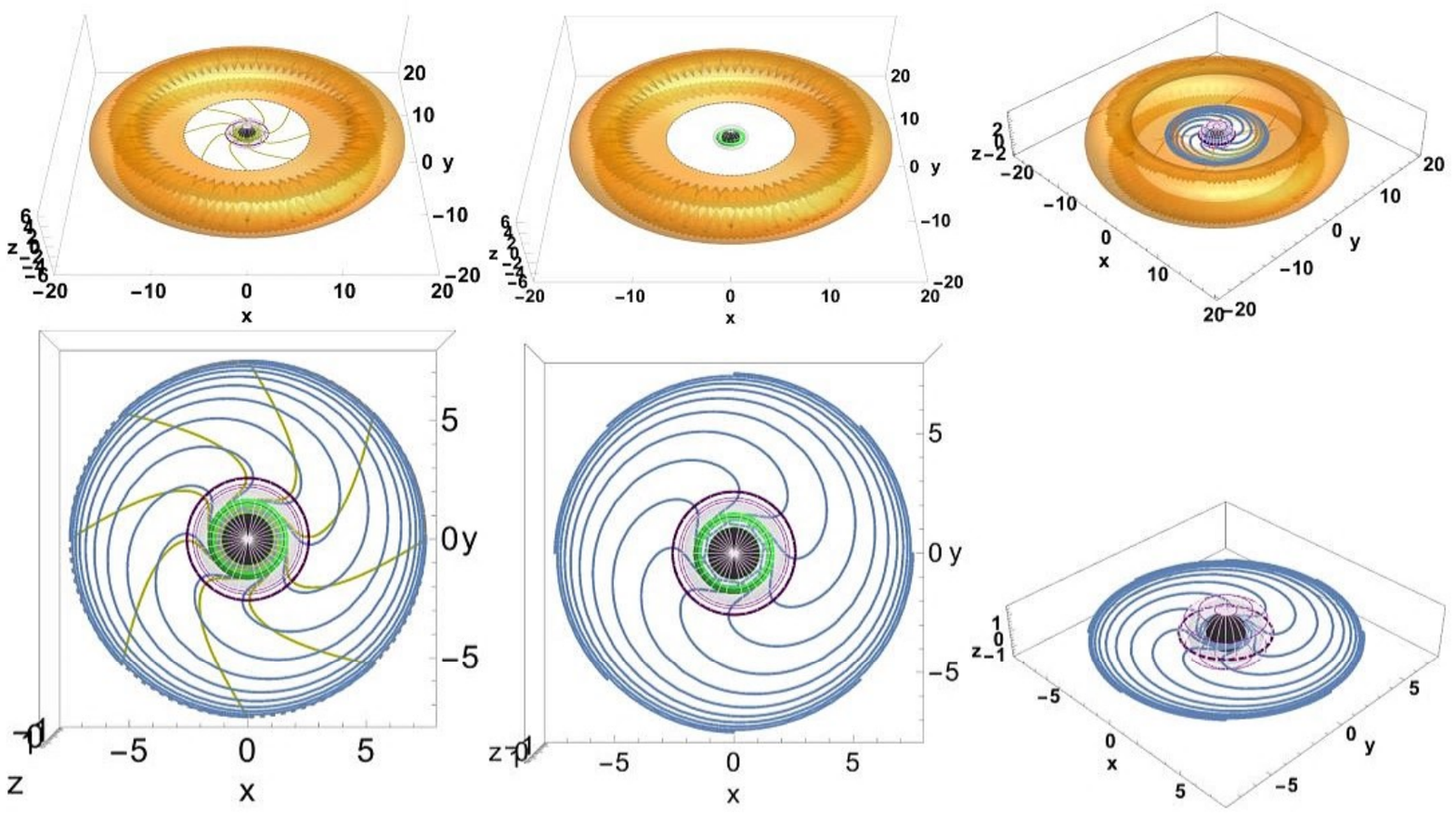}
   \caption{Tori driven counter--rotating  photons  inversion points in the \textbf{BH} spacetimes with dimensionless spin  $a=0.999$. The impact parameter (specific angular momentum) is    $\ell=-4.5$,  where there is $\{z=r \cos\theta,y=r \sin \theta \sin\phi,x=r \sin\theta \cos\phi\}$ {in dimensionless units}. The panels show a   view  of the  counter--rotating  flow stream from a torus inner edge (cusp-dashed gray curve)  to the central \textbf{BH} (black region $r<r_+$, radius $r_+$ is the outer horizon). {The torus is a cusped  equi-pressure surface  defined by the parameter  $\ell=-4.5$,}Orange outer (inner) surface is the counter--rotating (co--rotating) torus. Yellow (blue) curves are photons (test particles) trajectories. (See caption Fig.\il\ref{Fig:Ppoblotrte4a} for further details.)}\label{Fig:Plotrte4a}
\end{figure*}

Inversion  points are     defined by the  condition $u^{\phi}=0$ for particles and  photons--
\cite{retro-inversion}. It the context of the inversion points  it is convenient to
 use the notation
\emph{$u$counter--rotating} (\emph{$u$co--rotating}) for matter and photons with $u^\phi<0$ ($u^\phi>0$), along side with
 counter--rotating   for    $\ell<0$ (co--rotating   for    $\ell>0$).
Located out of the (outer) ergoregion, and totally embedding the central \textbf{BH}--see Fig.\il\ref{Fig:Plotrte4b}-- they are defined  for  counter--rotating trajectories    ($\ell<0$),  and    they can be interpreted as an effect of the Kerr  spacetime frame--dragging acting on the    $u$counter--rotating
particles (and   photons)  accreting into the central \textbf{BH}--Figs\il\ref{Fig:Ppoblotrte4a},\ref{Fig:Plotrte4a}.

\begin{figure*}
\centering
\includegraphics[width=5cm]{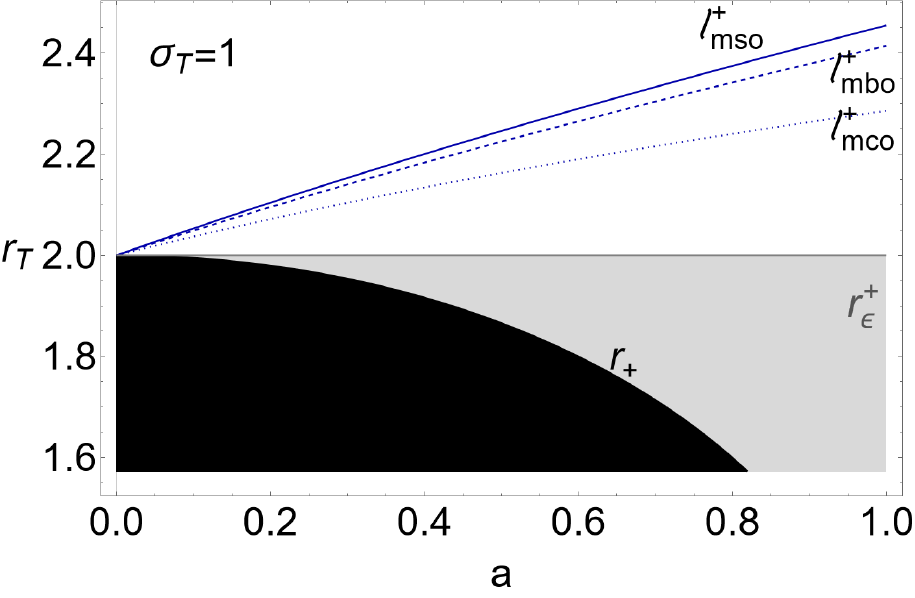}
\includegraphics[width=5cm]{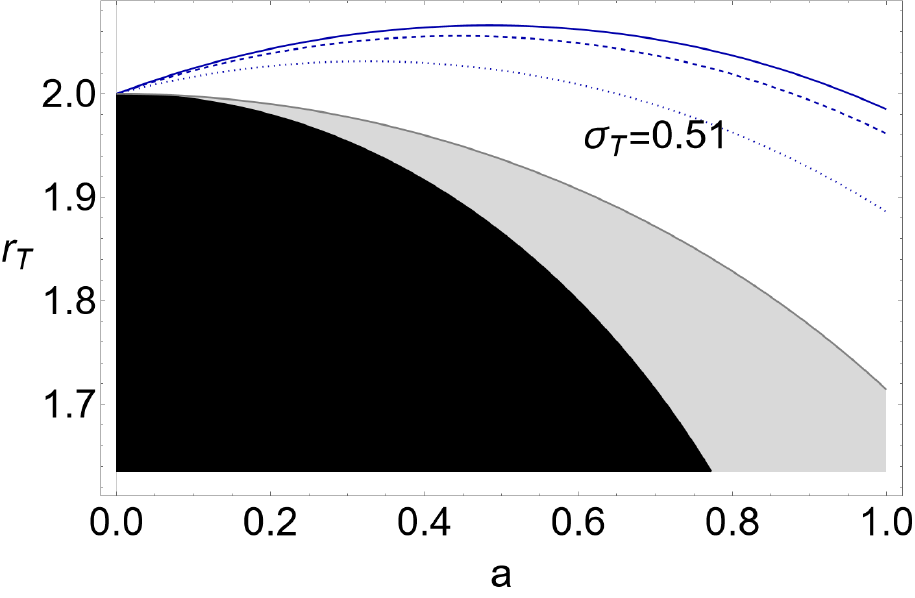}
\includegraphics[width=5cm]{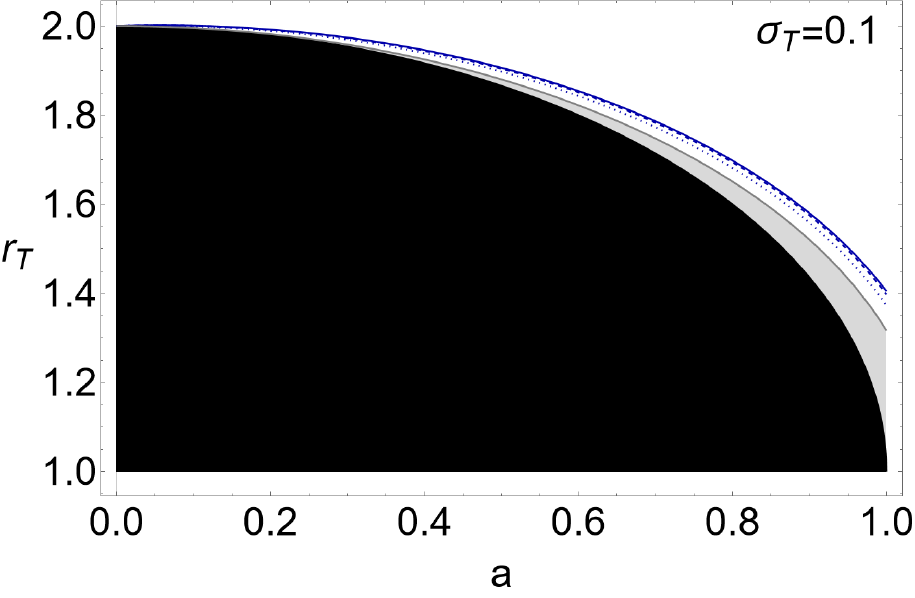}
\includegraphics[width=5cm]{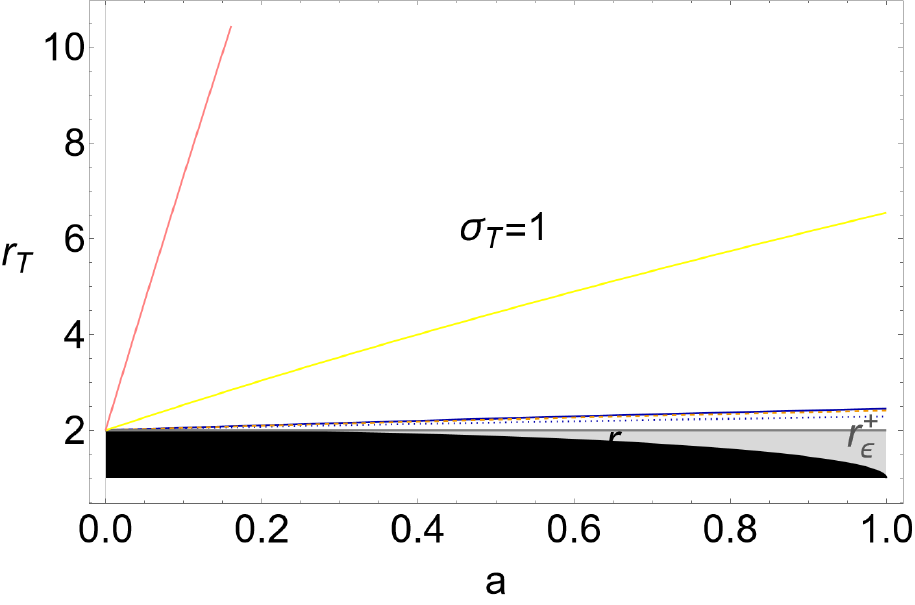}
\includegraphics[width=5cm]{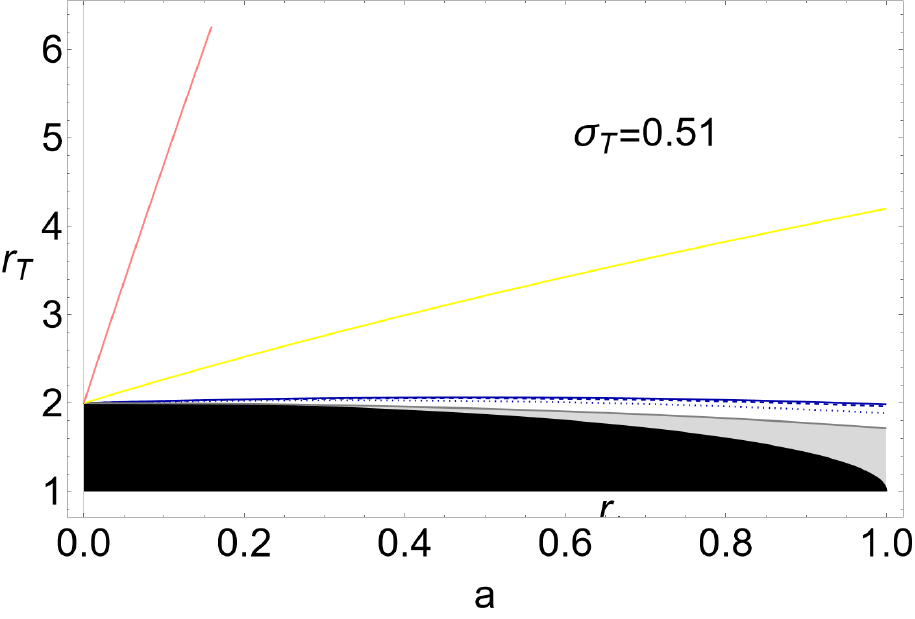}
\includegraphics[width=5cm]{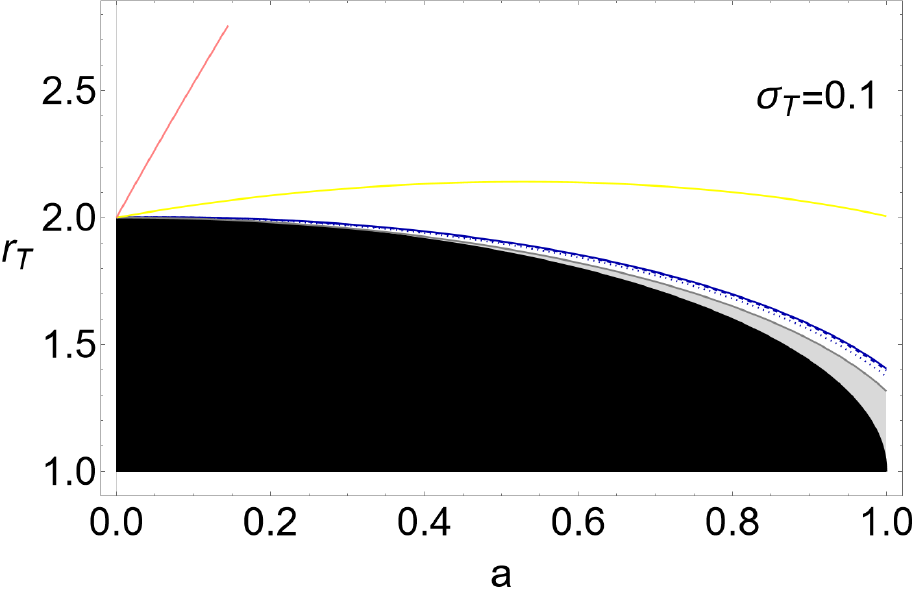}
\caption{ Inversion radius $r_\Ta$ (where $u^\phi$=0)  as function of the \textbf{BH} dimensionless  spin $a$ for different angles $\sigma\equiv \sin^2\theta$ signed on the panel and   different  $\ell^+$ (specific angular momentum) defined in Eq.\il(\ref{Eq:flo-adding}) for  counter--rotating $(+)$ orbits. There is  $\mathrm{Q}_\bullet$ for any quantity  evaluated at $r_\bullet$, where
$mbo$ is for marginally bound circular orbit, $mco$ is for marginally  circular orbit, $mso$ is for marginally stable circular orbit. Pink (yellow) curves are for  $\ell^+=\ell^+_{mso}/100$ ($\ell^+=\ell^+_{mso}/10$).  All quantities are dimensionless. Black region is the central \textbf{BH}, $r_+$ is the outer horizon, gray region is the outer ergoregion, radius $r_{\epsilon}^+$ is the outer ergosurface.}\label{Fig:PlotfondoT}
\end{figure*}

The inversion surfaces are defined by a general, necessary but not sufficient, condition,  $u^\phi=0$, on materials (including photons)  ingoing towards the \textbf{BH}, or outgoing or moving along the \textbf{BH} spinning axis\footnote{As clear  from  Figs\il\ref{Fig:PlotfondoT},\ref{Fig:Plotrte4b}, a particle, with  momentum $\ell^+$, moving along the \textbf{BH} vertical axis or, for sufficiently small angles  $\sigma$,  a particle  trajectory along  the $y$ axis  as  in Fig.\il\ref{Fig:Plotrte4b} could have, close to  \textbf{BH} poles, from two or even four inversion points, crossing in multiple points the inversion surface with  momentum $\ell^+$.  More in general,  multiple crossing points with  an inversion surface are possible-- \cite{retro-inversion}.  However, all the orbits  in the ergoregion must be $u$co--rotating  and therefore   $u$counter--rotating  materials, accreting with momentum $\ell^+$ into the central \textbf{BH}, will change toroidal velocity at the inversion surface  with momentum  $\ell^+$.}. However, an initially $u$counter--rotating flow of matter and  photons accreting into the central \textbf{BH} \emph{{will}}  have an inversion point at the inversion surface, as defined by its (conserved) specific angular momentum $\ell<0$.

At the inversion point there is the inversion radius  $r=r_\Ta$ on a plane $\sigma=\sigma_\Ta$:
\bea&&\label{Eq:add-equa-sigmata}
 r_{\Ta}\equiv \sqrt{a^2 \left(\sigma_\Ta +\frac{\sigma_\Ta ^2}{\ell^2}-1\right)-\frac{2 a \sigma_\Ta }{\ell }+1}-\frac{a \sigma_\Ta }{\ell }+1
 \quad
 \mbox{where}\quad
\lim\limits_{\ell\rightarrow -\infty} r_{\Ta}=r_{\epsilon}^+,\quad \lim\limits_{a\rightarrow 0} r_{\Ta}=\lim\limits_{\sigma\rightarrow 0} r_{\Ta}=r_{+},
\eea
(see  Figs\il\ref{Fig:PlotfondoT},\ref{Fig:Plotrte4b}).

In the following,  we use the notation $\mathrm{Q}_\Ta$  for any quantity $\mathrm{Q}$ considered at the inversion point, and therefore on the inversion point there is  $\ell=\ell_\Ta$, $\sigma=\sigma_\Ta$, $\dot{t} =\dot{t}_\Ta $ and
\bea\label{Eq:lLE}
&&\ell_\Ta\equiv \left.-\frac{g_{t\phi}}{g_{tt}}\right|_\Ta,\quad \Em_\Ta = -g_ {tt}(\Ta) \dot{t}_\Ta=\frac{a \sigma_\Ta  \dot{t}_\Ta}{a \sigma_\Ta -\ell },\quad \La_\Ta = g_ {t\phi}(\Ta)\dot{t}_\Ta=\frac{a \sigma_\Ta  \dot{t}_\Ta \ell }{a \sigma_\Ta -\ell }. 
\eea

At fixed $\ell<0$,  radius $r_\Ta(\sigma_\Ta)$ defines the \emph{inversion surface} $\Sa_\Ta$ of momentum $\ell$ and a region, \emph{inversion sphere},  embedding  the central attractor, upper  bounded by the inversion  surface and bottom bounded by the \textbf{BH} outer ergosurface. The photon or particle inversion  point $(r_\Ta,\theta_\Ta,\phi_\Ta)$ on the inversion surface  can be determined by the set of equations (\ref{Eq:eqCarter-full}) which also  relates $(r_\Ta,\sigma_\Ta)$ to the initial values $(r_0, \sigma_0)$,   depending  on the single particle trajectory.

Hence, the   region
defined by  the radii $r_\Ta(\ell^+_\bullet)$ and $r_{\epsilon}^+(a,\sigma)$ contains  all the inversion surfaces
$\Sa_\Ta(\ell^+_\star)$ with $\ell^+_\star\leq \ell^+_\bullet$.
 Therefore we  can  define an \emph{inversion  corona}, as the   shell   bounded  by  the   inversion surfaces of    radius $r_\Ta(\sigma_\Ta)$ in the range $[r_\Ta(\ell_{mbo}^+)],r_\Ta(\ell_{mso}^+)]$, for  \emph{ accretion driven  counter--rotating flows},  and  $[r_\Ta(\ell_{mbo}^+)],r_\Ta(\ell_{mco}^+)]$ for  \emph{proto-jets  driven counter--rotating flows}, where
 there is $r_{\Ta}(\ell_{mso}^+)>r_{\Ta}(\ell_{mbo}^+)>r_{\Ta}(\ell_{mco}^+)$--Fig.\il\ref{Fig:PlotfondoT}.
Surface $\Sa_\Ta$, as well as the radius $r_\Ta$, are uniquely   defined by the momentum $\ell^+_\Ta$, and we can use,  with no ambiguity, notation $r_{\Ta}(\ell_\Ta^+)$ for  the surface  $\Sa_\Ta$ with momentum
 $\ell_\Ta^+$.
 In the following with  radius $r_{\Ta}(\mathbf{L_\bullet^+})$ we also  indicate  any inversion surface $\Sa_\Ta$ (and inversion radius $r_\Ta$) defined by a momentum in the range $\mathbf{L_\bullet^+}$, and we equally indicate the entire set of inversion surfaces with momentum in $\mathbf{L_\bullet^+}$. Hence, according to this notation, there is
$r_{\Ta}(\mathbf{L_3^+})< r_{\Ta}(\mathbf{L_2^+})< r_{\Ta}(\mathbf{L_1^+})$.
A toroid $\pp_\bullet$  with  momentum in the range $\mathbf{L_\bullet}$, is   contained in (or embedded in, or internal  to)  an inversion surface  $r_\Ta^+$ if $\pp_\bullet<r_\Ta^+$, and it is external to  an inversion surface   if $\pp_\bullet>r_\Ta^+$.

A configuration, (entirely) contained in such  inversion surface,  is therefore shielded from impact with $u$counter--rotating material or photons  accreting into  the central \textbf{BH}.

The distance, $r_\Ta (\ell_
{mso}^+) -
 r_\Ta (\ell_ {mbo}^+) $, increases not monotonically  with the \textbf{BH} spin--mass ratio and with the plane $\sigma$--see Fig.\il\ref{Fig:PlotfondoT} and Figs\il\ref{Fig:Plotrte4a},\ref{Fig:Plotrte4b}--\cite{retro-inversion}.
Radii ($r_\Ta(\ell_{mco}^+),r_\Ta(\ell_{mbo}^+),r_\Ta(\ell_{mso}^+)$) vary   little with   the \textbf{BH} spin and plane $\sigma$.  Therefore,  the flow  in this region  is  actually located in a  restricted orbital range $(r_{\Ta},\sigma_{\Ta})$, localized  in the  orbital cocoon surrounding the central attractor outer ergosurface\footnote{The flow reaches the inversion surface at  different times $t_\Ta$ depending  on the initial data. It could  be  expected    the inversion   corona  to  be    observable    by an increase of   flow  temperature and  luminosity  (depending on the  $t_\Ta$ values).}) (see for example Figs\il\ref{Fig:Plotrte4b}--left panel).

All the co--rotating  toroids orbiting in the \textbf{BH} outer ergoregion (totally contained in the ergoregion) are embedded in all the  inversion surfaces and
will \emph{not} be  in contact with the $u$counter--rotating flows.
In Fig.\il\ref{Fig:corsozoom1}--left panel is an example of {inter--disks} inversion surface in the  system
$\cc^-<\Sa_\Ta(\ell^+)<\cc^+_\times$, where the inner co--rotating torus orbits  in the \textbf{BH} ergoregion\footnote{Fig.\il\ref{Fig:corsozoom1}--left panel  and Figs\il\ref{Fig:corso}--left panel show the co--rotating geodesic radii crossing the outer ergosurface, constraining this case.}.
In the next section  we will
explore the   case of a  co--rotating toroid  and an   inversion surface with momenta $\ell^+<\ell_{mso}^+$,   focusing  on  the  inter--surfaces inversion surfaces,  i.e.  the case $\pp^-<\Sa_\Ta<\pp^+$,
 the situation in which the inner  surface is co--rotating, and the counter--rotating fluids originate from (are related to) an outer counter--rotating accreting disk\footnote{We will  assume, in this case,   equal   specific momentum for the inversion surface, the accreting flow and the counter--rotating  accretion disk.}. In particular we shall examine the case of a quiescent or cusped  disks.
The case  of inversion surfaces in relation to  a double orbiting configurations places additional limits  on the \textbf{BH} spin when considered in relation to the presence of an inner  co--rotating surface.
However, we also discuss the case where one or two orbiting structures are proto--jets (inter proto-jets inversions surfaces).
\section{Inversion surfaces and inner co--rotating accretion disks}\label{Sec:center-Tori-crossing}
In this section we  focus on  the co--rotating  toroids orbiting a central Kerr \textbf{BH}   in relation to the spacetime  inversion surfaces.
The   analysis ultimately  provides the conditions when a co--rotating disk or proto--jet can collide with and be  replenished by  $u$counter--rotating (counter--rotating) matter and photons accreting into the central \textbf{BH} or, viceversa  when,  counter--rotating matter and photons impacting on the surface   will be $u$co--rotating.
This condition  is strictly constrained by
  the  flow parameters $\ell^\pm$, for the inversion surface, the orbiting toroids (disks or proto--jets), and the \textbf{BH} spin $a$.
In particular,  depending on the  parameters  $\ell$ and $K$ , disks and proto-jets  can be:  \emph{1.} embedded in an inversion surface ($\pp^-<\Sa_\Ta$);  \emph{2.}  located out of  the inversion surface ($\Sa_\Ta<\pp^-$);  or   \emph{3.} crossing  the inversion surfaces  in particular points (for example  the center, the geometrical maximum), being partially contained in  the inversion surface.

These three scenarios will be  considered as framework for the  analysis of Sec.\il(\ref{Sec:inversion-crossing}), where  we will investigate
 the  inversion surfaces crossing    the     inner co--rotating  toroids.
Some preliminary aspects  are examined  in Sec.\il(\ref{Sec:Aspects}), where characteristics  of the Kerr spacetimes  inversions surfaces will be  discussed  in relation to the  toroids orbiting  the \textbf{BH}, distinguishing  \textbf{BH} classes   by the properties of the inversion surfaces in relation to co--rotating toroids.
\subsection{BHs and inter--disks  inversion surfaces}\label{Sec:Aspects}
\begin{figure*}
\centering
\includegraphics[width=8cm]{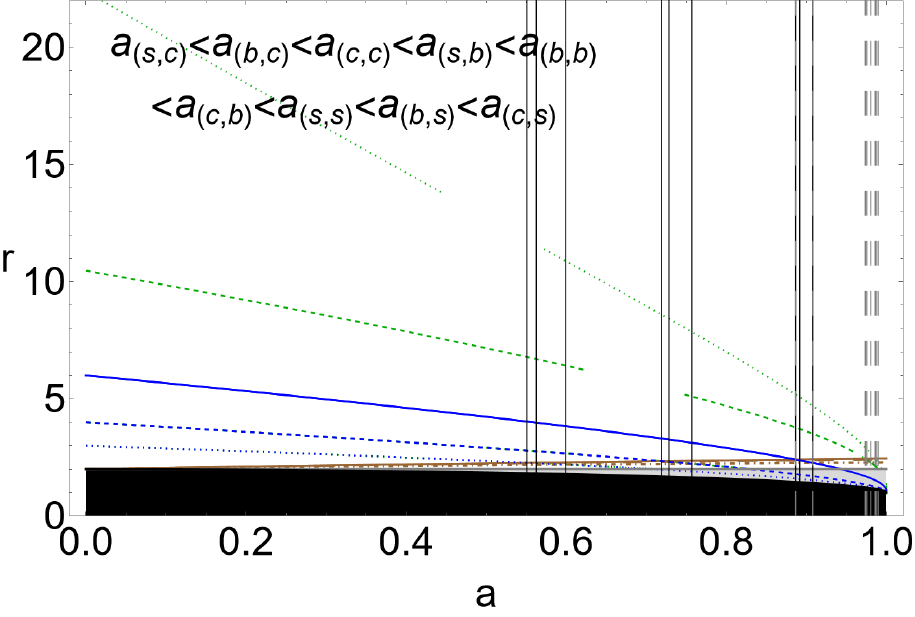}
\includegraphics[width=8cm]{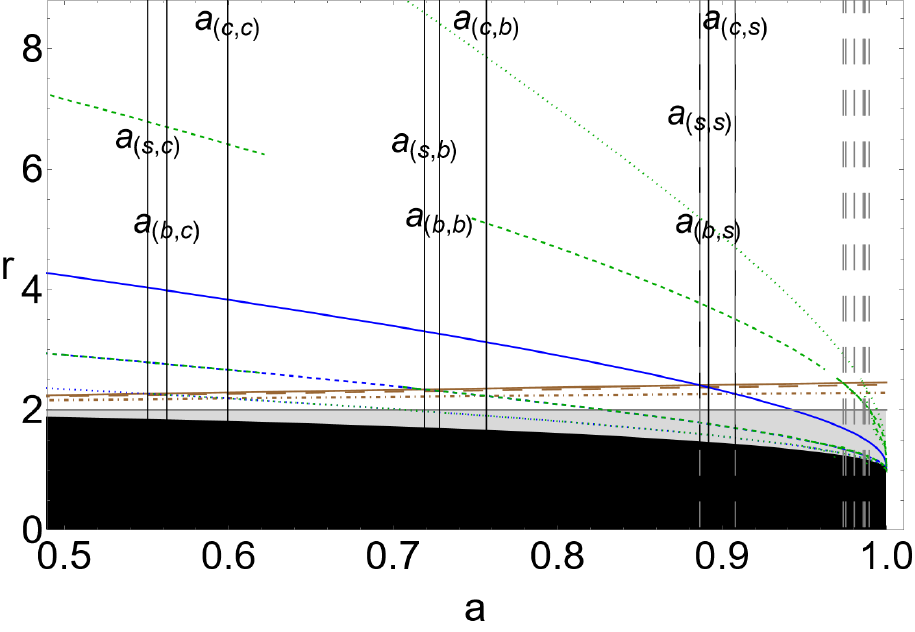}
\includegraphics[width=8cm]{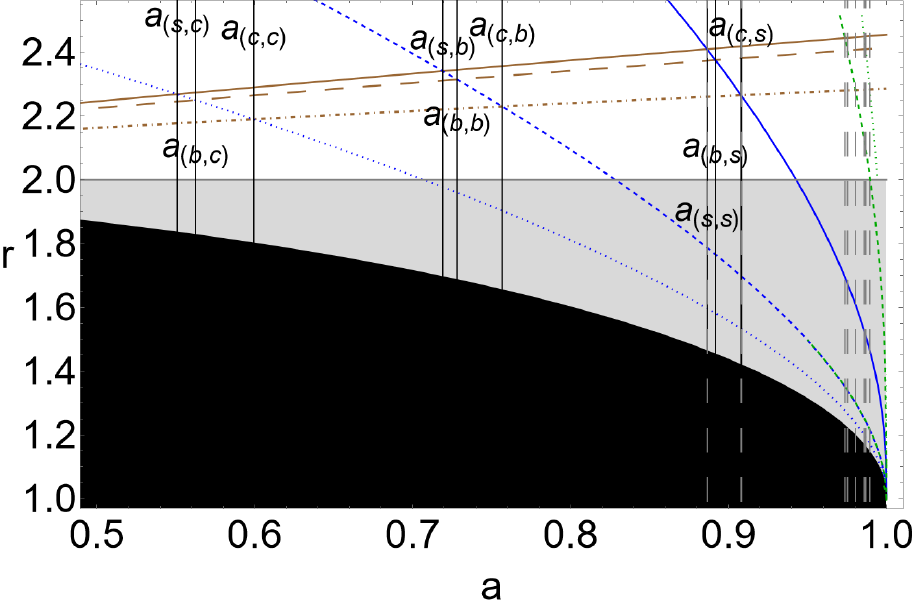}
\includegraphics[width=8cm]{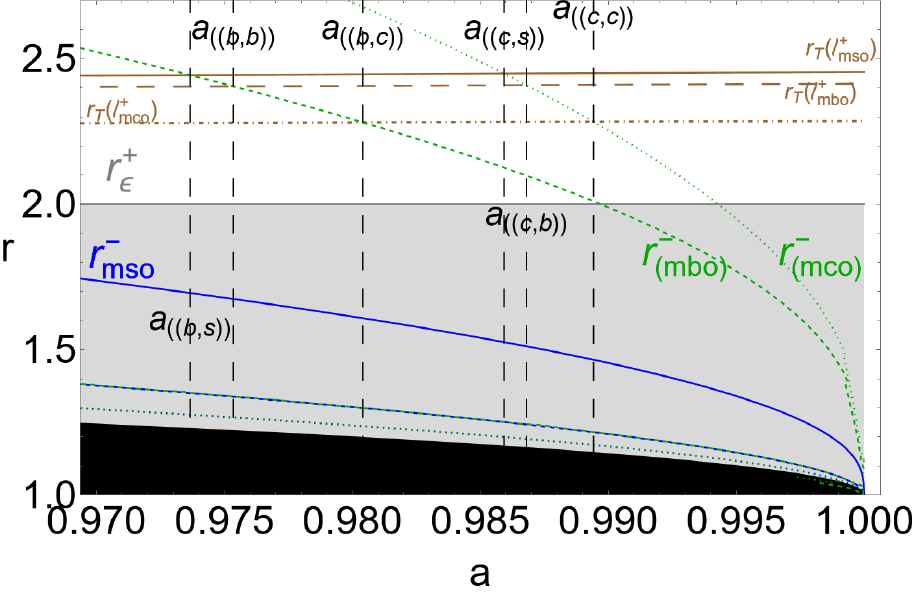}
\caption{Brown lines are the inversion radii $r_{\Ta}(\ell_{mso}^+)>r_{\Ta}(\ell_{mbo}^+)>r_{\Ta}(\ell_{mco}^+)$, where $\ell^\pm$ are the counter--rotating and co--rotating specific angular momentum respectively. Black region is the central \textbf{BH} with dimensionless spin $a$. Gray region is the outer ergoregion on the equatorial plane. (Radius $r_\epsilon^+$ is the outer ergosurface.)  Vertical lines are spins defined  in Table\il\ref{Table:spins} (see upper left panel for the vertical solid lines).  Upper right and bottom right and left panels are close--up view of the upper left panel. Blue dotted, blue dashed, blue solid and green dashed, green dotted are  the radii $r_{mco}^-<r_{mbo}^-<r_{mso}^-<r_{(mbo)}^-<r_{(mco)}^-$ defined in Eqs\il(\ref{Eq:def-nota-ell}). $mbo$  is for marginally bound circular orbit, $mco$   is for marginally  circular orbit, $mso$  is for marginally stable circular orbit. All quantities are dimensionless.}.\label{Fig:corso}
\end{figure*}

\begin{figure*}
\centering
\includegraphics[width=8cm]{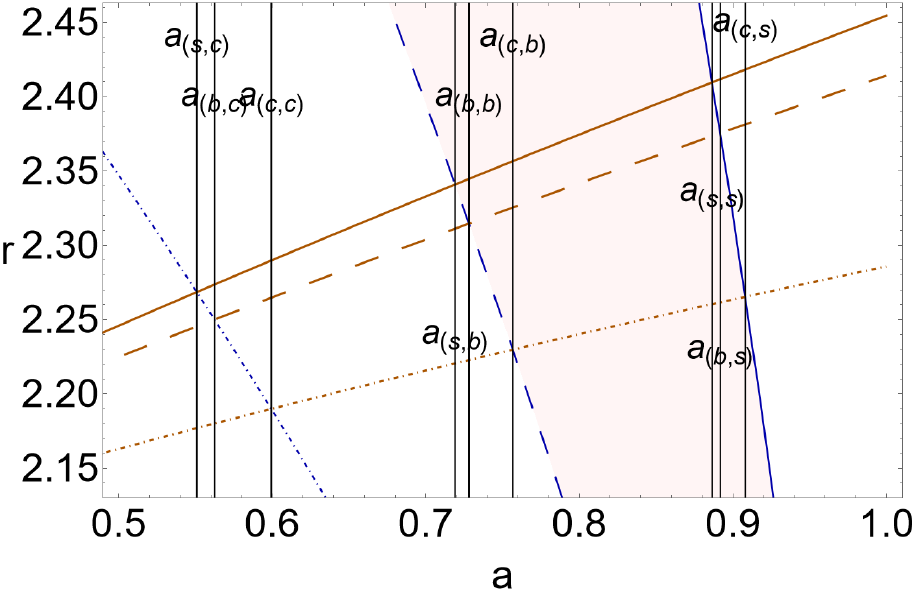}
\includegraphics[width=8cm]{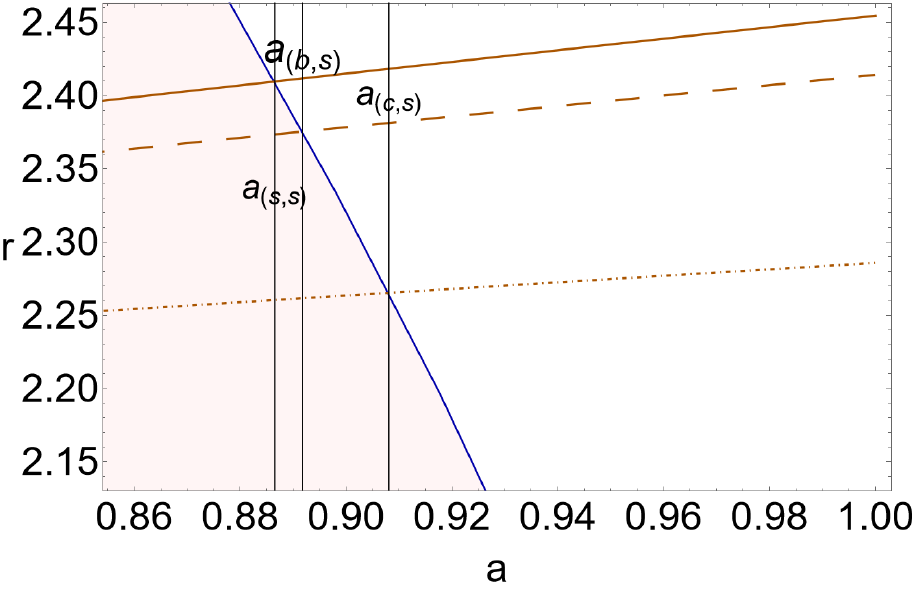}
\caption{Brown solid, dashed and dotted--dashed curves are the inversion radii $r_{\Ta}(\ell_{mso}^+)>r_{\Ta}(\ell_{mbo}^+)>r_{\Ta}(\ell_{mco}^+)$ respectively, where $\ell^\pm$ are the counter--rotating and co--rotating specific angular momentum respectively. Vertical lines are the spins defined in Table\il\ref{Table:spins}-(-see for further details also  caption of Fig.\il\ref{Fig:corso} ($mbo$  is for marginally bound circular orbit, $mco$   is for marginally  circular orbit, $mso$  is for marginally stable circular orbit.  Pink range is the range $[r_{mbo}^-,r_{mso}^-]$  for the location of the co--rotating   disk cusp. Blue dotted-dashed, dashed and solid lines are $r_{mco}^-<r_{mbo}^-<r_{mso}^-$. Right panel is a close--up view of the left panel.  All quantities are dimensionless.}.\label{Fig:corso1}
\end{figure*}

As there is
$r_\Ta(\ell_{mso}^+)<r_{mco}^+$--see Figs\il\ref{Fig:Plotfondo},\ref{Fig:corso},\ref{Fig:corso1}-- the cusped and  quiescent  counter--rotating  tori and  proto--jets are always external to  inversion surfaces\footnote{The discussion of the off--equatorial case constrains the condition  $\Sa_\Ta(\ell^+)<\pp^+$, on planes different from the equatorial. This case is investigated in Sec.\il(\ref{Sec:off-equatorial--crossing}).} (i.e. there is $r_\Ta(\ell^+)<\pp^+$).  Hence,  we  concentrate on  the  inner co--rotating toroids\footnote{The existence of an aggregate of orbiting toroids is constrained by the properties of the spacetime geodesic structures according to the constraints in
Table\il\ref{Table:tori}--Fig.\il\ref{Fig:Plotfondo}--\cite{ringed,letter,open,dsystem,long}.
As clear from Fig.\il\ref{Fig:Plotfondo}, depending on the relative parameters $(\ell^\pm,K^\pm)$,  there can always be  a double system with an inner co--rotating toroid  and an outer counter--rotating toroid. This situation is particularly characteristic of the fast spinning \textbf{BHs} spacetimes according to co--rotating and counter--rotating geodesic structure--Fig.\il\ref{Fig:Plotfondo}--\cite{dsystem}}
in a double orbiting system (that is the case   $\pp^-<\pp^+$).

\begin{figure*}
\centering
\includegraphics[width=8cm]{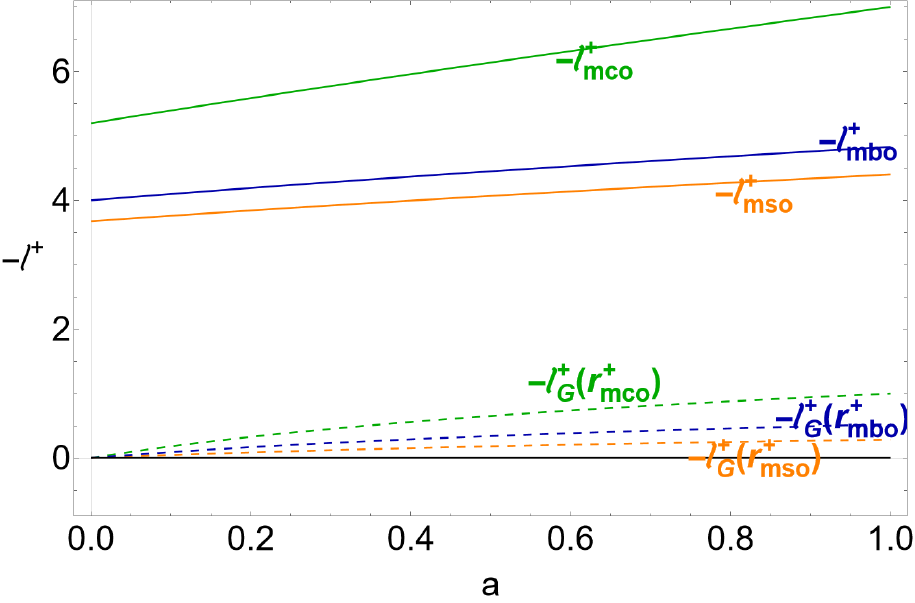}
\includegraphics[width=8cm]{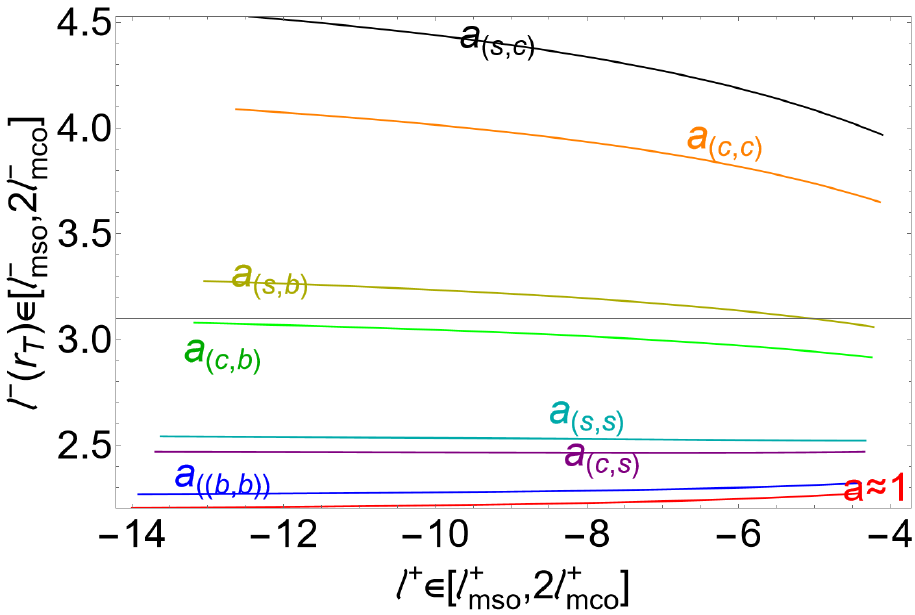}
\caption{Left panel: Specific angular momentum $\ell_G^+ $ defined in Eq.\il(\ref{Eq:LG}). $a$  is the \textbf{BH}  dimensionless spin.  $\ell^\pm$ are the counter--rotating and co--rotating specific angular momentum respectively.  $mbo$  is for marginally bound circular orbit, $mco$ is for marginally  circular orbit, $mso$  is for marginally stable circular orbit.  Right panel: momentum  $\ell^-$ is evaluated on the inversion radii $r_\Ta(\ell^+)$ versus $\ell^+$  for different spins defined in Table\il\ref{Table:spins} signed on the panel. The panel shows the momenta $\ell^\pm$ of the inversion surface and the inner co--rotating torus, where  the inversion surface, with momenta   $\ell^+$, crosses the co--rotating surface center or cusp at momenta $\ell^-$. All quantities are dimensionless. }\label{Fig:Plotsigi}
\end{figure*}

Our analysis is   limited to    the counter--rotating   flows and   the inversion surfaces  with  momenta in the range
 $-\ell^+>-\ell^+_{mso}$. However,  for small momentum  in magnitude,  the inversion surfaces  can always cross   the counter--rotating geodesic structure radii (and therefore can intersect  counter--rotating toroids or even be external to a counter--rotating toroid)--Fig.\il\ref{Fig:Plotfondo}.
This case occurs for  inversion surfaces momenta
 $\ell^+\in[\ell_G^+(r_{mco}^+),0[$,  where
\bea\label{Eq:LG}
 \ell_G^+(r)\equiv \frac{2 a}{2-r},\quad\mbox{with} \quad \ell_G^+(r): r=\left.r_\Ta^+\right|_{\sigma=1}
\eea
 where
  $\ell_G^+(r_{mco}^+)>\ell_{mso}^+$
-- (see Fig.\il\ref{Fig:Plotsigi} and  Figs\il\ref{Fig:Plotfondo},\ref{Fig:PlotfondoT},\ref{Fig:Plotrte4b} (pink and  yellow curves)).

On the other hand, part of our considerations  focuses  on the inversion surfaces properties   on the equatorial plane, discussing the location of the inversion radius $r_{\Ta}(\ell^+)$  in relation  to the (co--rotating)  geodesic structure.
However, conditions defined  on the equatorial plane may not be satisfied at  $\sigma<1$, for example this could be the case  for a  partially contained toroids, crossing  an inversion surface on planes other than the equatorial.
The off--equatorial properties are determined by the characteristics of the specific accretion disk model and regulated  by the  inversion surfaces characteristics,   which  are fixed exclusively by the background geometry. (The inversion surfaces  ``maximum  radial elongation" occurs on the equatorial plane, the maximum  radial distance from the central \textbf{BH} is $r\approx 2.45455$ --Fig.\il\ref{Fig:Plotrte4b}.)
 We address the off--equatorial case  in details in Sec.\il(\ref{Sec:off-equatorial--crossing}).

\begin{figure*}
\centering
\includegraphics[width=14cm]{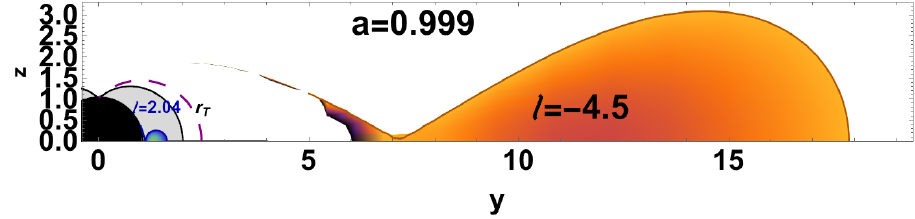}
\caption{Black region is the central \textbf{BH} with dimensionless spin $a=0.999$.
 Gray region is the outer ergoregion. Orange (blue) surface, is outer  counter--rotating (inner co--rotating) torus {(equi-potential surface  with fixed specific angular momentum parameter)}. $\ell$, signed on each toroid, is the fluid specific angular momentum.   Dashed curve  is  the inversion surface  $r_\Ta$ at fixed  torus counter--rotating angular momentum $\ell^+$.  There is   $r= \sqrt{y^2+z^2}$ and $\theta=\arccos({z}/r)$.   (All quantities are dimensionless.)}\label{Fig:corsozoom1}
\end{figure*}

The inversion surfaces radii $r_{\Ta}$ are shown in Fig.\il\ref{Fig:corso}  and Fig.\il\ref{Fig:corso1}  in relation to the  co--rotating geodesic structure for different  \textbf{BH} spins. We     distinguish sixteen \textbf{BHs}  classes, defined by the spins  in Table\il\ref{Table:spins} and  shown in Figs\il\ref{Fig:corsozoom1},\ref{Fig:corso},\ref{Fig:corso1}.
The boundary  spins  are defined by the intersection between the inversion surfaces radii $r_\Ta(\ell^+_\bullet)$ for  $\ell^+_\bullet\in\{\ell^+_{mso},\ell^+_{mbo},\ell^+_{mco}\}$ with the co--rotating  extended  geodesic structure --see Figs\il\ref{Fig:corso},\ref{Fig:corso1}.

Therefore, we introduce the notation
\bea\label{Eq:defini-spins-notation}
a_{(\hat{\bullet},\hat{\star})}: r_\Ta(\ell_\bullet^+)=r_{\star}^-\quad
\mbox{and}\quad
a_{((\hat{\bullet},\hat{\star}))}: r_\Ta(\ell_\bullet^+)=r_{(\star)}^-,\quad \mbox{where}\quad \{(\hat{\bullet},{\bullet}),(\hat{\star},{\star})\}\in\{(s,mso),(b,mbo),(c,mco)\},
\eea
 and   we can  define the  spins   $a_{(\bullet,s)}=\{a_{(s,s)},a_{(b,s)},a_{(c,s)}\}$, where the following relations hold:
\bea
a_{(s,\star)}<a_{(b,\star)}<a_{(c,\star)}\quad\mbox{and}\quad a_{((b,\star))}<a_{((c,\star))} \quad\mbox{for}\quad \star\in \{s,b,c\}.
 \eea
 \begin{table*}
 \centering
 \resizebox{.95\textwidth}{!}{%
\begin{tabular}{l|ll}
 \hline
 $a_{(c,s)}=0.908096:r_\Ta(\ell_{mco}^+)=r_{mso}^-$ & $
a_{(c,b)}=0.75673:r_\Ta(\ell_{mco}^+)=r_{mbo}^-$ & $
a_{(c,c)}=0.599433:r_\Ta(\ell_{mco}^+)=r_{mco}^-$
\\
$
a_{(b,s)}=0.891799:r_\Ta(\ell_{mbo}^+)=r_{mso}^-$ & $
a_{(b,b)}=0.728163:r_\Ta(\ell_{mbo}^+)=r_{mbo}^-$ & $
a_{(b,c)}=0.5625:r_\Ta(\ell_{mbo}^+)=r_{mco}^-$,
\\
$a_{(s,s)}=0.886532:r_\Ta(\ell_{mso}^+)=r_{mso}^-$ &
$a_{(s,b)}=0.719111:r_\Ta(\ell_{mso}^+)=r_{mbo}^-$ & $
a_{(s,c)}=0.550895:r_\Ta(\ell_{mso}^+)=r_{mco}^-$
\\
\hline
$a_{((c,c))}=0.989421: r_\Ta(\ell_{mco}^+)=r_{(mco)}^-$ & $
a_{((c,b))}=0.986806:r_\Ta(\ell_{mbo}^+)=r_{(mco)}^-$ & $
a_{((c,s))}=0.985935:r_\Ta(\ell_{mso}^+)=r_{(mco)}^-$
\\
$a_{((b,c))}=0.98041: r_\Ta(\ell_{mco}^+)=r_{(mbo)}^-$
&
 $
a_{((b,b))}=0.975354: r_\Ta(\ell_{mbo}^+)=r_{(mbo)}^-$
&
 $a_{((b,s))}=0.973659: r_\Ta(\ell_{mso}^+)=r_{(mbo)}^-$
\\
\hline
  \end{tabular}}
  \caption{Spins of Eqs\il(\ref{Eq:defini-spins-notation}) shown in Figs\il\ref{Fig:corsozoom1},\ref{Fig:corso},\ref{Fig:corso1}. $\ell^\pm$ is the specific fluid angular momentum for counter--rotating and co--rotating fluids respectively.
$mbo$ is for marginally bounded circular orbit, $mco$ is for marginally  circular orbit, $mso$ is for marginally stable circular orbit.   Radii   $r_{(mbo)}^{\pm}$ and $r_{(mco)}^{\pm}$ are defined in Eq.\il(\ref{Eq:def-nota-ell}).  $r_\Ta$ is the inversion radius.}
\label{Table:spins}
    \end{table*}
\begin{figure*}
\centering
\includegraphics[width=18cm]{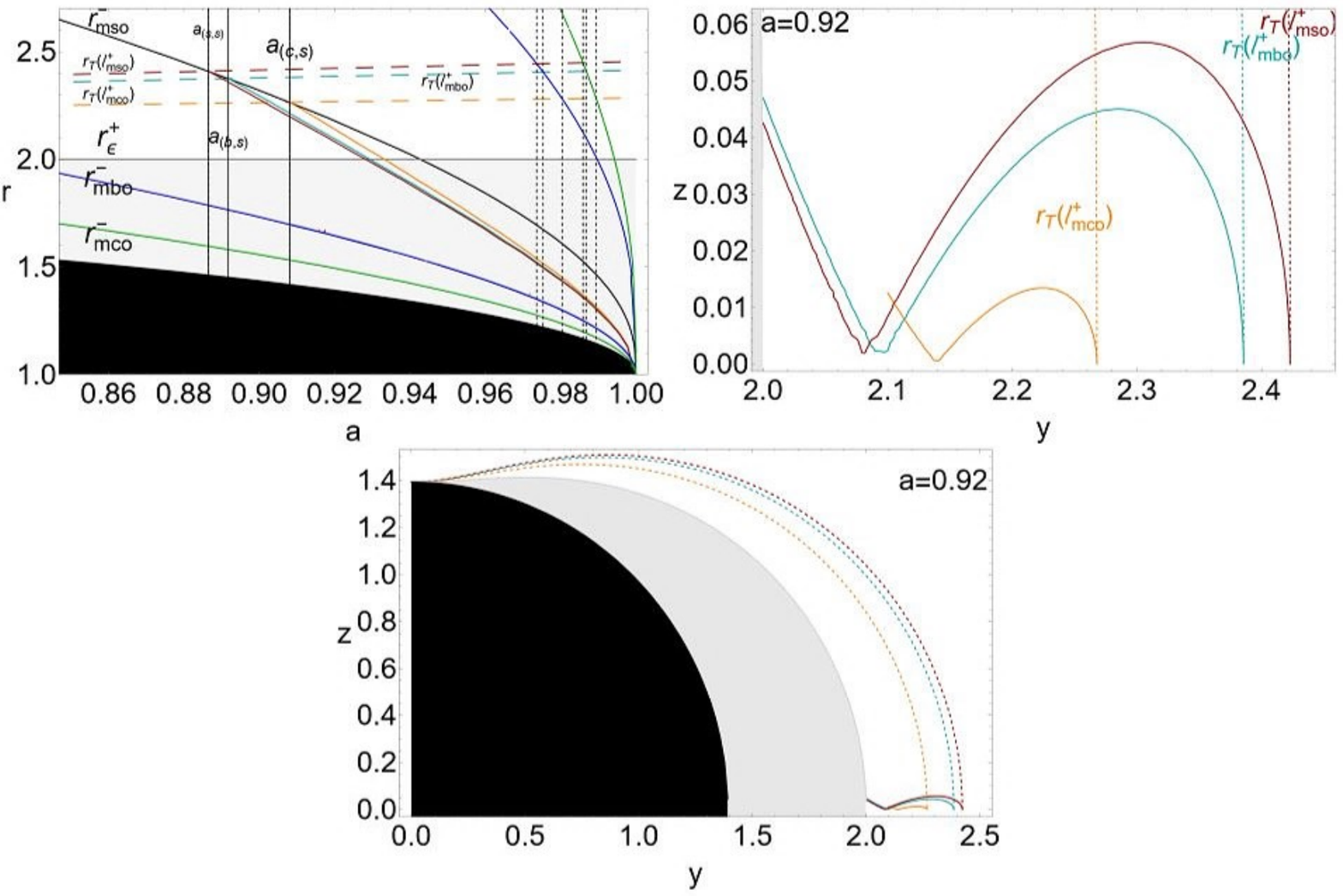}
\caption{Inversion surfaces crossing the co--rotating accretion disks   outer  edge. Black region is the central \textbf{BH}, with dimensionless spin $a$. There is   $r= \sqrt{y^2+z^2}$ and $\theta=\arccos({z}/r)$.  Gray region is the outer ergoregion ($r_\epsilon^+$ the outer ergosurface).  Dashed lines are the inversion surfaces  $r_\Ta$ at fixed counter--rotating angular momentum $\ell^+$.  Spins signed on the panels are defined in Table\il\ref{Table:spins}. All quantities are dimensionless. There is  $\mathrm{Q}_\bullet$ for any quantity  evaluated at $r_\bullet$, where
$mbo$ (blue curve) is for marginally bound circular orbit, $mco$ (green)  is for marginally  circular orbit, $mso$ (black curve) is for marginally stable circular orbit.  In the upper left panel the outer radii solutions
$\ell_{mco}^-=\ell^-(r)$ (green curve) and $\ell_{mbo}^-=\ell^-(r)$ (blue curve) are also shown on the \textbf{BH} equatorial plane--(see Eqs\il(\ref{Eq:def-nota-ell})). $\ell^-$ is the angular momentum for co--rotating fluids. Dotted vertical lines are spins in Fig.\il\ref{Fig:corso}.  Red, cyan and orange solid curves are the inner edges (cusps) correspondent to the  solutions of $r_{outer}^-=r_\Ta(\ell_*^+)$, where $r_{outer}^-$ is the outer edge of a co--rotating  cusped torus and  $\ell_*^+\in\{\ell_{mco}^+,\ell_{mbo}^+,\ell_{mso}^+\}$ (orange, cyan and red curves respectively). Upper right panel and bottom panel show the relative  co--rotating  tori (dashed curves are the inversion surfaces) {found as equi--potential surfaces from  $V_{eff}$ of  Eq.\il(\ref{Eq:scond-d})  with constant  $\ell_*^+\in\{\ell_{mco}^+,\ell_{mbo}^+,\ell_{mso}^+\}$ respectively}.  Upper right panel is a close--up view of the bottom panel. }.\label{Fig:corso2}
\end{figure*}
The spacetimes classes are characterized by the conditions listed in Table\il\ref{Table:Regions-BHs},  implying the  occurrence of a given configuration, composed by an inversion surface and a co--rotating toroid, \emph{or} the possibility of a given configuration constrained by the parameters $(\ell^\pm,K^-,a)$\footnote{As clear    from Fig.\il\ref{Fig:corso}, the \textbf{BHs} classes have different extensions,  from the smallest
 \textbf{(VIII)} class, with spin
$]a_{(s,s)},a_{(b,s)}[$, to   the large
 \textbf{(VII)} class with spin
$]a_{(c,b)},a_{(s,s)}[$
or   \textbf{(XVI)}  class with spin $]a_{((c,c))},1[$.}.

 \begin{table*}
 \centering
 \resizebox{.851\textwidth}{!}{%
\begin{tabular}{l|l}
 \hline
\textbf{(I)} $a\in]0,a_{(s,c)}]$&
 $r_{mco}^->r_\Ta(\ell_{mso}^+)$
\\
\textbf{(II)} {$a\in ]a_{(s,c)},a_{(b,c)}[$}&
$r_{mco}^-\in ]r_\Ta(\ell_{mbo}^+),r_\Ta(\ell_{mso}^+)[$
.
It can be
\textbf{(}$r_\bullet^-(\mathbf{L_2^-})<r_\Ta(\mathbf{L_1^+})$\textbf{) }
\\
\textbf{(III)} \textbf{$a\in ]a_{(b,c)},a_{(c,c)}[$}& $r_{mco}^-\in ]r_\Ta(\ell_{mco}^+),r_\Ta(\ell_{mbo}^+)[$. It can be
$r_\bullet^-(\mathbf{L_2^-})< r_{\Ta}(\mathbf{L_2^+})$
\\
\textbf{(IV)}  $a\in ]a_{(c,c)},a_{(s,b)}[$
&
$r_{mco}^-<r_\Ta(\ell_{mco}^+)$,
$r_{mbo}^->r_\Ta(\ell_{mso}^+)$
\\
&It can  be
$r_\bullet^-(\mathbf{L_2^-})< r_{\Ta}(\mathbf{L_3^+})$
\\
\textbf{(V)} $a \in ]a_{(s,b)},a_{(b,b)}[$&
$r_{mbo}^-\in ]r_\Ta(\ell_{(mbo)}^+),r_\Ta(\ell_{(mso)}^+)[$
\\
&It is
$r_\bullet^-(\mathbf{L_2^-})<r_\Ta^+(\ell_{mso}^+)$
\\
&
It can be $
r_\bullet^-(\mathbf{L_2^-})<r_\Ta^+(\mathbf{L_3^+})$,
$r_\bullet^-(\mathbf{L_1^-})<r_\Ta^+(\mathbf{L_1^+})$
\\
\textbf{(VI)} $a\in ]a_{(b,b)},a_{(c,b)}[$
&
$r_{mbo}^-\in ]r_\Ta(\ell_{(mco)}^+),r_\Ta(\ell_{(mbo)}^+)[$
\\
&
It is
$r_\bullet^-(\mathbf{L_2^-})<r_\Ta^+(\ell_{mbo}^+)$, $
r_\bullet^-(\mathbf{L_2^-})<r_\Ta^+(\mathbf{L_1^+})$].
\\
&
It can be ($
r_\bullet^-(\mathbf{L_2^-})<r_\Ta^+(\mathbf{L_3^+})$,
$r_\bullet^-(\mathbf{L_1^-})<r_\Ta^+(\mathbf{L_2^+}))$
\\
\textbf{(VII)} $a\in ]a_{(c,b)},a_{(s,s)}[$&
$r_{mbo}^-<r_{\Ta}(\ell_{mco}^+)
$,  $r_{mso}^-> r_{\Ta}(\ell_{mso}^+)$
\\
&
It is
$r_\bullet^-(\mathbf{L_2^-})<r_\Ta^+(\ell_{mco}^+)$, $
r_\bullet^-(\mathbf{L_2^-})<r_\Ta^+(\mathbf{L_2^+})$
\\
&
It can be
$ r_\bullet^-(\mathbf{L_2^-})<r_\Ta(\mathbf{L_3^+})$,
$r_\bullet^-(\mathbf{L_1^-})<r_\Ta(\mathbf{L_3^+})$
\\
\textbf{(VIII)} $a\in]a_{(s,s)},a_{(b,s)}[$&
$r_{mso}^-\in ]r_\Ta(\ell_{mbo}^+),r_\Ta(\ell_{mso}^+)[$.
It can be
$r_\bullet^-(\mathbf{L_1^-})<r_\Ta(\mathbf{L_3^+})$
\\
&
$r_{center}^-(\mathbf{L_1^-})<r_\Ta(\mathbf{L_1^+})$
\\
 \textbf{(IX)} $a\in]a_{(b,s)},a_{(c,s)}[$ &  $r_{mso}^-\in ]r_\Ta(\ell_{mco}^+),r_\Ta(\ell_{mbo}^+)[$
 \\
&
{It is}
 $r_\bullet^-(\mathbf{L_1^-})<r_\Ta(\mathbf{L_1^+})$
It can be
$r_\bullet^-(\mathbf{L_1^-})<r_\Ta(\mathbf{L_3^+})$
\\
&
$r_{center}^-(\mathbf{L_1^-})<r_\Ta(\mathbf{L_2^+})$
\\
 \textbf{(X)}  $a\in]a_{(c,s)},a_{((b,s))}[$
&
$
r_{mso}^-<r_\Ta(\ell_{mco}^+)$,
$ r_{(mbo)}^->r_\Ta(\ell_{mso}^+)$
\\
&
{It is}
  $r_\bullet^-(\mathbf{L_1^-})<r_\Ta(\mathbf{L_3^+})$
\\
&
It can be
$r_{center}^-(\mathbf{L_1^-})<r_\Ta(\mathbf{L_3^+})$
\\
 \textbf{(XI)}  $a\in ]a_{((b,s))},a_{((b,b))}[$
&
 $r_{(mbo)}^-\in ]r_\Ta(\ell_{mbo}^+),r_\Ta(\ell_{mso}^+)[$
 \\
&
{It is} $ r_{center}^-(\mathbf{L_1^-})<r_\Ta(\ell_{mso}^+)$
\\
&
 It can be $r_{center}^-(\mathbf{L_1^-})<r_\Ta(\mathbf{L_3^+})$, $r_{center}^-(\mathbf{L_2^-})<r_\Ta(\mathbf{L_1^+})$
\\
 \textbf{(XII)}   $a\in]a_{((b,b))},a_{((b,c))}[$
&
 $r_{(mbo)}^-\in ]r_\Ta(\ell_{mco}^+),r_\Ta(\ell_{mbo}^+)[$
 \\
&
{It is} $ r_{center}^-(\mathbf{L_1^-})<r_\Ta(\mathbf{L_1^+})$
\\
&
It can be
$r_{center}^-(\mathbf{L_1^-})<r_\Ta(\mathbf{L_3^+})$,  $r_{center}^-(\mathbf{L_2^-})<r_\Ta(\mathbf{L_2^+})$
\\
 \textbf{(XIII)}   $a\in ]a_{((b,c))},a_{((c,s))}[$ &
$r_{(mbo)}^-<r_\Ta(\ell_{mco}^+) $, $r_{(mco)}^->r_\Ta(\ell_{mso}^+) $
 \\
&
{It is} $ r_{center}^-(\mathbf{L_1^-})<r_\Ta(\mathbf{L_2^+})$,  $r_{center}^-(\mathbf{L_1^-})< r_\Ta(\ell_{mco}^+)$.
\\
&
It can be $r_{center}^-(\mathbf{L_2^-})<
r_\Ta(\mathbf{L_3^+})$,
$r_{center}^-(\mathbf{L_1^-})<r_\Ta(\mathbf{L_3^+})$
\\
 \textbf{(XIV)}    $a\in ]a_{((c,s))},a_{((c,b))}[$ & $r_{(mco)}^-\in ]r_\Ta(\ell_{mbo}^+),r_\Ta(\ell_{mso}^+)[$
\\
&
It is $ r_{center}^-(\mathbf{L_1^-})<r_\Ta(\mathbf{L_2^+})
$,  $ r_{center}^-(\mathbf{L_1^-})< r_\Ta(\ell_{mco}^+)$,
 $ r_{center}^-(\mathbf{L_2^-})< r_\Ta(\ell_{mso}^+)$.
 \\
 &
It can be  $r_{center}^-(\mathbf{L^-_3})<r_\Ta(\mathbf{L^+_1})$, $r_{center}^-(\mathbf{L_2^-})
<r_\Ta(\mathbf{L_3^+})$
\\
&
$r_{center}^-(\mathbf{L_1^-})<r_\Ta(\mathbf{L_3^+})$
\\
 \textbf{(XV)}  $a\in ]a_{((c,b))},a_{((c,c))}[$
&
 $r_{(mco)}^-\in]r_\Ta(\ell_{mco}^+),r_\Ta(\ell_{mbo}^+)[$
\\
&
{It is} $r_{center}^-(\mathbf{L_1^-})<r_\Ta(\mathbf{L_2^+})$,  $r_{center}^-(\mathbf{L_1^-})< r_\Ta(\ell_{mco}^+)$,
\\
&
 $r_{center}^-(\mathbf{L_2^-})< r_\Ta(\ell_{mbo}^+)$,  $r_{center}^-(\mathbf{L_2^-})<r_\Ta(\mathbf{L_1^+})$
 \\
 &
It can be $r_{center}^-(\mathbf{L_2^-})<r_\Ta(\mathbf{L_3^+})$, $r_{center}^-(\mathbf{L_3^-})<r_\Ta(\mathbf{L_2^+})$
\\
 \textbf{(XVI)}   $a\in ]a_{((c,c))},1[$
&
$r_{(mco)}^-<r_\Ta(\ell_{mco}^+)$
\\
&
{It is} $r_{center}^-(\mathbf{L_1^-})<r_\Ta(\mathbf{L_2^+})$,
 $r_{center}^-(\mathbf{L_1^-})< r_\Ta(\ell_{mco}^+)$,
 \\
 &
 $r_{center}^-(\mathbf{L_2^-})< r_\Ta(\ell_{mco}^+)$,  $r_{center}^-(\mathbf{L_2^-})<r_\Ta(\mathbf{L_2^+})$.
 \\
 &
It can be $r_{center}^-(\mathbf{L_2^-})<r_\Ta(\mathbf{L_3^+})$, $r_{center}^-(\mathbf{L_3^-})<r_\Ta(\mathbf{L_3^+})$
\\
\hline
  \end{tabular}}
  \caption{ \textbf{BHs}  classes \textbf{(I)}--\textbf{(XVI)}  defined by the spins $a_{(\bullet,\star)}$ and $a_{((\bullet,\star))}$ in Table\il\ref{Table:spins} and  shown in Figs\il\ref{Fig:corsozoom1},\ref{Fig:corso},\ref{Fig:corso1}, in terms of the  co--rotating toroids and inversion surfaces. $r_\Ta$ is the inversion radius.  $r_{center}$   is the torus center.  $\ell$ is the specific fluid angular momentum.
$mbo$ is for marginally bounded circular orbit, $mco$ is for marginally  circular orbit, $mso$ is for marginally stable circular orbit.  $\pm$ is for counter--rotating and co--rotating fluids respectively. Radii   $r_{(mbo)}^{\pm}$ and $r_{(mco)}^{\pm}$ are defined in Eq.\il(\ref{Eq:def-nota-ell}).
 It is   $\ell^+\leq \ell_{mso}^+$ for the   counter--rotating specific angular momentum. It is  $r^-_\bullet\in\{r_{inner}^-,r_J^-,r_\times^-\}$  according to the momentum parameter ranges $\mathbf{L_i^\pm}$ of  Table\il\ref{Table:tori}  for the inner edge of a quiescent torus, the proto-jet cusp and the torus cusp respectively. There is $r_{\Ta}(\mathbf{L_3^+})< r_{\Ta}(\mathbf{L_2^+})< r_{\Ta}(\mathbf{L_1^+})$.  $r_{\Ta}(\mathbf{L_i^+})$ is for any inversion surface in the momentum range $\mathbf{L_i^+}$ (and, for example,  $r_{center}^-(\mathbf{L_3^-})$  is for a co--rotating toroid center with  momentum
in the range $\mathbf{L_3^-}$). }\label{Table:Regions-BHs}
    \end{table*}

In  the spacetimes where
$r_\Ta(\ell^+_\bullet)\in]r_{mbo}^-,r_{mso}^-[$,  for example,   a co--rotating  proto--jet  cusp   \emph{is} internal to the inversion surface con momentum $\ell_\bullet^+$, while a toroid cusp can be external to the inversion surface.
In the spacetimes where  $r_\Ta(\ell_\bullet^+)<r_{mco}^-$   the co--rotating toroids  \emph{are} external to the  inversion surface  defined by the momentum $\ell_\bullet^+$.
On the other hand,  with   $r_\Ta(\ell_\bullet^+)\in  ]r_{mco}^-,r_{mbo}^-[$,   the co--rotating toroids  $(\pp_1^-,\pp_3^-)$  are external to the
 inversion surface with  momentum $\ell_\bullet^+$, but  the inner region of a $\pp_2^-$ toroid, with its inner edge $r_{inner}^-$   bottom bounded by the radii $r_J^-<r_{inner}^-$, can be internal to the inversion surface with radius $r_\Ta(\ell_\bullet^+)$  (the $\cc^-_2 $  torus outer part, $[r_{center}^-,r_{outer}^-]$, has to be external to the inversion surface).

Note, for a torus external to an inversion surface $r_\Ta(\ell_\bullet^+)$  with momentum
$\ell_\bullet^+$  there is  $r_{outer}^-<r_\Ta(\ell_\bullet^+)$.
On the other hand,  for toroids  $\pp_1^-$, having momenta in   $\mathbf{L_1^-}$,  there is  $r_{\times}^-<r_{inner}^-$,   for  toroids  $\pp_2^-$, having momenta in   $\mathbf{L_2^-}$,  there is  $r_{J}^-<r_{inner}^-$,  while for    $\cc_3^-$ quiescent tori, with  momenta in   $\mathbf{L_3^-}$ , described in Table\il\ref{Table:Regions-BHs} for fast attractors, there is no  bottom boundary on the torus inner edge, and  these surfaces can be constrained considering  the relative location of the  inversion surfaces  with the radius   $r_{(mco)}^-$, as there is $r_{center}^-(\mathbf{L_3^-})>r_{(mco)}^-$.

Condition
  $r_\Ta(\ell_\bullet^+)\in  ]r_{(mbo)}^-,r_{(mco)}^-[$   ensures that the center
   $r_{center}^-(\mathbf{L_1^-})$  of a  (quiescent or cusped)  torus with momentum in  $\mathbf{L_1^-}$   is internal to  the
  inversion surface $r_\Ta(\ell_\bullet^+)$  with momentum
$\ell_\bullet^+$  then,  for small momentum $\ell^-$ (ie. $\gtrsim\ell_{mso}^-$)   or for
small    disks
(with $K^-\gtrapprox K_{center}^-$, these co--rotating  disks could be restricted, according to the \textbf{BH} spin, to quiescent tori only)  the torus is entirely contained in the
  inversion surface $r_\Ta(\ell_\bullet^+)$  with momentum
$\ell_\bullet^+$.
viceversa, condition
  $r_\Ta(\ell_\bullet^+)\in  ]r_{(mbo)}^-,r_{(mco)}^-[$  implies that the center $r_{center}^-(\mathbf{L_2^-})$  of a (quiescent) torus with any momentum in the range  $\mathbf{L_2^-}$  could be external to the
  inversion surface $r_\Ta(\ell_\bullet^+)$, while  $r_{center}^-(\mathbf{L_3^-})$  \emph{is} external to
     $r_\Ta(\ell_\bullet^+)$.
%
Hence, condition
$r_{\Ta}(\ell^+_\bullet)\in]r_{mso}^-,r_{(mbo)}^-[$, for example,  implies that a quiescent disk  $\cc_1^-$ could be external, depending on the parameters  $\ell$ and  $K$,
to the inversion surface  $r_{\Ta}(\ell^+_\bullet)$, but a cusped torus  $\cc_\times^-(\mathbf{L_1^-})$ is partially included,
 with its inner region, in the  inversion surface $r_{\Ta}(\ell^+_\bullet)$.
As there is
$r_{\Ta}(\mathbf{L_3^+})< r_{\Ta}(\mathbf{L_2^+})< r_{\Ta}(\mathbf{L_1^+})$, a condition $r_\bullet^-(\mathbf{L_\bullet^-})< r_{\Ta}(\mathbf{L_3^+})$ implies
$r_\bullet^-(\mathbf{L_\bullet^-})<r_{\Ta}(\mathbf{L_3^+})< r_{\Ta}(\mathbf{L_2^+})< r_{\Ta}(\mathbf{L_1^+})$.

Therefore, classes of Table\il\ref{Table:Regions-BHs}, delimited by the spins $a_{(\bullet,\star)}$ are defined by the  inversion surfaces properties in relation to the cusps $\{r_\times^-,r_J^-\}$  or the inner edge $r_{inner}^-$ of a quiescent torus, and therefore by the possibility that part of the inner  region of a torus is external  or internal to  the inversion surface. (Note, there is always  $r_{\times}^-(\ell_1^-)<r_{inner}^-(\ell_1^-)$ and  $r_{J}^-(\ell_2^-)<r_{inner}^-(\ell_2^-)$  for equal specific angular momentum $\ell^-_1\in\mathbf{ L_1^-}$ and    $\ell^-_2\in\mathbf{L_2^-}$ respectively).

For spins  $a\leq a_{(s,s)}$,  the  centers $r_{center}^-$  of the co--rotating toroids are external to  the inversion surfaces, i.e. $r_{center}^->r_\Ta(\ell^+)$,  and in this range we investigate  the possibility that a disk or proto-jet cusp could be  internal or external to  an inversion surface--i.e. $r_{\bullet}^-\lessgtr r_\Ta(\ell^+)$ respectively, where $r_\bullet^-\in\{r_\times^-,r_J^-\}$.
For  $a\geq a_{(s,s)}$  the toroid center can be contained in an inversion surface i.e. it can be   $r_{center}^-<r_\Ta(\ell^+)$, and therefore there can be $r_{center}^-<r_{outer}^-<r_\Ta(\ell^+)$, i.e. it is  entirely embedded in the inversion surface with momentum $\ell^+$,  for small co--rotating specific angular momentum $\ell^-$.
In this range, and in the ranges with higher spins defined by  $a\geq a_{((\bullet,\star))}$,
we  investigate the possibility that the  torus center and the entire torus could be embedded in an inversion surface.
Therefore,
for  $a<a_{(s,s)}$,  co--rotating toroids could be only partially included  (with the inner region) in an inversion surface.
As clear from Table\il\ref{Table:Regions-BHs}, in the spacetimes with $a\in]0,a_{(s,c)}]$ the  inversion surfaces are always (for  $\ell^+\leq \ell_{mso}^+$) internal (on the equatorial plane)  to the outer co--rotating  toroids.
For  $a\in]a_{(s,c)}a_{(b,s)}[$, tori  can be only partially included in an inversion surface.
For  $a>a_{(s,s)}$  quiescent tori or, for the faster spinning attractors,
cusped tori,  can be or are internal to  the inversion surfaces  (note some of these tori are  in the ergoregion), according to the conditions in  Table\il\ref{Table:Regions-BHs} see also Figs\il\ref{Fig:corso},\ref{Fig:corso1}.

These results are confirmed also by  the analysis of  Fig.\il\ref{Fig:Plotsigi}--right panel  showing the co--rotating surfaces specific momentum  $\ell^-(r_\Ta(\ell^+))$, evaluated on the inversion surface  $r_\Ta(\ell^+)$ of momentum $\ell^+$  for different spins of  Table\il\ref{Table:spins}.
The momenta $\ell^\pm$ are in all the  ranges $\{\mathbf{L^\pm_1},\mathbf{L^\pm_2},\mathbf{L^\pm_3}\}$. The inversion surfaces with momenta   $\ell^+$ crosses the co--rotating toroid center or cusp (the critical points of pressure) at momenta $\ell^-$.
For co--rotating specific angular momentum $\ell^-$ at larger spins, as shown in Fig.\il\ref{Fig:Plotfondo}, for extreme Kerr \textbf{BHs}, the  co--rotating torus specific angular momentum $\ell^-$ is small, increasing with the flow counter--rotating momentum $\ell^+$. It also is noted the limiting spin $a_{(s,s)}$. Decreasing the \textbf{BH} spin, the co--rotating specific angular momentum increases,  decreasing with the increase of the inversion surface and flows  specific angular momentum.  %
\subsection{Inversion surfaces crossing the  inner co--rotating  toroids}\label{Sec:inversion-crossing}
In this section we investigate
 the inter--disks  inversion surfaces crossing   different regions of  the     inner co--rotating  toroids which are embedded, external  or partially contained  in the inversion surfaces.
In here we focus mainly on the cusped configurations.
We examine the crossing with  the   tori outer edges in   Sec.\il(\ref{Sec:outer--edge-crossing}), the inversion surfaces crossing the  toroids cusps in Sec.\il(\ref{Sec:cusp-crossing}),  the toroids  center in  Sec.\il(\ref{Sec:torus-center-crossing}),  the critical points of pressure and  tori geometrical maxima   in Sec.\il(\ref{Sec:geometric--crossing}). The crossings with  tori on    planes different from the equatorial are investigated in Sec.\il(\ref{Sec:off-equatorial--crossing}).
\subsubsection{The co--rotating torus outer edge}\label{Sec:outer--edge-crossing}
In Fig.\il\ref{Fig:corso2}  inversion surfaces are shown crossing  the co--rotating  disks  outer edgee. In this case the co--rotating tori are  totally embedded  in the inversion surfaces\footnote{In this case the co--rotating toroid, with  $K^-\in[K_{center}^-,K^-_\times]$, hence with $r_{inner}^-\in[r_\times^-,r_{center}^-]$ for a torus,  or $r_{inner}^-\in[r_J^-,r^-_{center}]$, for a proto-jet,   can cross the inversion surface on planes different from the equatorial. This case is discussed in Sec.\il(\ref{Sec:off-equatorial--crossing}).}.

In Fig.\il\ref{Fig:corso2} we illustrate the solutions of  the problem:
\bea\label{Eq.:outer-rrT}
r_{outer}^-(\ell^-,K,a)=r_\Ta(\ell_\bullet^+)
\eea
for co--rotating cusped tori.    Eq.\il(\ref{Eq.:outer-rrT}) is verified  when
$r_{mso}^-<r_{center}^-<r_{outer}^-(\ell^-,K,a)=r_\Ta(\ell_\bullet^+)$,
implying  $a>a_{(\bullet,s)}$, according to the different counter--rotating momenta $\ell_\bullet^+\in\{\ell_{mco}^+,\ell_{mbo}^+,\ell_{mso}^+\}$.
 (On the other hand, part of the inner toroids orbiting these faster spinning \textbf{BHs}  can be entirely contained in the \textbf{BHs} ergoregion.)

Radius $r_{outer}^-(\ell^-,K^-,a)$ depends also on the parameter $K$, however
for $\ell^-\in \mathbf{L_1^-}$ we consider the upper bound
for  the cusped disk $K\in]K_{center}^-,K_\times^-]$  in Fig.\il\ref{Fig:corso2}.
For  $\ell^-\in \mathbf{L_2^-}$  it  has to be
$r_{outer}^->r_{center}^-\in[r_{(mbo)}^-,r_{(mco)}^-]$,
hence $r_\Ta=r_{outer}^->r_{(mbo)}^-$,  occurring  for
$a>a_{((b,\bullet))}$--(see Fig.\il\ref{Fig:corso2}--upper left panel).

\begin{figure*}
\centering
\includegraphics[width=5cm]{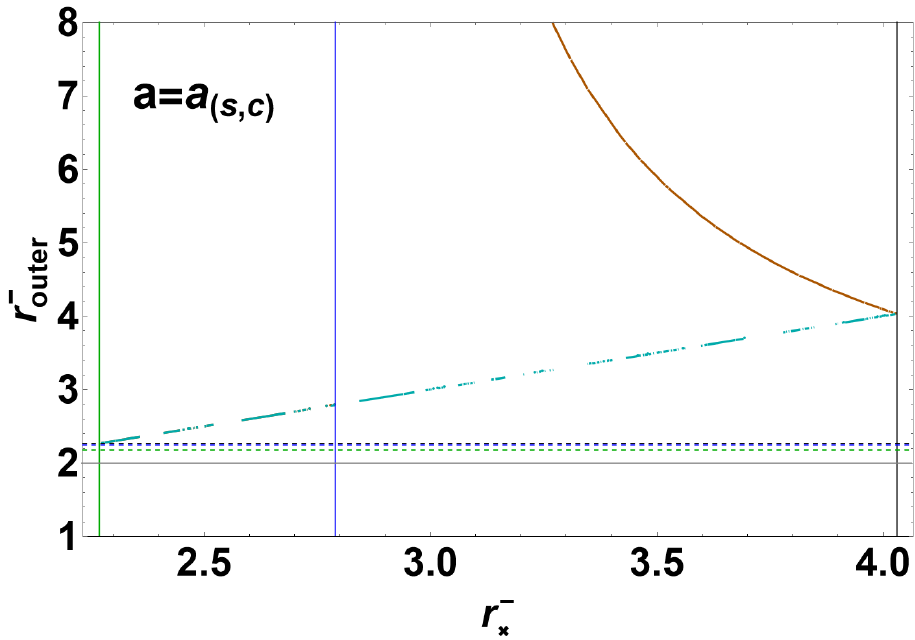}
\includegraphics[width=5cm]{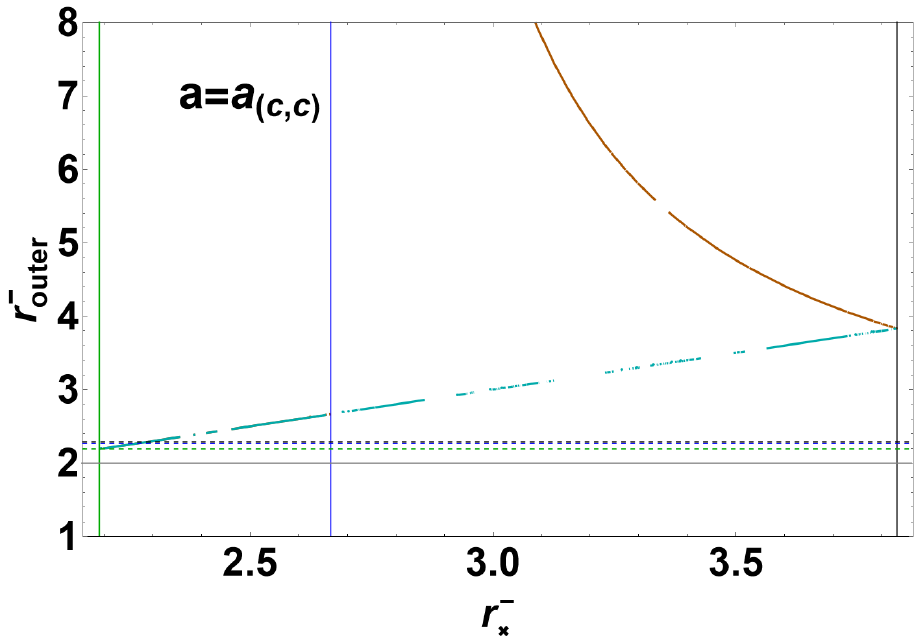}
\includegraphics[width=5cm]{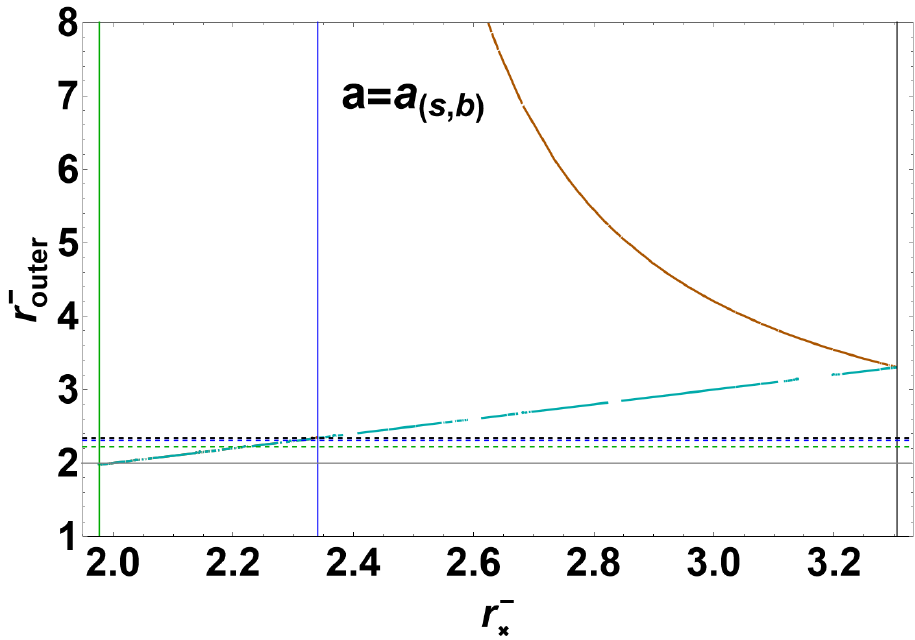}
\includegraphics[width=5cm]{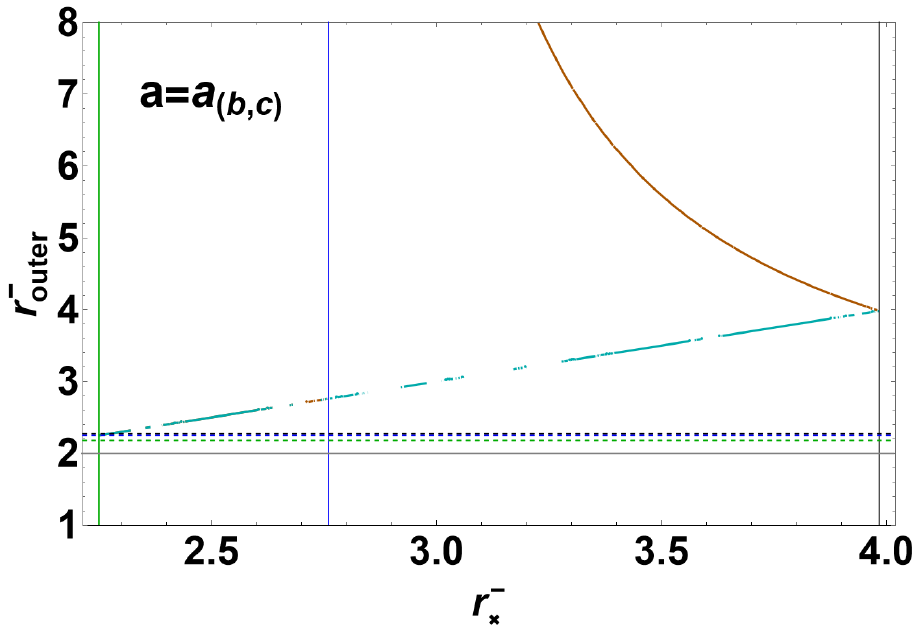}
\includegraphics[width=5cm]{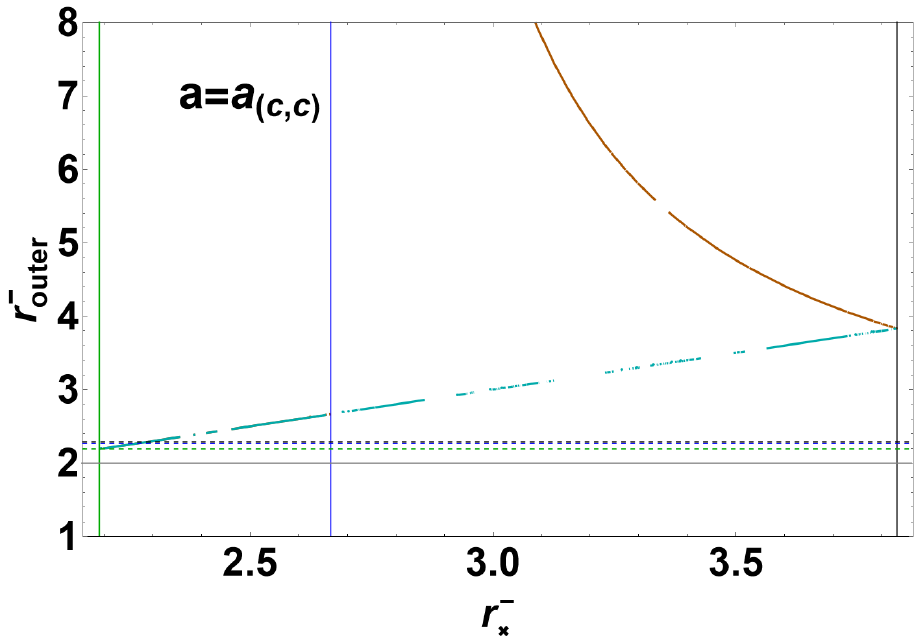}
\includegraphics[width=5cm]{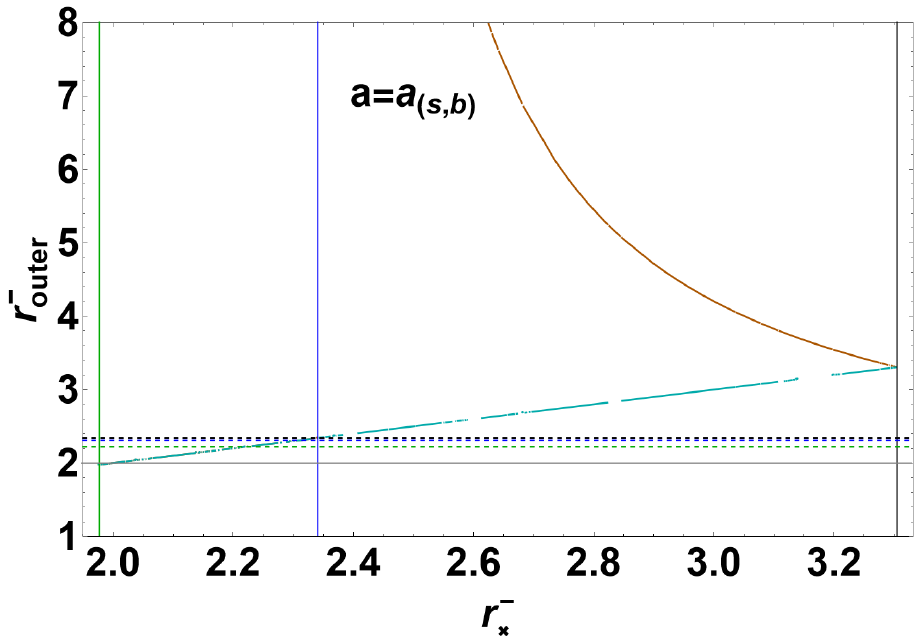}
\includegraphics[width=5cm]{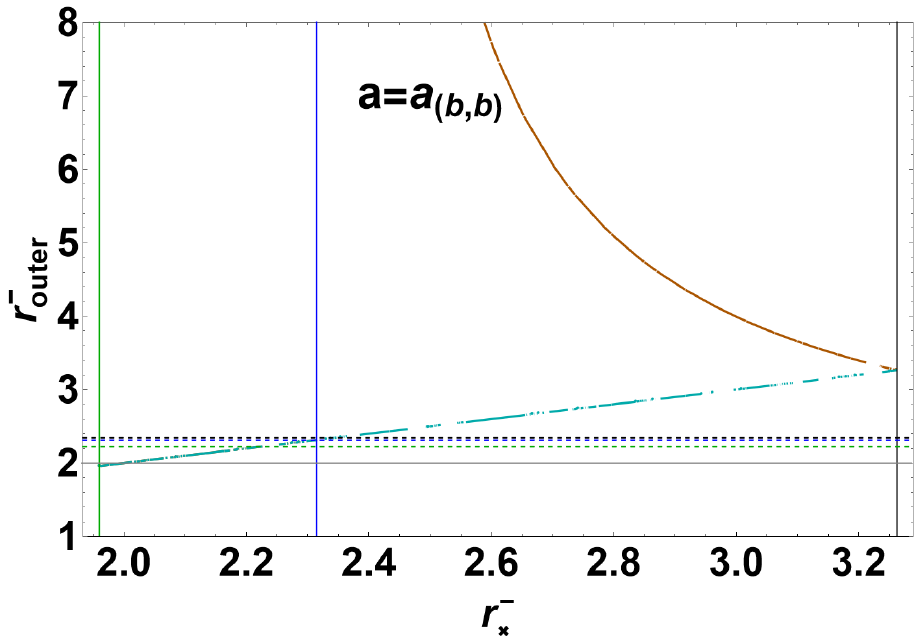}
\includegraphics[width=5cm]{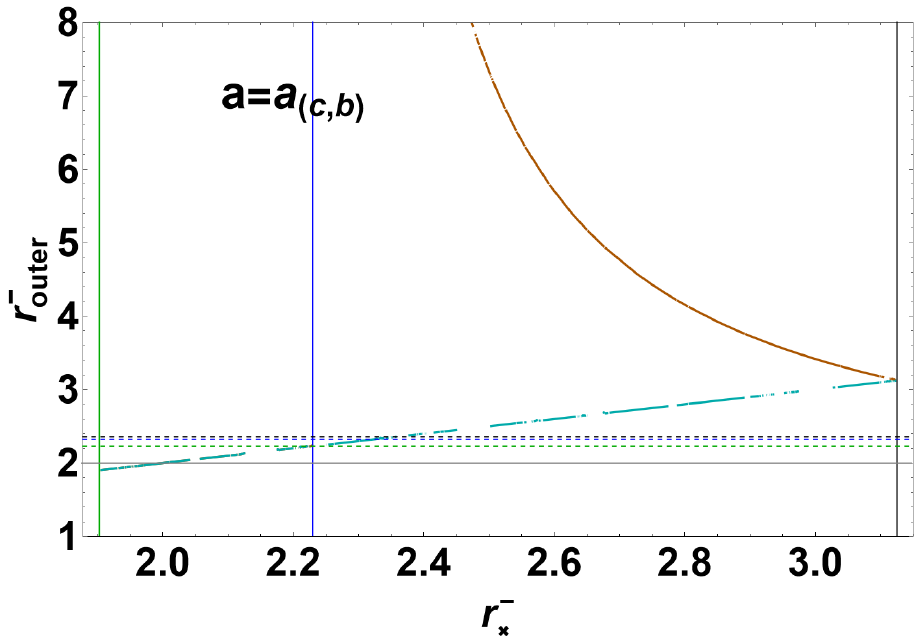}
\includegraphics[width=5cm]{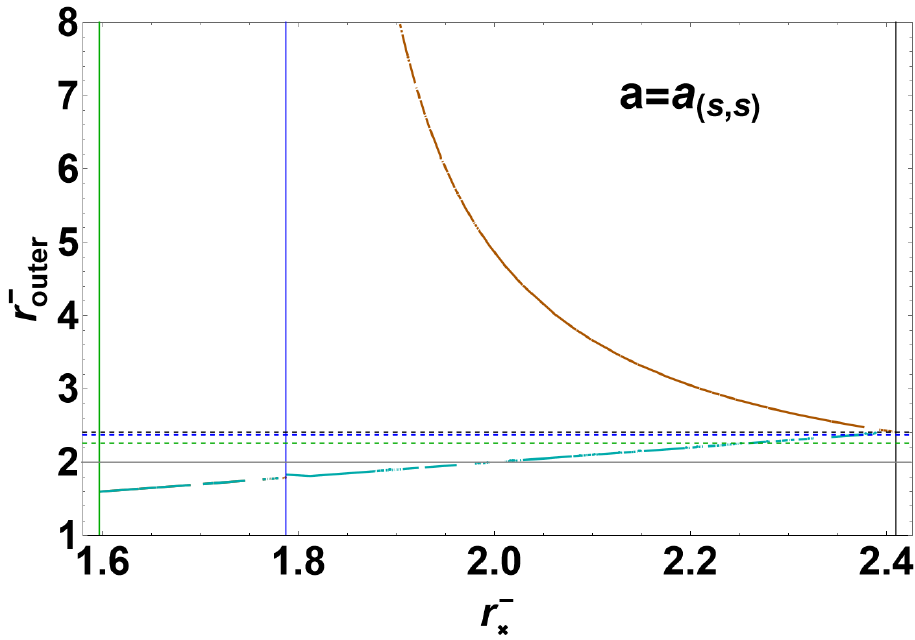}
\includegraphics[width=5cm]{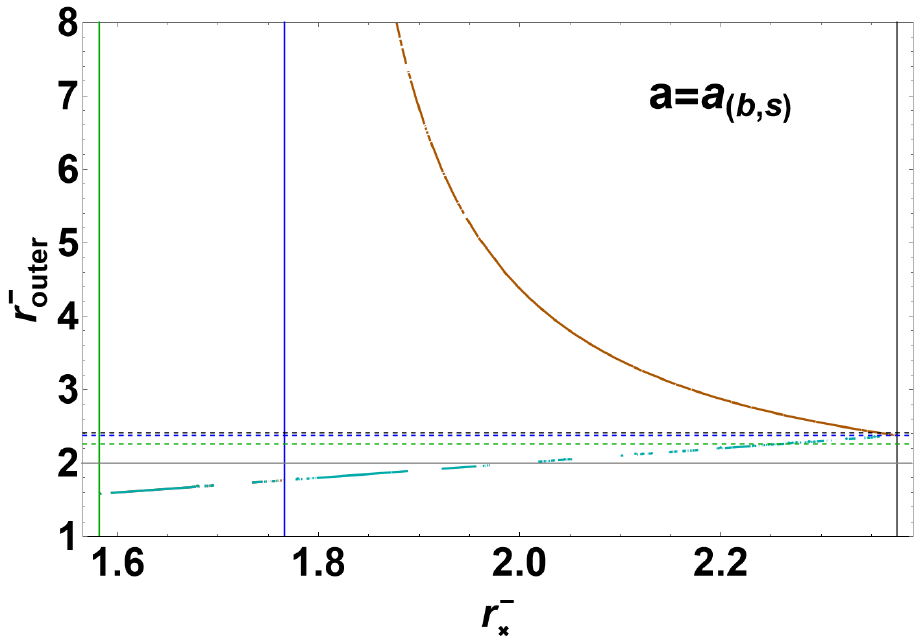}
\includegraphics[width=5cm]{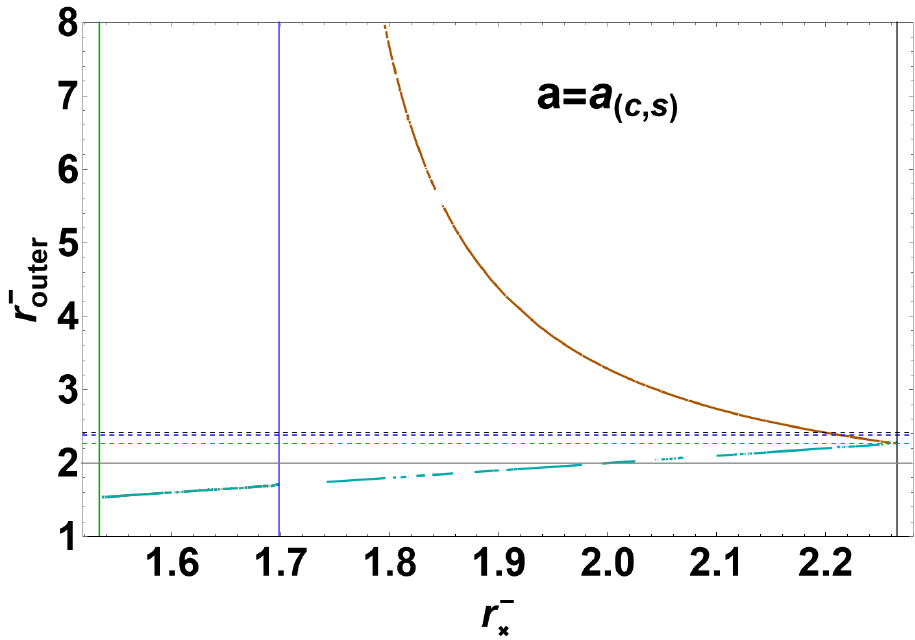}
\includegraphics[width=5cm]{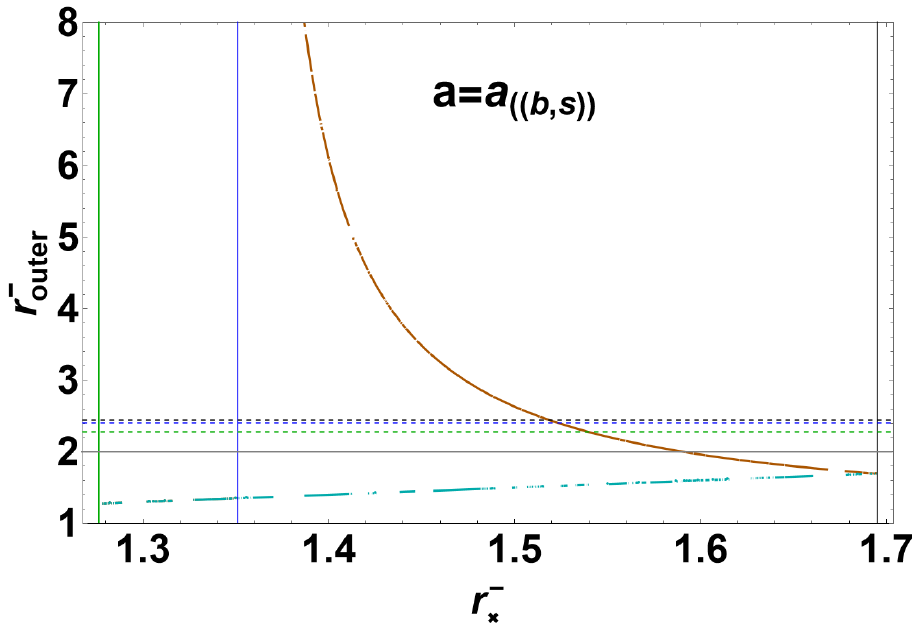}
\includegraphics[width=5cm]{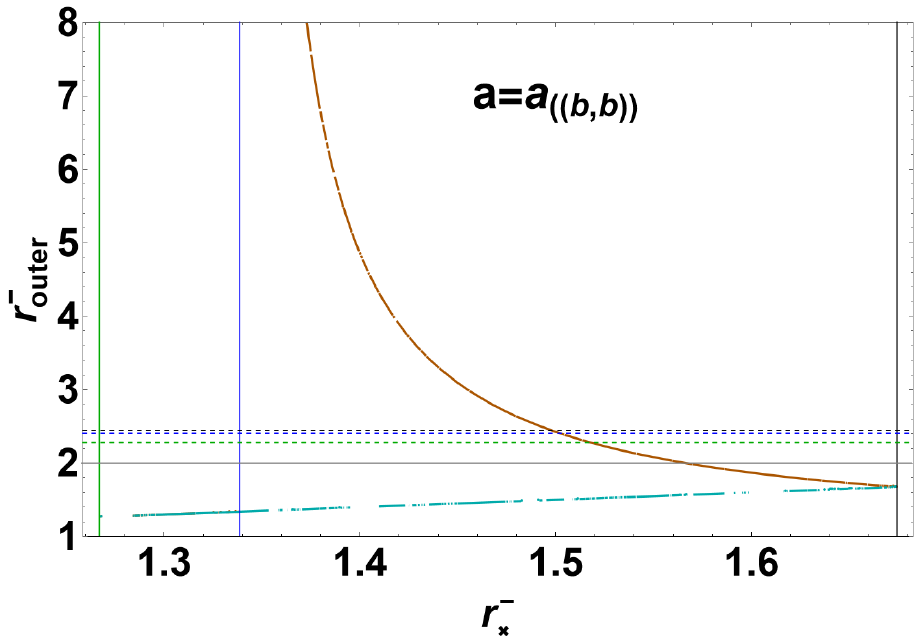}
\includegraphics[width=5cm]{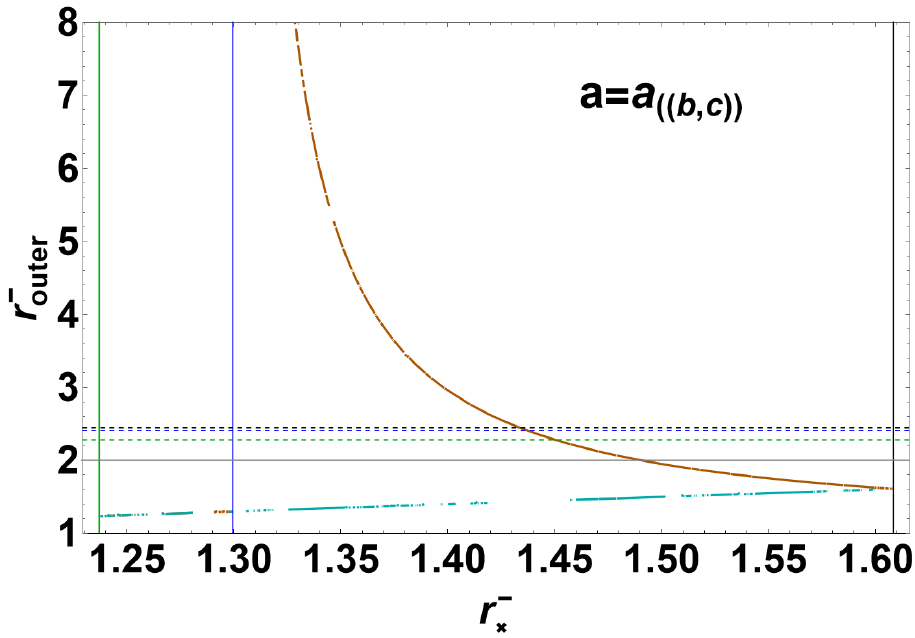}
\includegraphics[width=5cm]{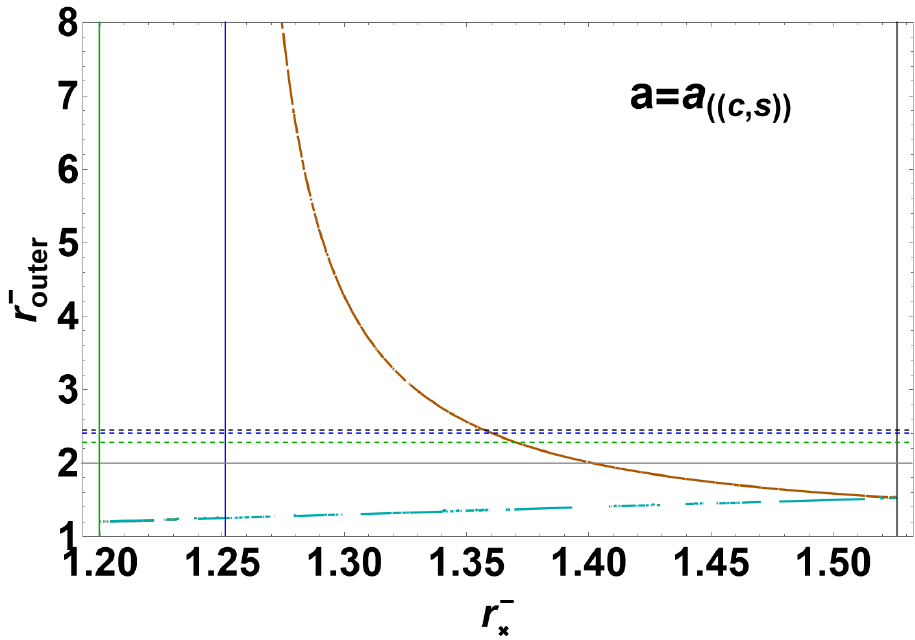}
\includegraphics[width=5cm]{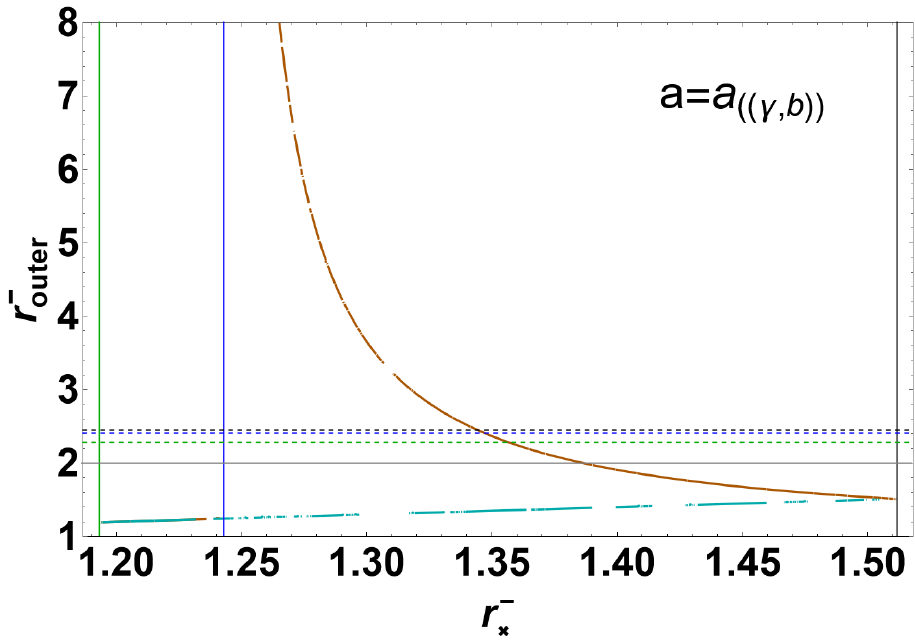}
\includegraphics[width=5cm]{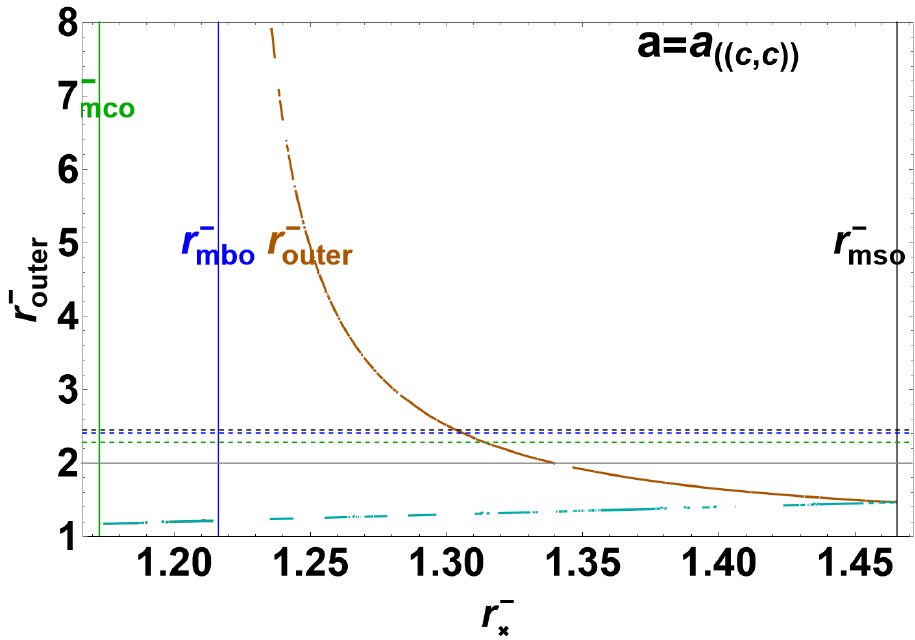}
\caption{Outer edge $r_{outer}^-$ (brown curve) versus cusp $r_{\times}^-$ (cyan curve) of the co--rotating cusped  disks (and proto-jets) for different spins  of Table\il\ref{Table:spins} signed on the panel. Vertical blue (black)(green) line is the marginally bound (stable) (circular) co--rotating  orbit $r_{mbo}^-$ ($r_{mso}^-$)  ($r_{mco}^-$). All quantities are dimensionless. The outer ergosurface on the equatorial plane is $r=2$ (gray line). Dotted lines are  $r_{\Ta}(\ell_{mso}^+)>r_{\Ta}(\ell_{mbo}^+)>r_{\Ta}(\ell_{mco}^+)$,  $r_\Ta$ is the inversion radius.}.\label{Fig:PlotLibC}
\end{figure*}

For $\ell^-\in \mathbf{L_3^-}$,   there is
$r_{outer}^->r_{center}^->r_{(mco)}^-$, then
 $r_\Ta=r_{outer}^->r_{(mco)}^-$, hence  these cases occur for
$a>a_{((c,\bullet))}$.
In Fig.\il\ref{Fig:corso2} we show, for  fixed  \textbf{BH} spin,  the inner and outer edges of the cusped toroids in relation to the inversion surfaces, confirming   that the crossing can occur for small momenta in magnitude of the inversion surface and large \textbf{BH} spin.
Focusing on the case $\ell\in \mathbf{L_1^-}$, occurring in the  spacetimes with
$a>a_{(\bullet,s)}$, we note that
the (inner part of the co--rotating) disk increases with the \textbf{BH} up to a maximum,   decreasing when the disk outer part (range $[r_{center}^-,r_{outer}^-]$) can be increasingly large, increasing with decreasing    $\ell_{mso}^+$ in magnitude --Fig.\il\ref{Fig:PlotLibC}.
Finally,
the outer  edge of the  $\cc_1^-$  toroids  is upper  bounded by the cusped tori outer edge, which is a function  of the parameter $\ell^-$ only.  However, for quiescent disks,
$(\cc_1^-,\cc_2^-,\cc_3^-)$, the outer edge is a function of the   parameters $\ell^-$ and $K^-$ and, for  the $(\cc_2^-,\cc_3^-)$ toroids there is no upper  bound on the outer edge provided by the configuration cusp.
   Hence, for a quiescent torus  the outer edge $r_{outer}^-(\ell^-,K^-,a)$ depends on the torus parameter $K^-$, and  we could consider the bottom  bound on the outer  edge provided by the torus center $r_{center}^-$. A torus  outer (and inner) edge could  be also very close to the torus center, to the limit of a  thin disk. This case will be considered in Sec.\il(\ref{Sec:torus-center-crossing}) constraining  the disk center.
\begin{figure*}
\centering
\includegraphics[width=18cm]{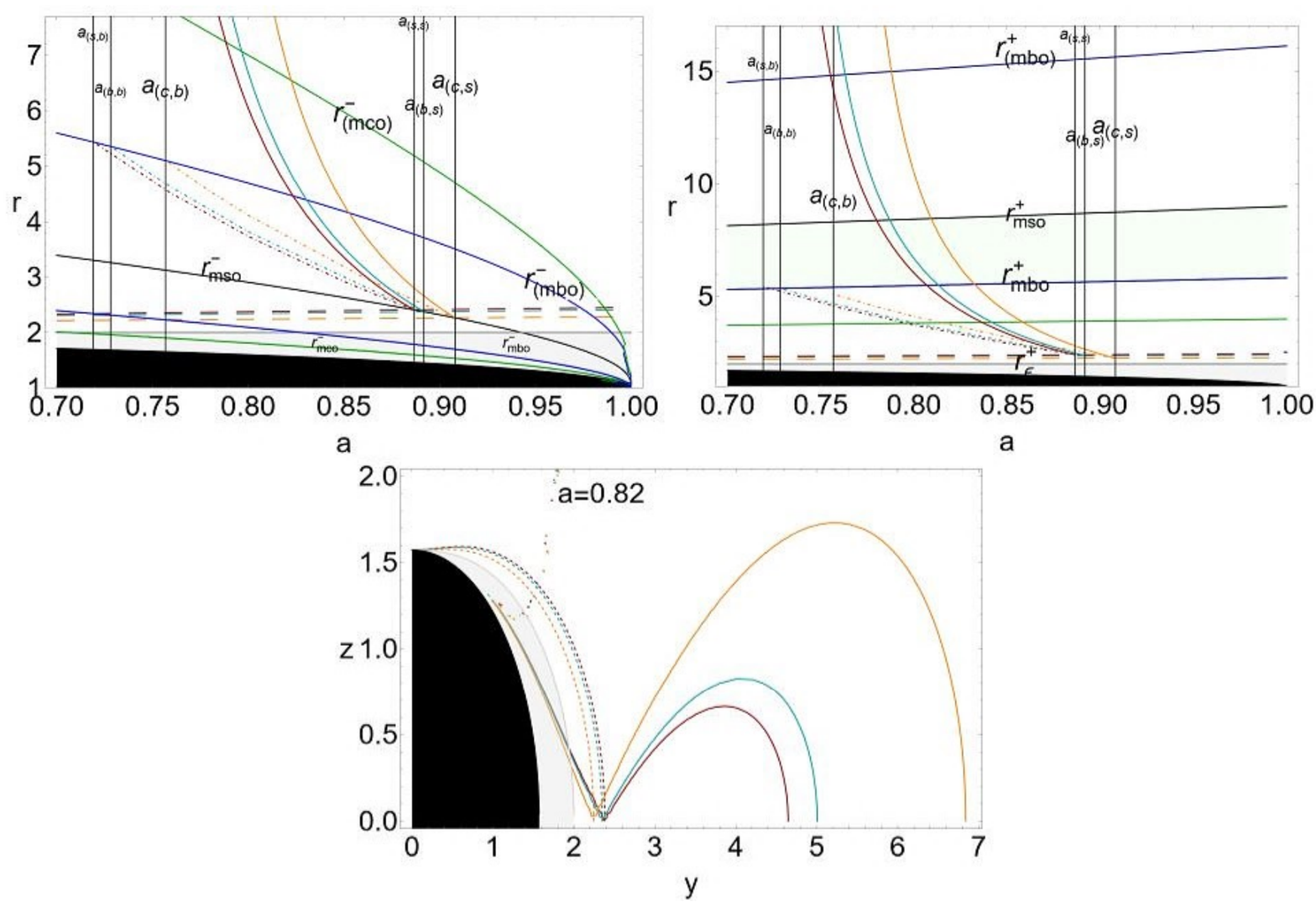}
\caption{Inversion surfaces crossing the disks cusp. Black region is the central \textbf{BH}, with dimensionless spin $a$. There is   $r= \sqrt{y^2+z^2}$ and $\theta=\arccos({z}/r)$.  Gray region is the outer ergoregion (radius $r_\epsilon^+$ is the outer ergosurface).
 Dashed lines are the inversion surfaces  $r_\Ta$ at fixed counter--rotating angular momentum $\ell^+$.  Spins signed on the panels (vertical black lines) are defined in Table\il\ref{Table:spins}. There is  $\mathrm{Q}_\bullet$ for any quantity  evaluated at $r_\bullet$, where
$mbo$ (blue curve) is for marginally bound circular orbit, $mco$ (green curve)  is for marginally  circular orbit, $mso$ (black curve) is for marginally stable circular orbit.  In the upper left panel the outer  solutions
$\ell_{mco}^-=\ell^-(r)$ (green curve) and $\ell_{mbo}^-=\ell^-(r)$ (blue curve) are also shown--(see Eqs\il(\ref{Eq:def-nota-ell})). $\ell^-$ is the  co--rotating angular momentum.  Red, cyan and orange solid  (dotted-dashed) curves  are  the outer edge (center) correspondent to the solutions of $r_{\times}^-=r_\Ta(\ell_*^+)$, where $r_{\times}^-$ is here the co--rotating disk or proto-jet cusp and the counter--rotating angular momentum is  $\ell_*^+\in\{\ell_{mco}^+,\ell_{mbo}^+,\ell_{mso}^+\}$ (orange, cyan and red curves respectively).   Upper left  panel is a close--up view of the upper right   panel.  Green region is the location range   of the accretion disk cusp.   Bottom panel shows the relative   tori {equi--potential surfaces} and the inversion surfaces.  All quantities are dimensionless. }.\label{Fig:PlotenIpno5finner}
\end{figure*}
\subsubsection{The co--rotating torus  inner edge}
\label{Sec:cusp-crossing}
We  explore the situation where a toroid  cusp crosses an inversion surface. In this case the disks are external to the  inversion surface as shown  in  Fig.\il	 \ref{Fig:PlotenIpno5finner} with     the related torus center and outer edge.
Constraints, according to the \textbf{BHs} spin, have been also discussed for the \textbf{BH} classes of Table\il\ref{Table:Regions-BHs}.
The condition
\bea\label{Eq.:ouinnerter-rrT}
r_{inner}^-(\ell^-,K^-,a)=r_\Ta(\ell_\bullet^+)
\eea
is  satisfied  for
$r_{mbo}^-<r_\Ta(\ell_\bullet^+) =r_\times^-(\ell^-)<r_{mso}^-$,
implying  $a\in [a_{(\bullet,b)},a_{(\bullet,s)}[$, for  the torus cusp, or  $r_{mco}^-<r_\Ta(\ell_\bullet^+) =r_J^-(\ell^-)<r_{mbo}^-$, implying  $a\in [a_{(\bullet,c)},a_{(\bullet,b)}[$, for the proto-jet cusp, according to the different counter--rotating momenta $\ell_\bullet^+\in\{\ell_{mco}^+,\ell_{mbo}^+,\ell_{mso}^+\}$.

However, we should note that  for the  toroid inner edge it has to be,
 $r_{inner}^-(\ell^-,K^-,a)\in [r_\times^-(\ell^-),r_{center}^-(\ell^-)[$ for  $\pp_1^-$ toroids,
  $r_{inner}^-(\ell^-,K^-,a)\in [r_J^-(\ell^-),r_{center}^-(\ell^-)[$  for    $\pp_2^-$ toroids, and  $r_{inner}^-(\ell^-,K^-,a)<r_{center}^-(\ell^-)$ for  $\cc_3^-$ tori.
  Hence, for a quiescent torus,  the inner edge $r_{inner}^-(\ell^-,K^-,a)$ depends on the torus parameter $K^-$, but it is upper bounded by the toroid cusp for
 $(\pp_1^-, \pp_2^-)$ configurations, which is the case considered in this section.
 However, for the $\cc_3^-$ tori and, more in general, for a quiescent torus, we could consider the upper bound on the inner edge provided by the torus center $r_{center}^-$. A torus inner (and outer) edge could possibly be also very close to the torus center up to the limit of a very thin torus, this case will be considered in Sec.\il(\ref{Sec:torus-center-crossing}).

\begin{figure*}
\centering
\includegraphics[width=8cm]{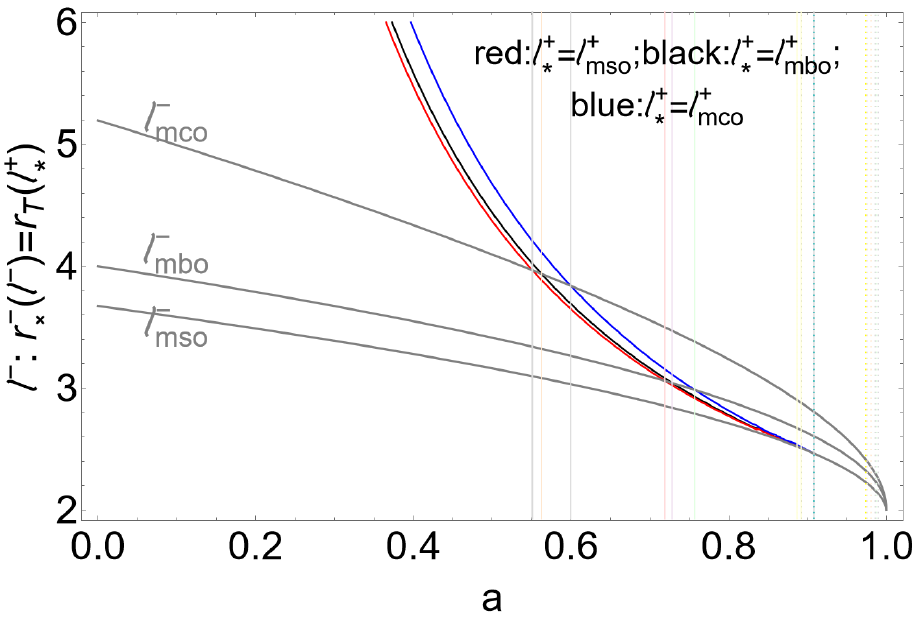}
\includegraphics[width=8cm]{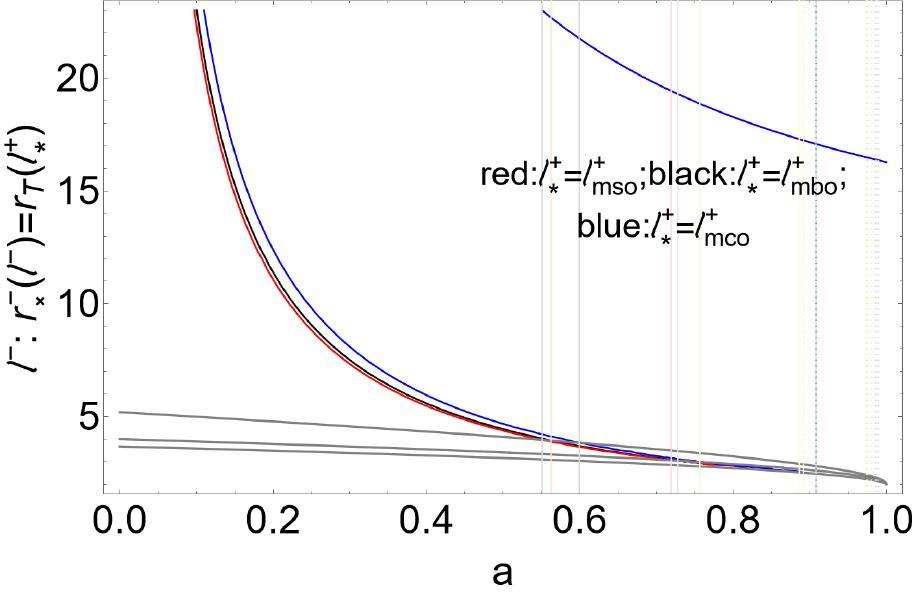}
\caption{Co--rotating  specific angular momentum $\ell^-$ solution of the equation $r_\times^-(\ell^-)=r_\Ta(\ell_*^+)$,  as function of the \textbf{BH} dimensionless  spin $a$ for different counter--rotating specific angular momentum $\ell_*^+$ signed on the panel. There is  $\mathrm{Q}_\bullet$ for any quantity  evaluated at $r_\bullet$, where
$mbo$ is for marginally bound circular orbit, $mco$ is for marginally  circular orbit, $mso$ is for marginally stable circular orbit. $r_\times^-$ is the cusp of a co--rotating disk or proto-jets. $r_\Ta$ is the inversion radius  (where $u^\phi$=0).  Dotted vertical lines are spins of Table\il\ref{Table:spins}. Left panel is a close up vew of the right panel. All quantities are dimensionless. }\label{Fig:Plotgsnedibubble}
\end{figure*}

The disk inner region and the proto-jet  range $[r_J^-,r_{center}^-]$, decrease with the
increasing of  the \textbf{BH} spin.
See also  Fig.\il\ref{Fig:PlotLibC} and   Fig.\il\ref{Fig:Plotgsnedibubble}, where Eq.\il(\ref{Eq.:ouinnerter-rrT})
 is solved in terms of the co--rotating specific angular momentum $\ell^-$, i.e. as  solution of the equation $r_\times^-(\ell^-)=r_\Ta(\ell_*^+)$,  seen as function of the \textbf{BH} dimensionless  spin $a$ for different counter--rotating specific angular momentum $\ell_*^+$ of the inversion surfaces.
\subsubsection{The co--rotating torus center}\label{Sec:torus-center-crossing}
\begin{figure*}
\centering
\includegraphics[width=8cm]{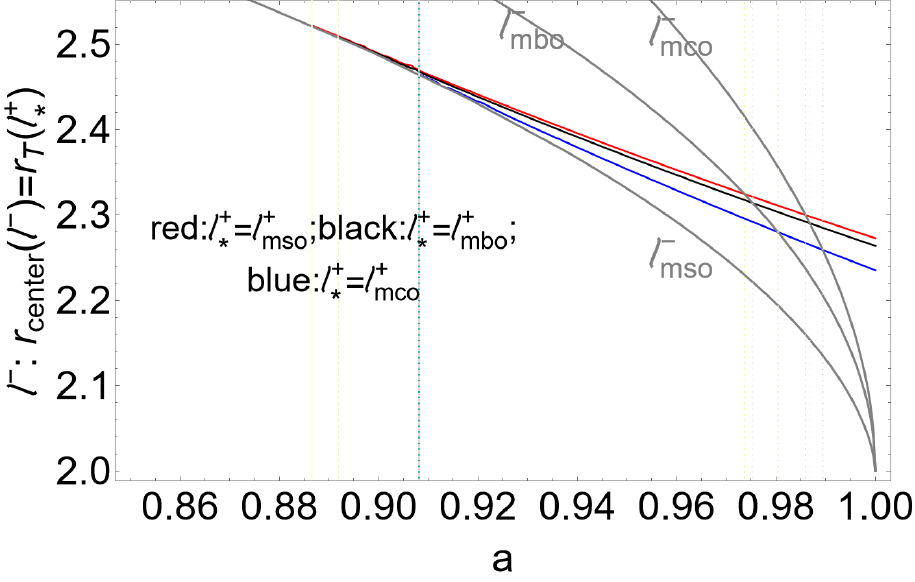}
\caption{Co--rotating  specific angular momentum $\ell^-$ solution of the equation $r_{center}^-(\ell^-)=r_\Ta(\ell_*^+)$,  as function of the \textbf{BH} dimensionless  spin $a$ for different counter--rotating specific angular momentum $\ell_*^+$ signed on the panel. There is  $\mathrm{Q}_\bullet$ for any quantity  evaluated at $r_\bullet$, where
$mbo$ is for marginally bound circular orbit, $mco$ is for marginally  circular orbit, $mso$ is for marginally stable circular orbit. $r_{center}^-$ is the center  of a co--rotating disk or proto-jets. $r_\Ta$ is the inversion radius  (where $u^\phi$=0).  (Dotted vertical line  are spins of Table\il\ref{Table:spins}). All quantities are dimensionless.}.\label{Fig:PlotgsnedibubbleC}
\end{figure*}
In Fig.\il\ref{Fig:PlotgsnedibubbleC}   some aspects of the analysis of the co--rotating toroids centers crossing an inversion surface are shown.
In general for these systems, the  relation
\bea\label{Eq:cross-center}
r_{center}^-(\ell^-)=r_\Ta(\ell_\bullet^+)
\eea
 holds. (Note,  the toroid  center $r_{center}^-(\ell^-)$ depend on the $\ell$ parameter only--see Sec.\il(\ref{Sec:thinkc}).)  In Fig.\il\ref{Fig:PlotgsnedibubbleC}    the
co--rotating   specific angular momentum $\ell^-$, solution of  Eq.\il(\ref{Eq:cross-center}), is shown   as function of the \textbf{BH} dimensionless  spin $a$ for different counter--rotating specific angular momentum $\ell_\bullet^+$ of the inversion surfaces.  Results for   $\cc_3^-$ toroids, having momentum in $\mathbf{L_3^-}$, are also  shown. As discussed in Table\il\ref{Table:Regions-BHs}, for the spacetimes classes, co--rotating  disks or proto--jets are partially embedded in the  inversion surfaces  for large \textbf{BH} spin ($a>a_{(b,s)}$) (and can be partially embedded with their  centers for $a_{(s,s)}$).
 For these systems the toroids inner region (including the cusp) is embedded in the inversion surface, while
  the tori outer region can be  external to the inversion surface.  However  Eq.\il(\ref{Eq:cross-center}) can also be seen as a limiting case  for  the configuration where the  torus center is internal or external to the inversion surface.
  As discussed in Sec.\il(\ref{Sec:cusp-crossing}) and
  Sec.\il(\ref{Sec:outer--edge-crossing}), this case will also serve as a constraint for the location of   the inner and outer edges of a quiescent torus. A quiescent torus edges depend also on the torus parameter $K^-$, and for $K^-\geq K_{center}^-(\ell^-)$  the torus can be very small, with the edges close to the disks centers.
  (This condition can also be verified for cusped tori with momenta $\ell^-\geq\ell_{mso}^-$.)
\begin{figure*}
\centering
\includegraphics[width=5cm]{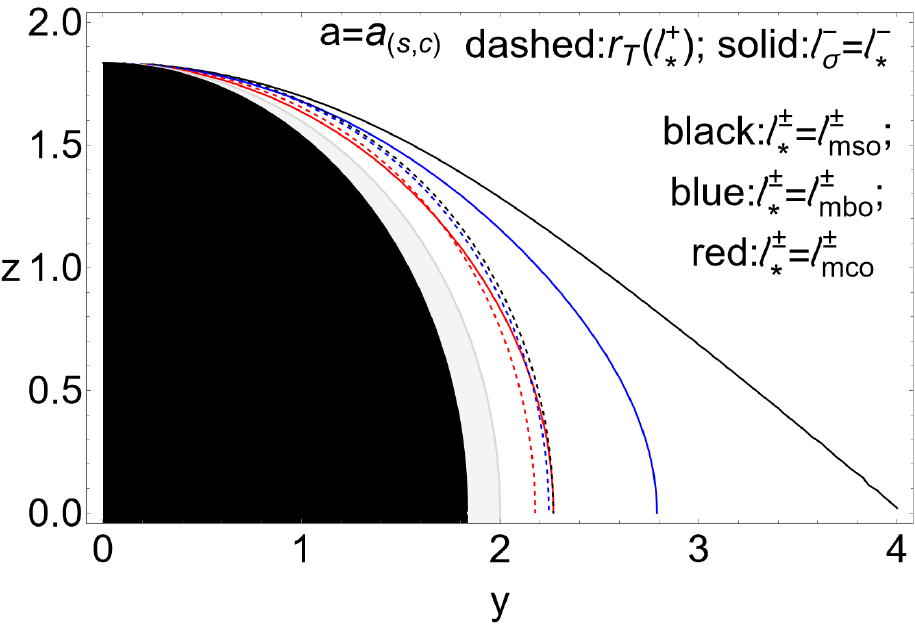}
\includegraphics[width=5cm]{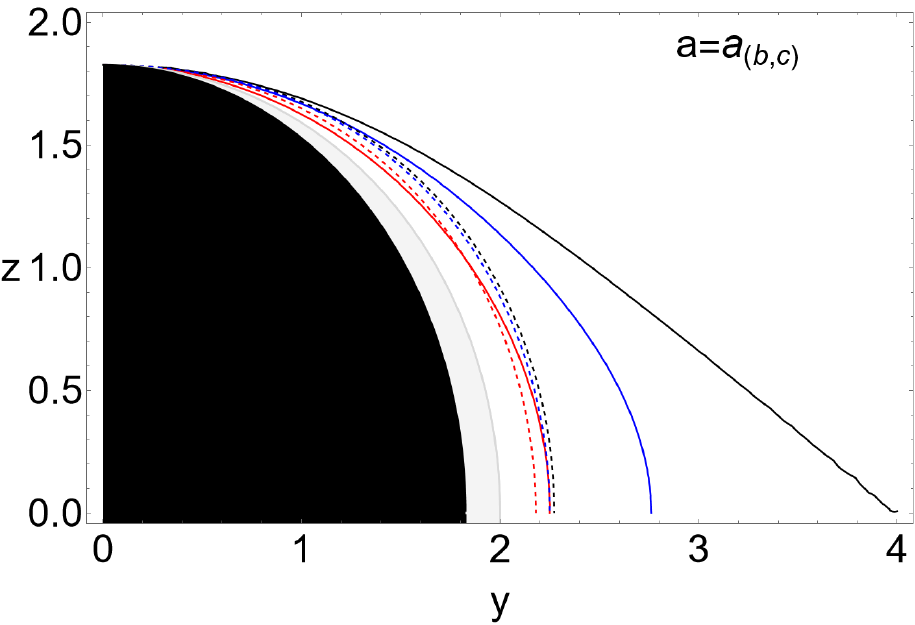}
\includegraphics[width=5cm]{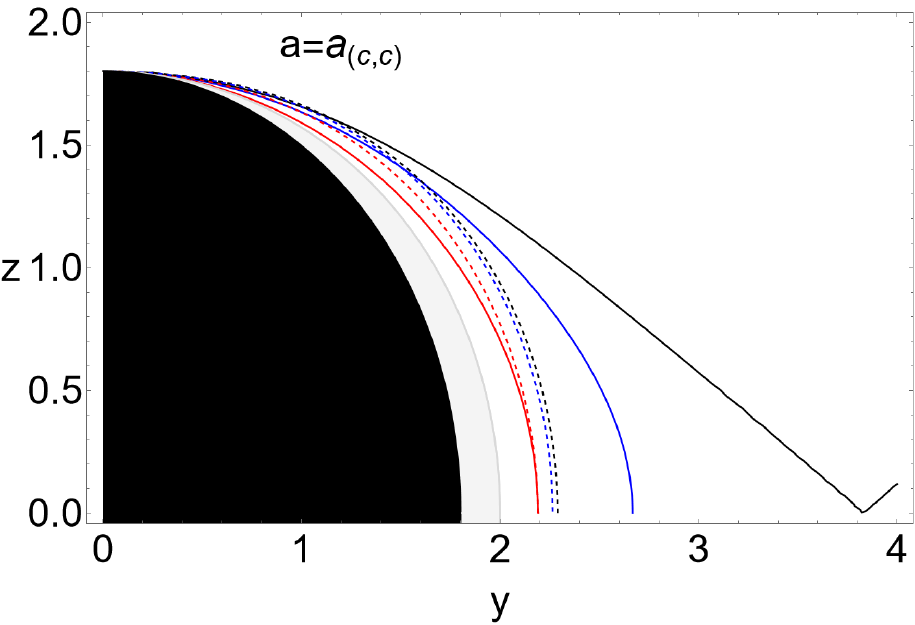}
\includegraphics[width=5cm]{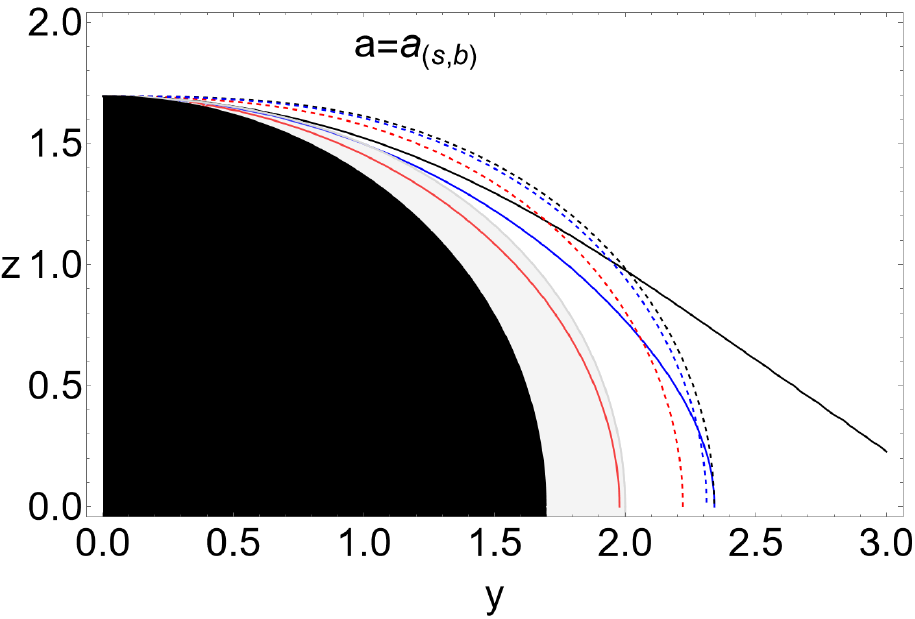}
\includegraphics[width=5cm]{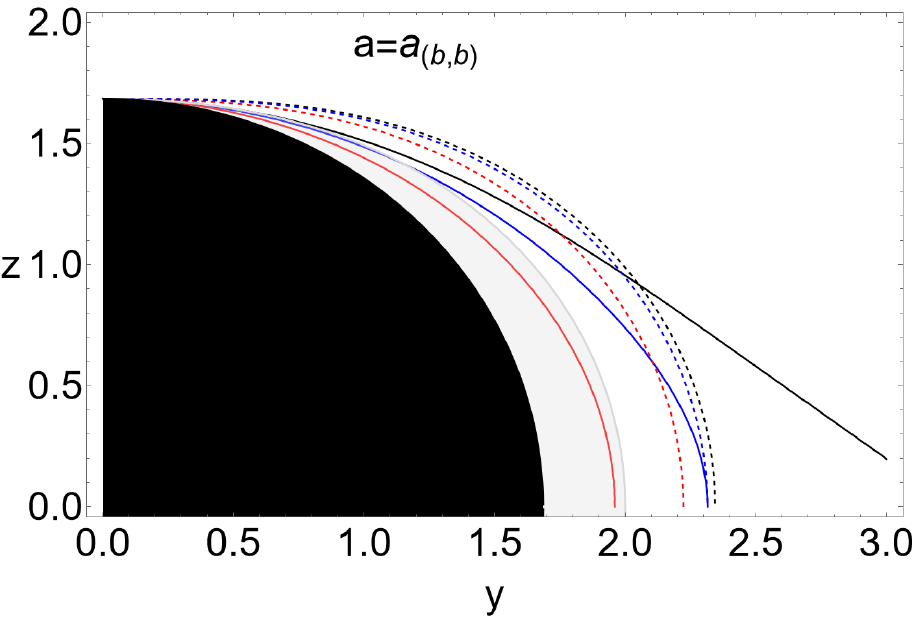}
\includegraphics[width=5cm]{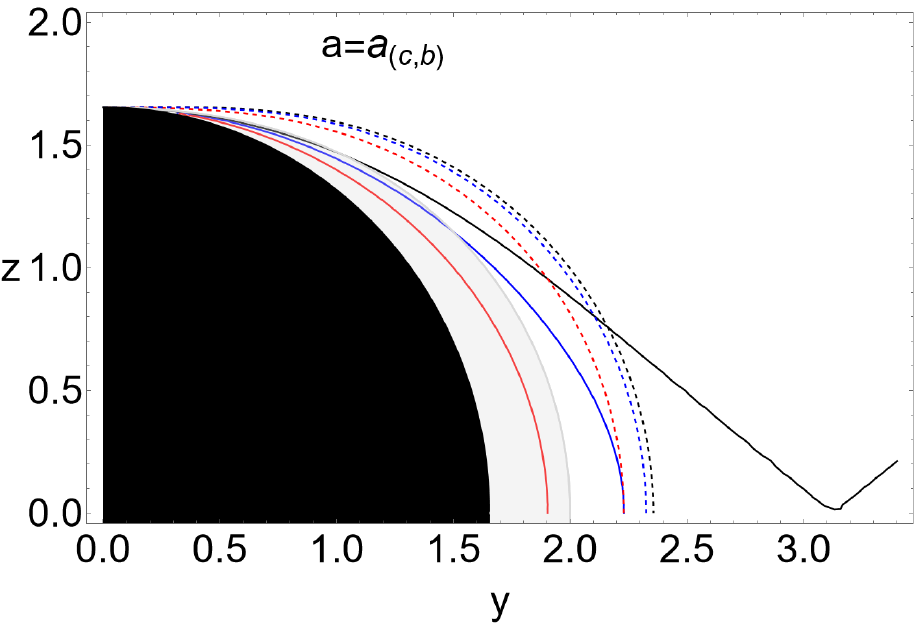}
\includegraphics[width=5cm]{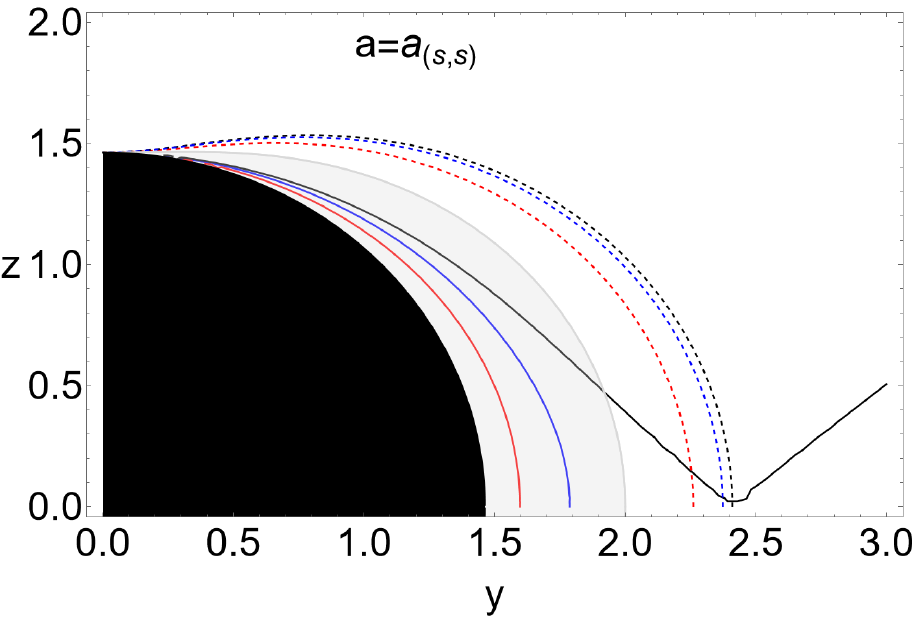}
\includegraphics[width=5cm]{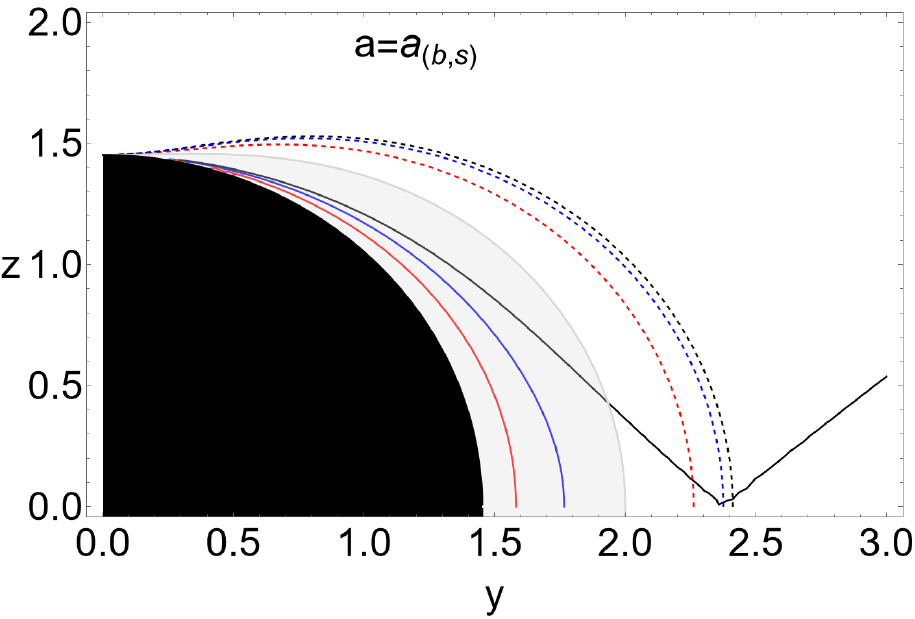}
\includegraphics[width=5cm]{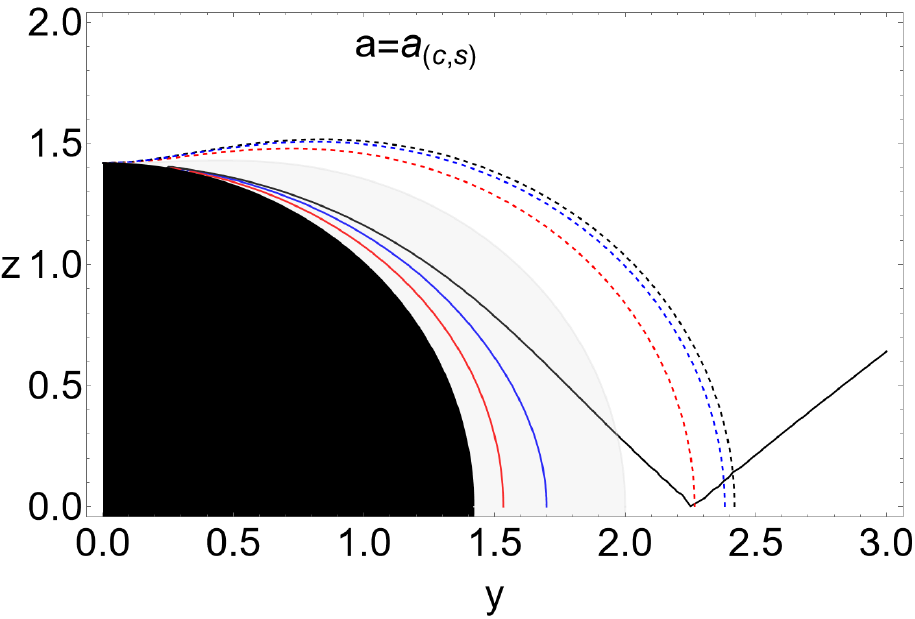}
\includegraphics[width=5cm]{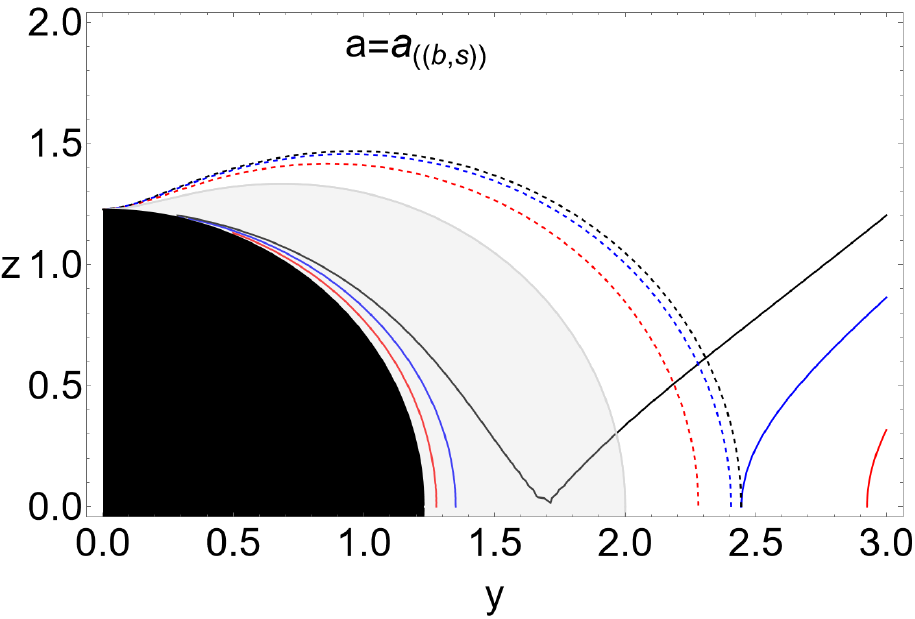}
\includegraphics[width=5cm]{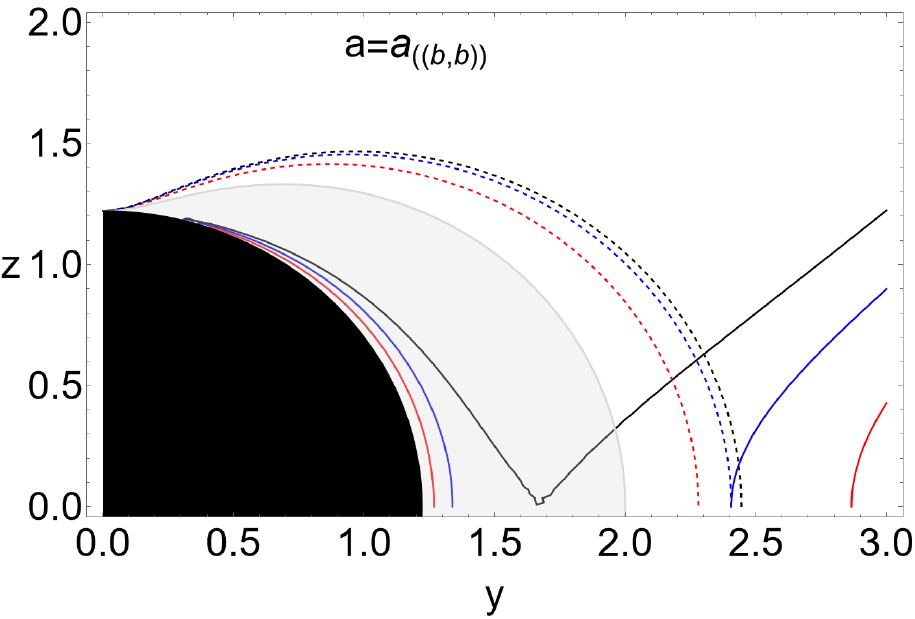}
\includegraphics[width=5cm]{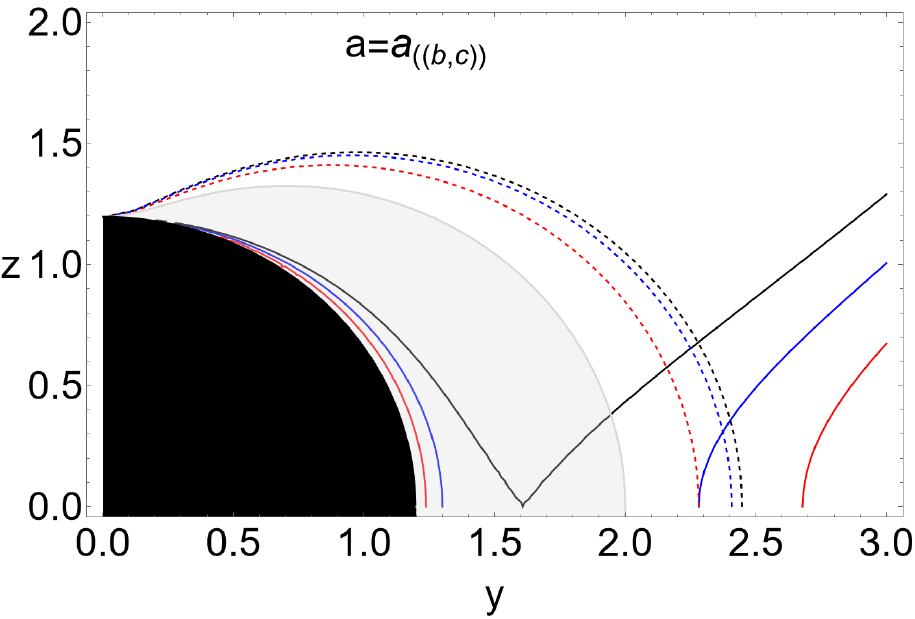}
\includegraphics[width=5cm]{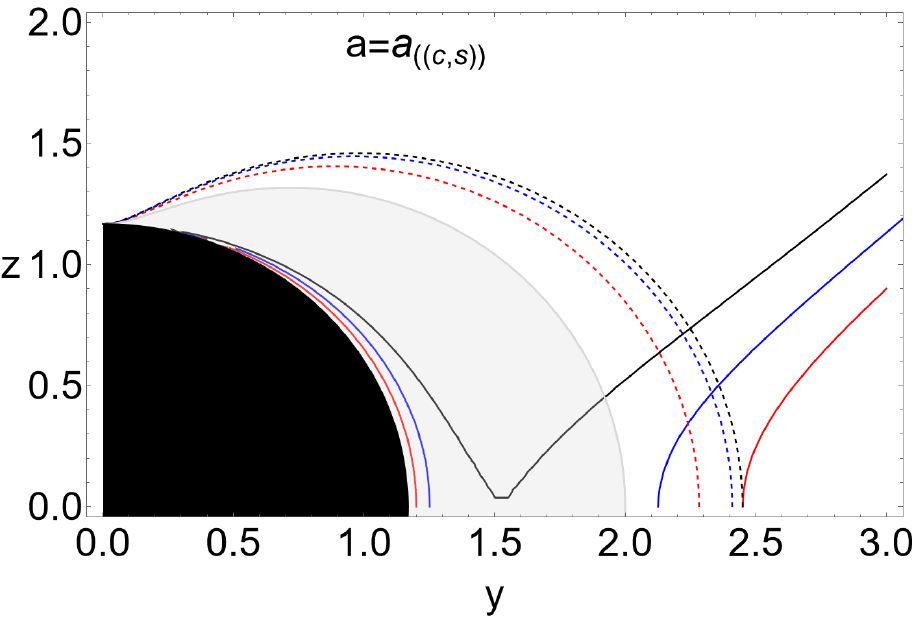}
\includegraphics[width=5cm]{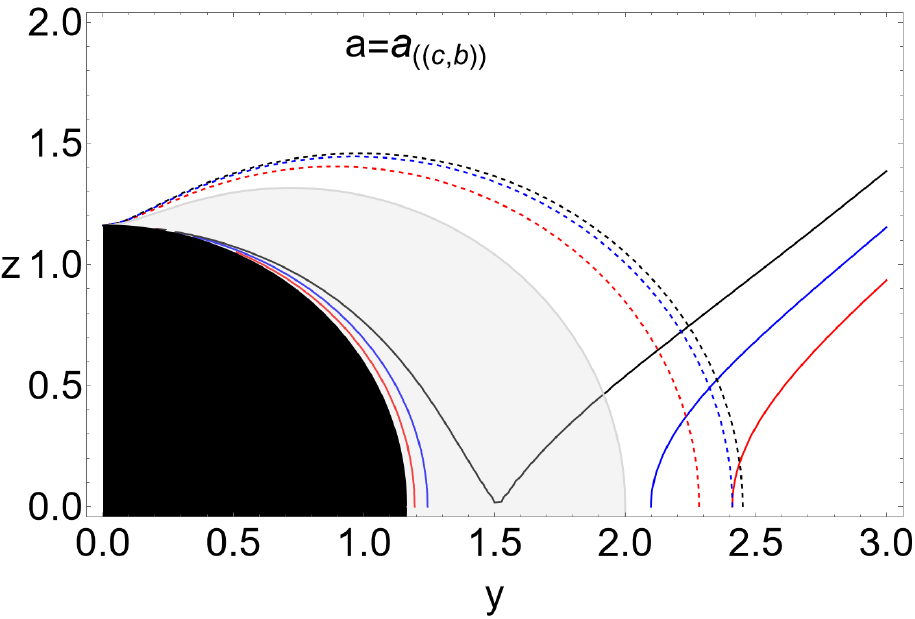}
\includegraphics[width=5cm]{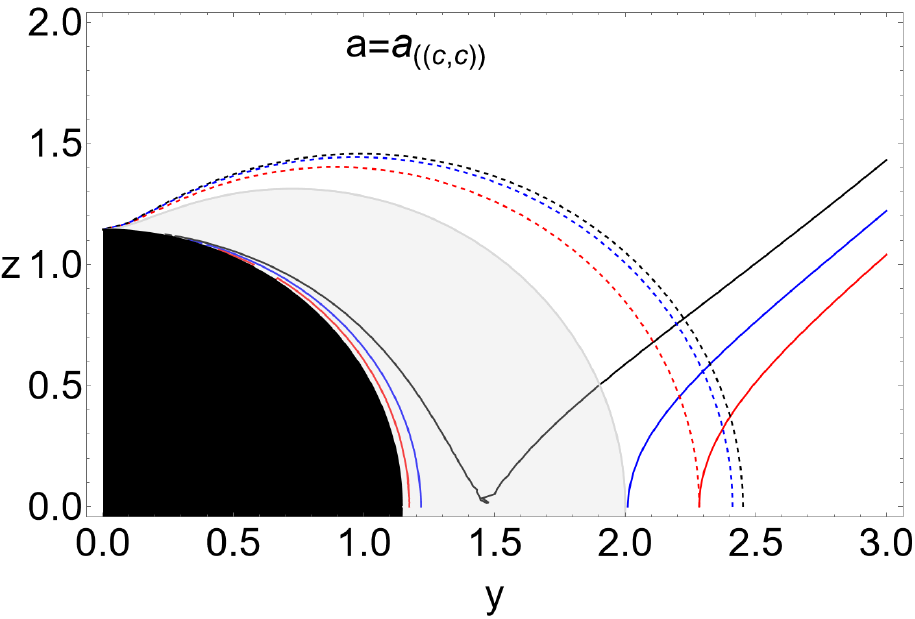}
\caption{Inversion surfaces crossing the toroids  pressure and geometrical extremes. Black region is the central \textbf{BH}, with dimensionless spin $a$. There is where  $r= \sqrt{y^2+z^2}$ and $\theta=\arccos({z}/r)$. Gray region is the outer ergoregion.  Dashed curves are the inversion surfaces  $r_\Ta$ at fixed counter--rotating angular momentum $\ell^+$. Solid  curves are  the momenta $\ell_\sigma^-=$constant for co--rotating surfaces defined  in Sec.\il(\ref{Sec:thinkc})  connecting the   toroids  pressure (and density) extremes  and geometrical extremes. Spins on the panels are defined in Table\il\ref{Table:spins}. There is  $\mathrm{Q}_\bullet$ for any quantity  evaluated at $r_\bullet$, where
$mbo$ is for marginally bound circular orbit, $mco$ is for marginally  circular orbit, $mso$  is for marginally stable circular orbit.  All quantities are dimensionless.    %
}\label{Fig:PlotMaxcrit099}
\end{figure*}
\begin{figure*}
\centering
\includegraphics[width=8cm]{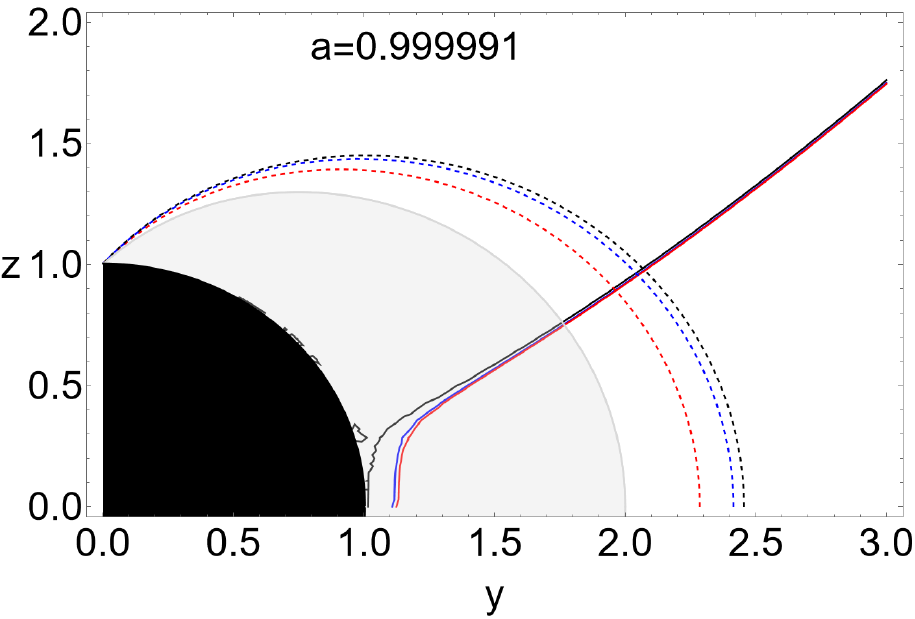}
\includegraphics[width=8cm]{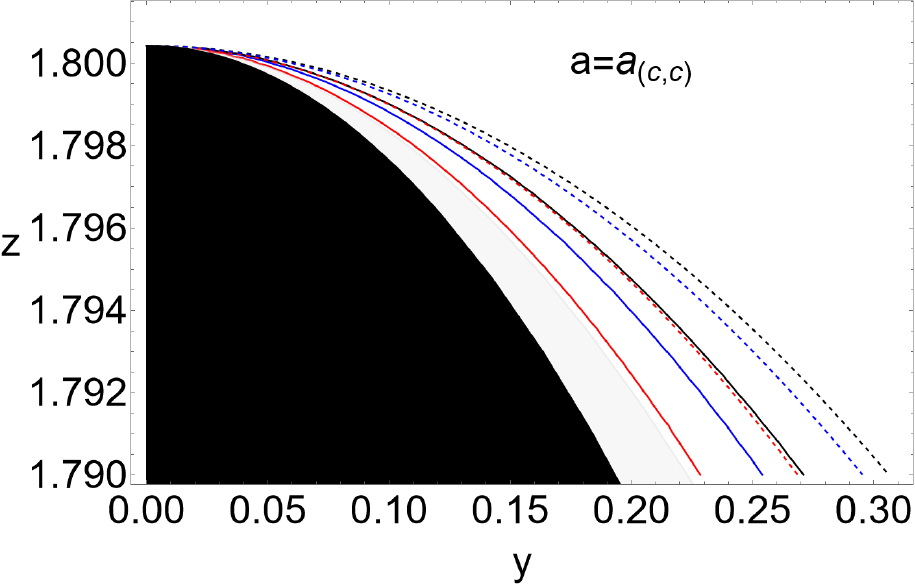}
\includegraphics[width=8cm]{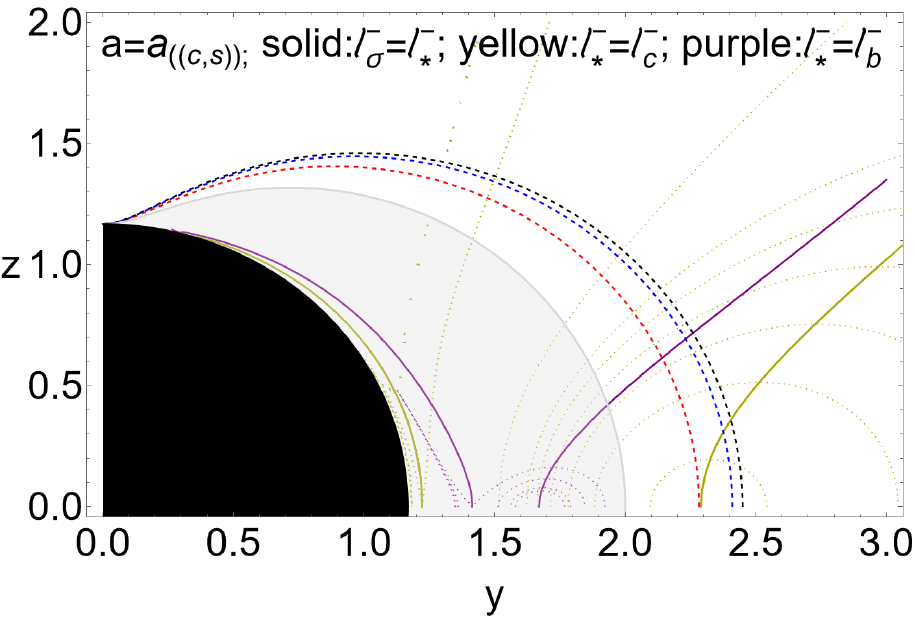}
\includegraphics[width=8cm]{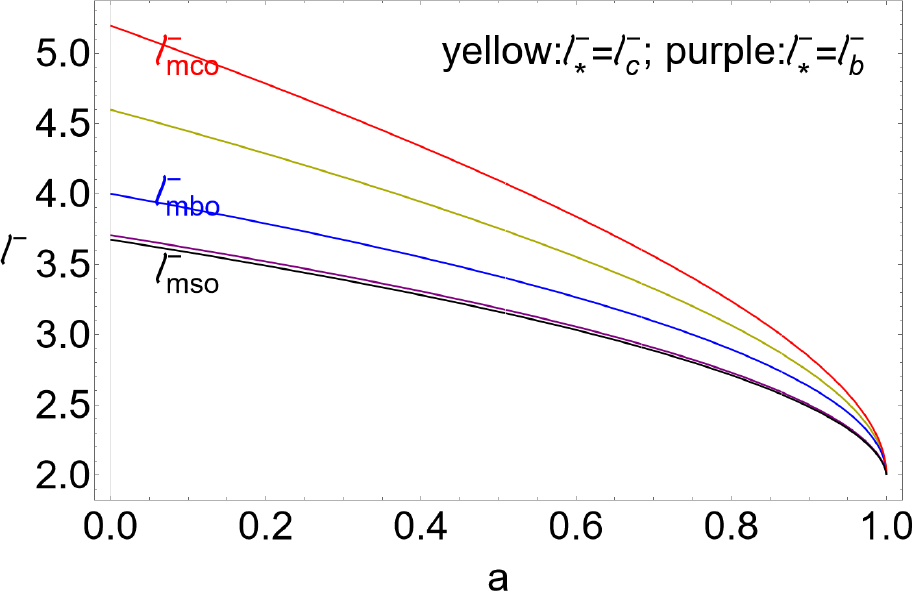}
\caption{Inversion surfaces crossing the geometrical and density  surfaces critical points. Black region is the central \textbf{BH} with dimensionless spin $a$. There is where  $r= \sqrt{y^2+z^2}$ and $\theta=\arccos({z}/r)$. Gray region is the outer ergoregion.  Dashed curves are the inversion surfaces, with radius $r_\Ta$, at fixed counter--rotating angular momentum $\ell^+$. (For further details see also caption of  Fig.\il\ref{Fig:PlotMaxcrit099}).  Upper line:   Analysis  of Fig.\il\ref{Fig:PlotMaxcrit099} for spin $a=0.999991$ (left panel),  and a close up-view for spin $a=a_{(c,c)}$ (right panel). Below left panel:   Analysis  of Fig.\il\ref{Fig:PlotMaxcrit099} for spin $a=a_{((c,s))}$ and angular momentum, shown  in the right panel as function of $a$, signed on the panel. Dotted lines are the orbiting surfaces.  Dashed lines are the inversion surfaces. All quantities are dimensionless.}.\label{Fig:PlotMaxcrit099sbcz}
\end{figure*}
\begin{figure*}
\centering
\includegraphics[width=5cm]{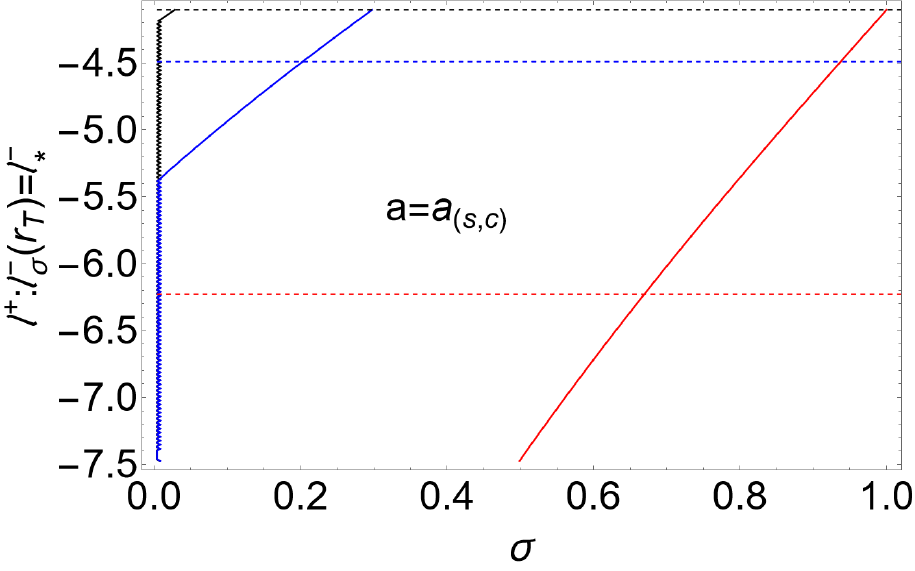}
\includegraphics[width=5cm]{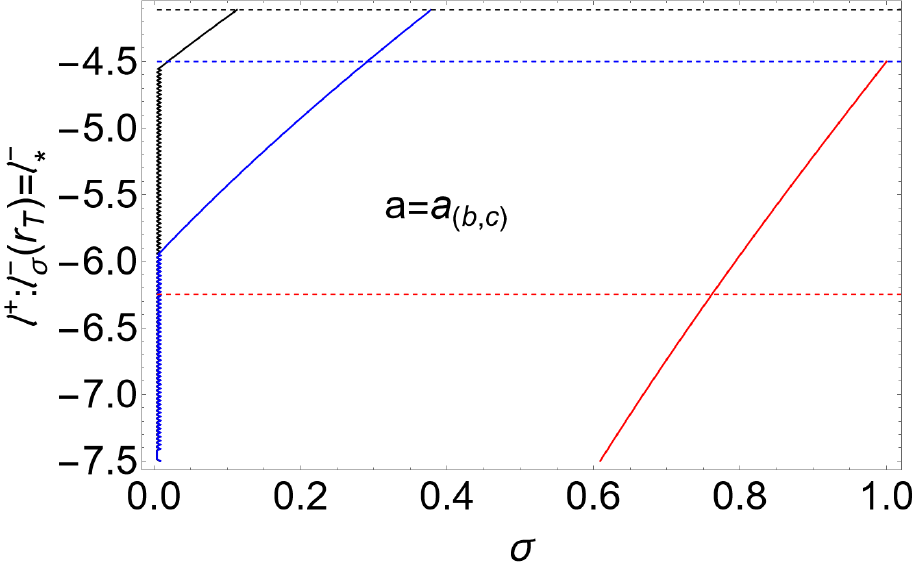}
\includegraphics[width=5cm]{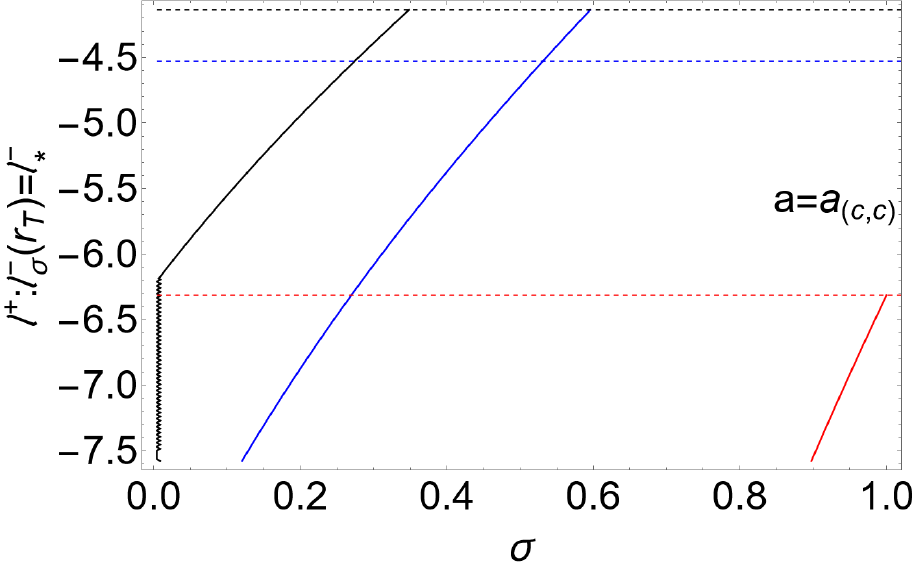}
\includegraphics[width=5cm]{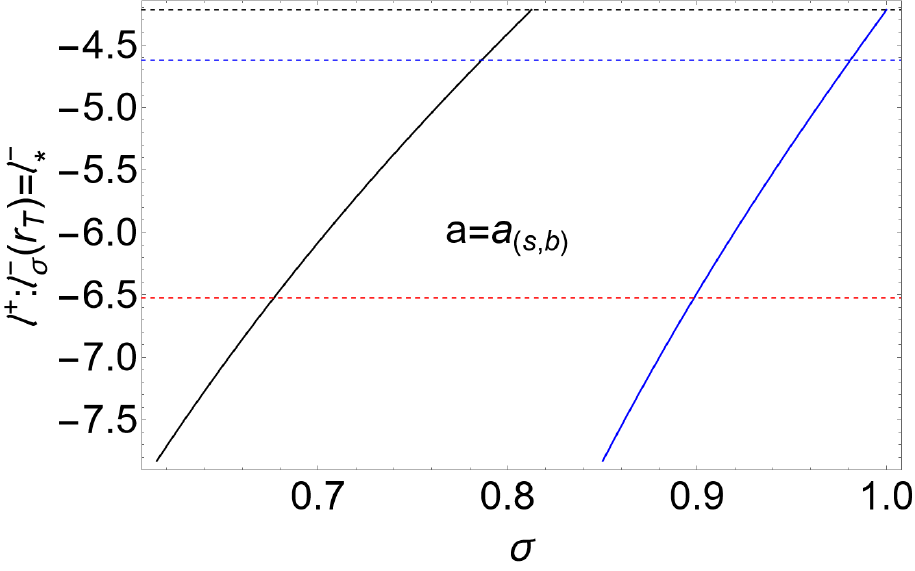}
\includegraphics[width=5cm]{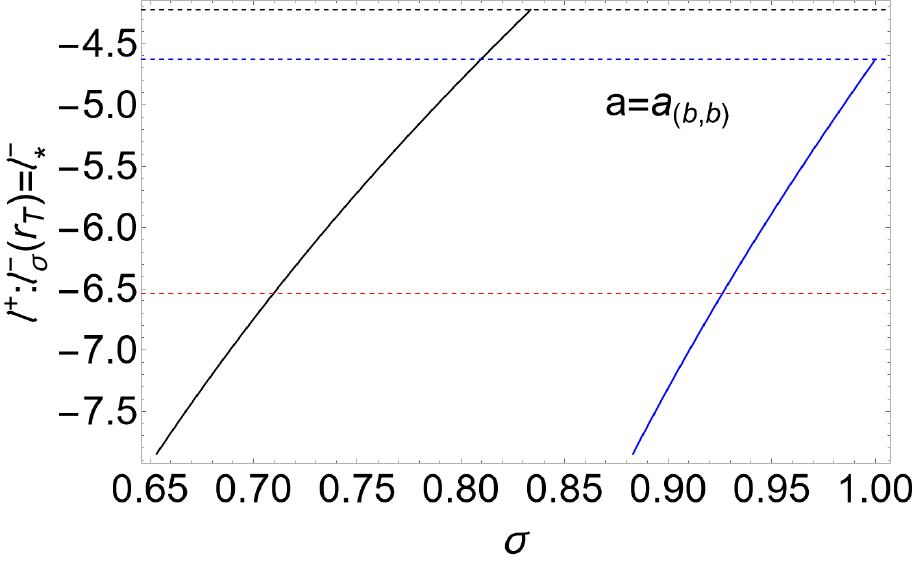}
\includegraphics[width=5cm]{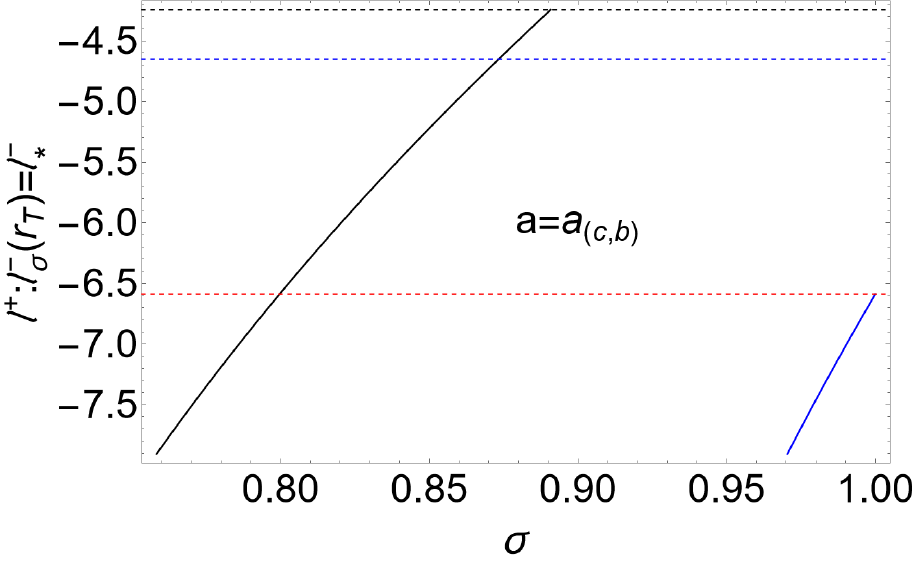}
\includegraphics[width=5cm]{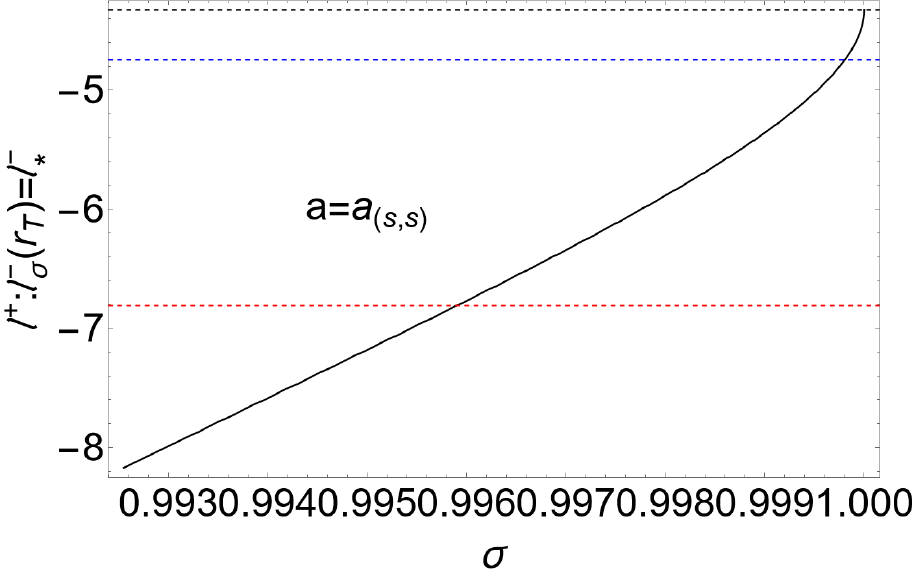}
\includegraphics[width=5cm]{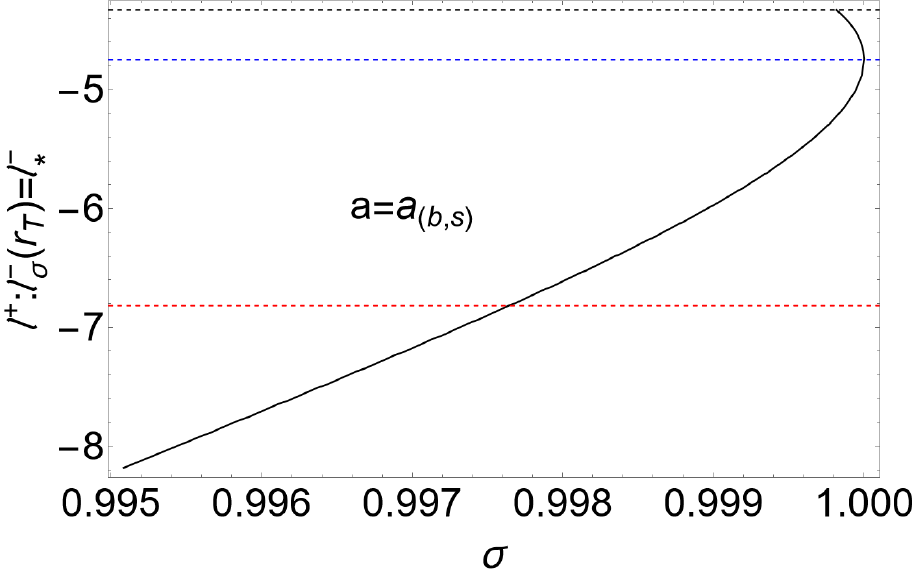}
\includegraphics[width=5cm]{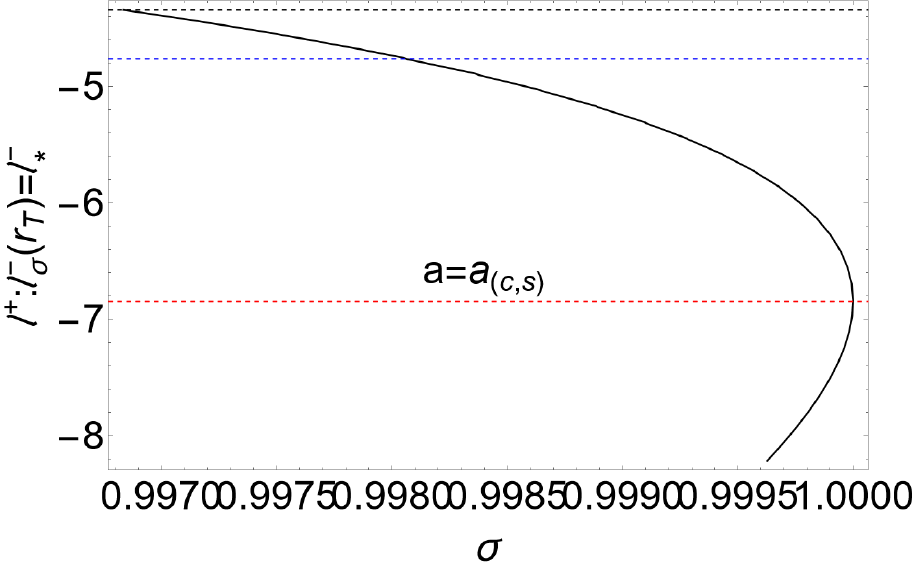}
\includegraphics[width=5cm]{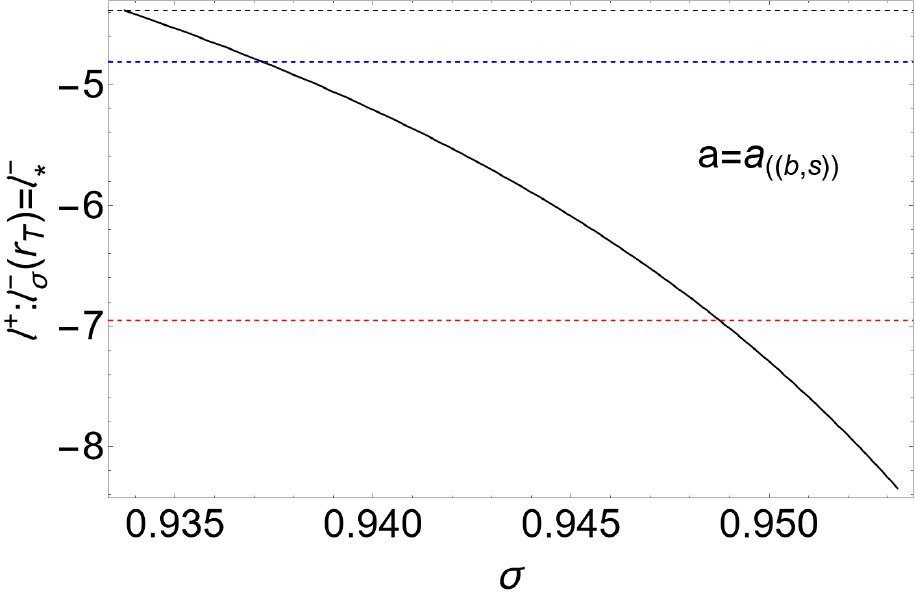}
\includegraphics[width=5cm]{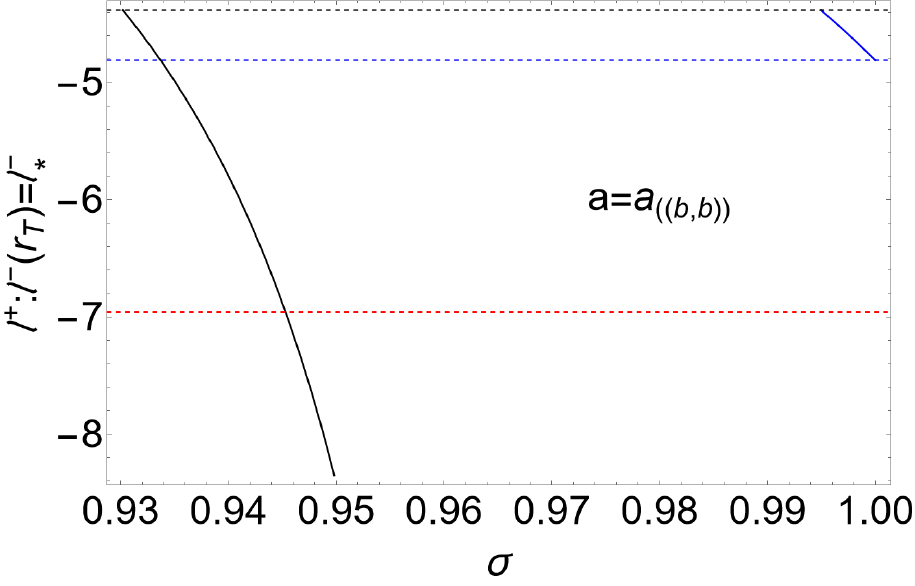}
\includegraphics[width=5cm]{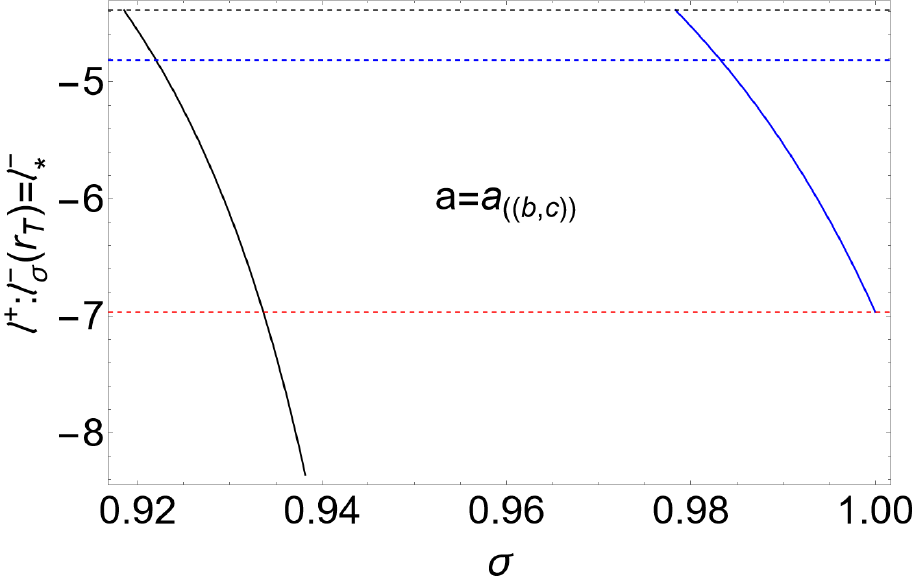}
\includegraphics[width=5cm]{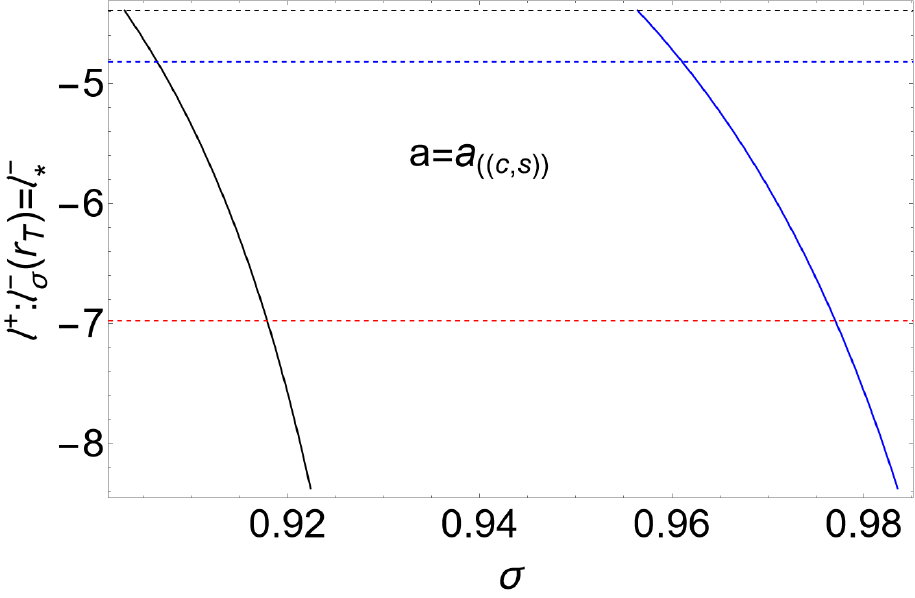}
\includegraphics[width=5cm]{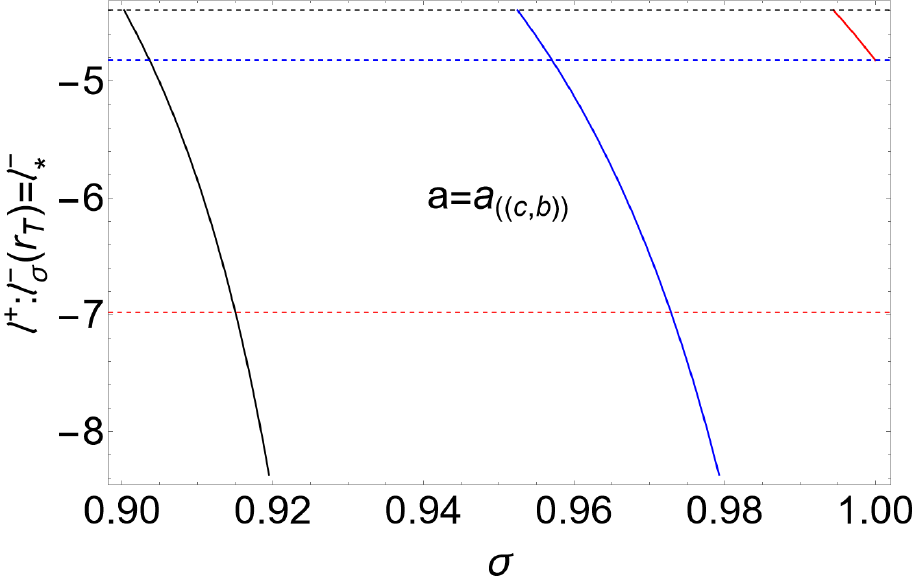}
\includegraphics[width=5cm]{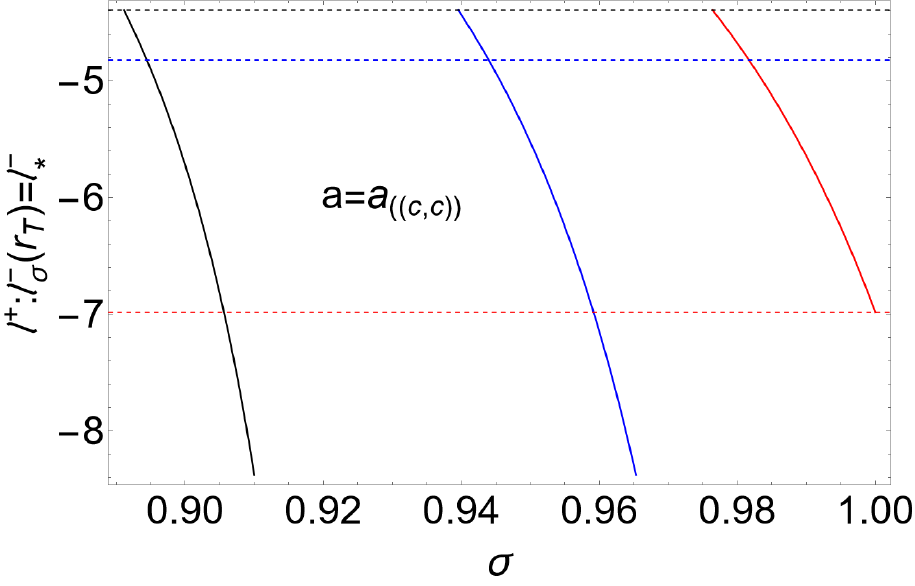}
\includegraphics[width=5cm]{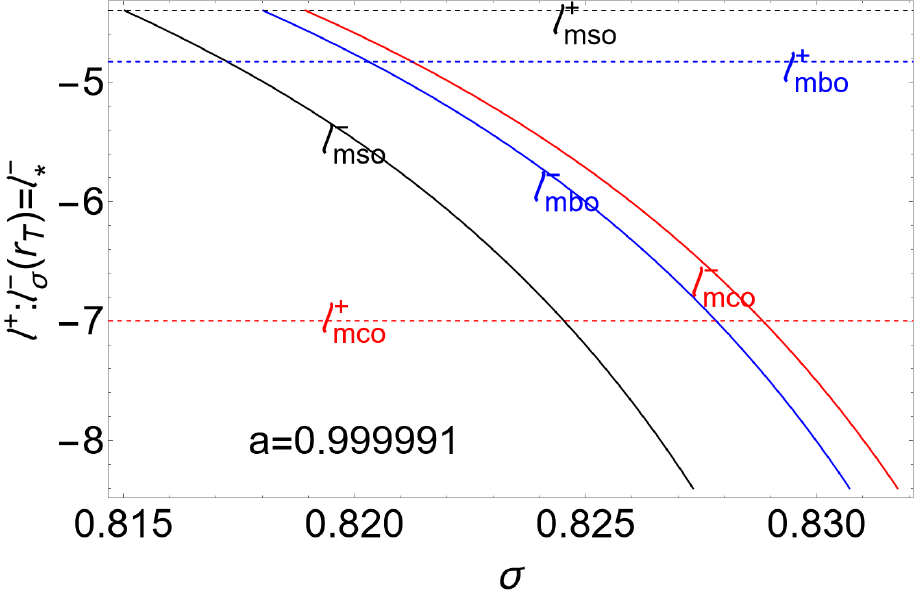}
\caption{Counter--rotating angular momentum $\ell^+$, solution of the equation $\ell_\sigma^-(r_\Ta(\ell^+))=\ell_\star^-$, versus the angle $\sigma\equiv\sin^2\theta\in[0,1]$ for  different spins, defined in Table\il\ref{Table:spins}, where the co--rotating specific angular momentum
$\ell_\star^-$ is signed on the (bottom)  panel.  The co--rotating angular momentum $\ell_\sigma^-$ is defined in Sec.\il(\ref{Sec:thinkc}) .   Radius $r_\Ta(\ell^+)$ is the inversion surface.  There is  $\mathrm{Q}_\bullet$ for any quantity  evaluated at $r_\bullet$, where
$mbo$ is for marginally bound circular orbit, $mco$  is for marginally  circular orbit, $mso$  is for marginally stable circular orbit.     All quantities are dimensionless. (For further details see also  caption of  Fig.\il\ref{Fig:PlotMaxcrit099}.)}\label{Fig:PlotMaxcrit099M}
\end{figure*}
\begin{figure*}
\centering
\includegraphics[width=8cm]{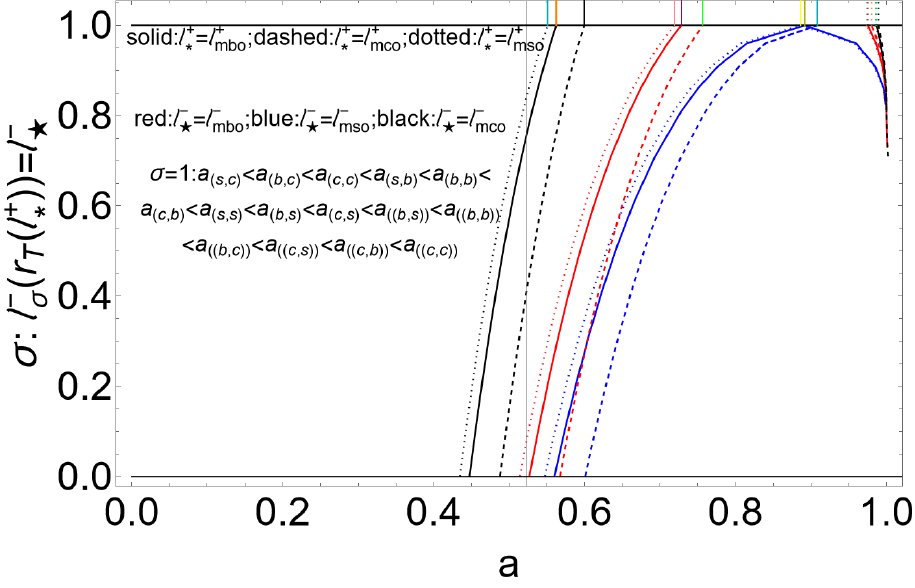}
\caption{Solution of the equation
 $\sigma: \ell_\sigma^-((r_\Ta)=\ell_\star^-$ versus the dimensionless \textbf{BH}  spin $a$. There is  $\sigma\equiv\sin^2\theta\in[0,1]$. The co--rotating angular momentum $\ell_\sigma^-$ is defined in Sec.\il(\ref{Sec:thinkc}) .  $r_\Ta(\ell^+)$ is the  inversion surface radius.  Solutions express the angle $\sigma$ for   the inversion surface  crossing  the curves  correspondent to  the   toroids extreme points  (centers, cusps and geometrical maxima). The figure shows the solutions for different counter--rotating  (co--rotating) angular momentum
$\ell_*^+$   ($\ell_\star^+$).  There is  $\mathrm{Q}_\bullet$ for any quantity  evaluated at $r_\bullet$, where
$mbo$ is for marginally bound circular orbit, $mco$ is for marginally  circular orbit, $mso$  is for marginally stable circular orbit.
Solid vertical lines in the region $\sigma>1$ mark the spins defined in Table\il\ref{Table:spins}.
  All quantities are dimensionless.}\label{Fig:Plotgrimmunificatione}
\end{figure*}
\begin{figure*}
\centering
\includegraphics[width=8cm]{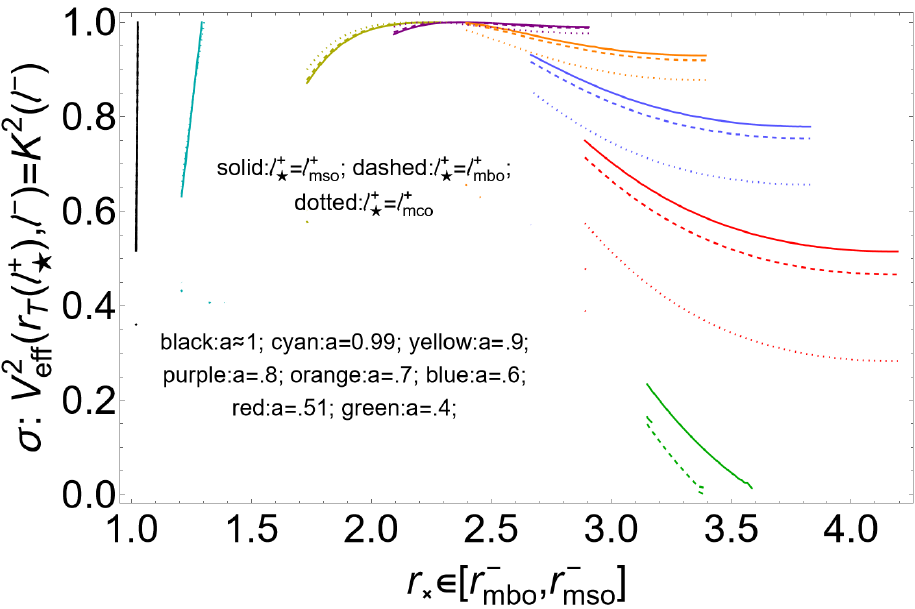}
\includegraphics[width=8cm]{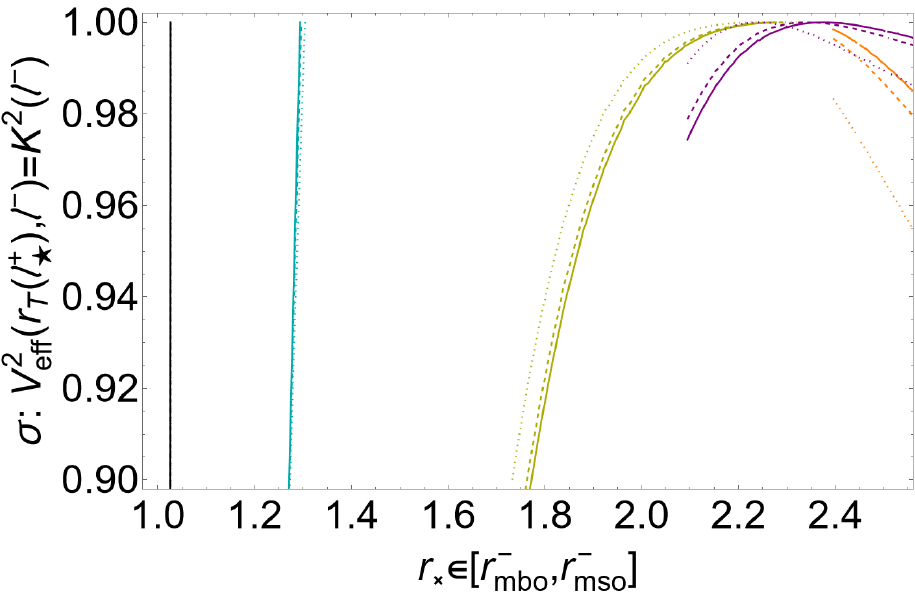}
\caption{Solution of the equation
 $\sigma: V_{eff}^2((r_\Ta,\ell^-)=K^2$ versus the co--rotating disk cusp $r_\times$. There is  $\sigma\equiv\sin^2\theta\in[0,1]$. $ V_{eff}$ is the torus effective potential function. There is $K\equiv V_{eff}(r,\ell(r))$.  $r_\Ta(\ell^+)$ is the  inversion surface radius. Right panel is a  close--up view of the left panel. The figures show the solutions for different counter--rotating angular momentum
$\ell_\star^+$,   for different spins  signed on the panel.  There is  $\mathrm{Q}_\bullet$ for any quantity  evaluated at $r_\bullet$, where
$mbo$ is for marginally bound circular orbit, $mco$  is for marginally  circular orbit, $mso$  is for marginally stable circular orbit.
  All quantities are dimensionless. Solutions express the crossing of the inversion surface with the  co--rotating  toroids orbiting the central Kerr  \textbf{BH}.}.\label{Fig:Plotgsnder}
\end{figure*}
It can be $r_{center}^-(\mathbf{L_1^-})< r_\Ta(\mathbf{L_1^+})$ for $a\in]a_{(s,s)},a_{(b,s)}[$,
it can be $r_{center}^-(\mathbf{L_1^-})< r_\Ta(\mathbf{L_2^-})$ for $a\in]a_{(b,s)},a_{(c,s)}[$,
 It can be  $r_{center}^-(\mathbf{L_1^-})< r_\Ta(\mathbf{L_3^-})$ for $a\in]a_{(c,s)},a_{((b,s))}[$.
 Within these conditions it can be also
 $r_{center}^-(\mathbf{L_2^-})< r_\Ta(\mathbf{L_1^-})$ for $a\in]a_{((b,s))},a_{((b,b))}[$, while for
$a\in]a_{((b,b))},a_{((b,c))}[$  it must be
 $r_{center}^-(\mathbf{L_1^-})< r_\Ta(\mathbf{L_1^+})$
 other conditions and faster spinning \textbf{BHs}  are detailed in Table\il\ref{Table:Regions-BHs}
\subsubsection{The critical points of pressure (disks centers and cusps) and the tori geometrical maximum}\label{Sec:geometric--crossing}
The  toroids  critical points of pressure ({the extreme points of the effective potential function $V_{eff}$  of Eq.\il(\ref{Eq:scond-d})}), which are the disks centers and cusps, are located on the equatorial plane, and depend on the parameter $\ell^-$ only. The torus  geometrical maximum , which is located on plane different from the equatorial,  depends  on the parameter $\ell^-$ only, for the  cusped tori  $\cc_\times^-$, and on  the   $(\ell^-,K^-)$ parameters   for quiescent tori. The critical points of pressure and the  geometrical maximum are  points of
  the curve $\ell^-_\sigma=$constant, where
$\ell^-_\sigma:\partial_y V_{eff}^2=0$ (for  $\{z=r \cos\theta,y=r \sin \theta \sin\phi,x=r \sin\theta \cos\phi\}$),
where  $ \left.\ell_\sigma^-\right|_{\sigma=1}=\ell^-:\left.\partial_r V_{eff}\right|_{\sigma=1}=0$,  providing,  on the equatorial plane, the critical points of pressure--see Fig.\il\ref{Fig:PlotMaxcrit099sbcz}--bottom left panel.

In this section we constrain the inversion surfaces crossing  the tori critical points of pressure and  geometrical maximum
by considering   the solutions of the equation
\bea\label{Eq:sigma:ellellstar}
 \sigma(a): \ell_\sigma^-(r_\Ta(\ell^+_\bullet))=\ell_\star^-
 \eea
for different angular momenta $(\ell^+_\bullet,\ell^-_\star)$,
  versus the dimensionless \textbf{BH}  spin $a$. Results are shown in Fig.\il\ref{Fig:Plotgrimmunificatione}.
 Solutions provide   the angle $\sigma$  where    the inversion surface with momentum $\ell^+_\bullet$   crosses the curves  correspondent to   extreme points  (centers, cusps and geometrical maxima) of the   toroids with angular momentum $\ell_\star^-$.

We also solved Eq.\il(\ref{Eq:sigma:ellellstar}), for   the  counter--rotating angular momentum $\ell^+$ of the inversion surfaces,  obtaining the solutions
\bea\label{Eq:sigma:ellellstars}
 \ell^+(\sigma): \ell_\sigma^-(r_\Ta(\ell^+_\bullet))=\ell_\star^-
 \eea
  versus the angle $\sigma$ for  different spins defined in Table\il\ref{Table:spins}, providing  the momentum  $\ell^+_\bullet$  of   the inversion surface   crossing  the curves  correspondent to   extreme points  (centers, cusps and geometrical maxima) of the   toroids with angular momentum $\ell_\star^-$. Results are shown in  Fig.\il\ref{Fig:PlotMaxcrit099M}.
For  a torus with parameter $(\ell^-,K^-)$  there is
 $r^-_{inner}(\ell^-,K^-)<r_{center}^-(\ell^-)<r_{\max}^-(\ell^-,K^-)<r_{outer}^-(\ell^-,K^-)$,    where $r_{\max}^-(\ell^-)$ is the projection on the equatorial plane of the torus  geometric  maximum.    In Fig.\il\ref{Fig:PlotMaxcrit099} and  Fig.\il\ref{Fig:PlotMaxcrit099sbcz}    we showed
  the curves  with momenta $\ell_\sigma^-$ crossing (in the plane $(y,z)$) the inversion surfaces $r_\Ta$ at fixed counter--rotating angular momentum $\ell^+$.  From Fig.\il\ref{Fig:PlotMaxcrit099}   it is clear how the  solutions $\ell_\sigma^-=\ell_\star^-$  on plane $(y,z)$  are two curves, which are  joint on  $r=r_{mso}^-$ for $\ell_\star^-=\ell^-_{mso}$. The outer  curve contains the torus center (on the equatorial plane) and the geometrical maximum (on $\sigma<1$).  The inner curve is  related to the accretion flows onto \textbf{BH}, and on the equatorial plane it coincides with  the toroid cusp (see Fig.\il\ref{Fig:PlotMaxcrit099sbcz}-bottom right panel).
 For $\ell_\star^-=\{\ell^-_{mbo},\ell^-_{mco}\}$ the outer and inner curves are disjoint.
 The curves cross the inversion surfaces in the spacetimes with  \textbf{BH} spins  $a\geq a_{(s,s)}$.
 From Fig.\il\ref{Fig:PlotMaxcrit099} it is clear that the relevant spins for the inversion surface   $r_\Ta(\ell^+_{mso})$ are $a_{((\bullet,s))}$ and
  $a_{(s,s)}$, according to the different co--rotating momenta $\ell_\bullet^-\in\{\ell_{mco}^-,\ell_{mbo}^-\}$.
  For the   inversion surface   $r_\Ta(\ell^+_{mbo})$ the discriminant spins are   $a_{((\bullet,b))}$ and
  $a_{(b,s)}$, according to the different co--rotating momenta $\ell_\bullet^-\in\{\ell_{mco}^-,\ell_{mbo}^-\}$, indicating the spacetimes characterized by the crossing of the inversion surfaces with the curves of tori pressure and geometrical extremes for fixed momenta $\ell^\pm$,  and  for the inversion surface   $r_\Ta(\ell^+_{mco})$ we note the role of the  \textbf{BH} spins  $a_{((\bullet,c))}$ and
  $a_{(c,s)}$.
  On the other hand, focusing on the  co--rotating toroids curves $\ell_\sigma^-=\ell_\star^-=\ell^-_{mso}$,
   spins $a_{(\bullet,s)}$ (for the different counter--rotating momenta $\ell_\bullet^+\in\{\ell_{mco}^+,\ell_{mbo}^+,\ell_{mso}^+\}$) are relevant for the crossings with the inversion surfaces,   while the spin for the curves  for
  $\ell_\star^-=\ell^-_{mbo}$ there are the  limiting spins $a_{((b,\bullet))}$.
 For
  $\ell_\star^-=\ell^-_{mco}$, there are the spins  $a_{((c,\bullet))}$, limiting  for the crossings with the co--rotating surfaces with $\ell^-=\ell^-_{mco}$.
The surfaces
   $\ell_\sigma^-=\ell_\star^-$ are very close for faster spinning \textbf{BHs} (Fig.\il\ref{Fig:PlotMaxcrit099sbcz}--upper left panel).
   The inner curves regulate therefore also the accretion flows with respect to the inversion surfaces close to the \textbf{BH} poles (see Fig.\il\ref{Fig:PlotMaxcrit099sbcz}--upper right panel). For materials accreting into   the central Kerr \textbf{BH} the  toroidal velocity of the $u$counter--rotating  accretion flows  changes sign on the inversion surfaces.
  It is  evident from  Fig.\il\ref{Fig:PlotMaxcrit099M} the discriminant rule   of the spins $a_{(\bullet,s)}$, relating $\ell^\pm$ of the  inversion surfaces and the co--rotating toroids respectively.
\subsubsection{The off--equatorial crossing of the  co--rotating toroids with  an inversion surface}\label{Sec:off-equatorial--crossing}
In this section we investigate the  co--rotating cusped toroids (proto-jets and tori)   partially contained in an inversion surface, considering the  toroids crossing the inversion surfaces on planes different from the equatorial.
Part of this problem has been faced in Sec.\il(\ref{Sec:off-equatorial--crossing}), where  we concentrated on the torus  geometrical maximum.
In here, we narrow the analysis to  cusped co--rotating  toroids.
 Results of this analysis are in Figs\il\ref{Fig:Plotgrimmunificatione},\ref{Fig:Plotgsnder} and Figs\il\ref{Fig:PlotMaxcrit099},\ref{Fig:PlotMaxcrit099sbcz},\ref{Fig:PlotMaxcrit099M}.

In Fig.\il\ref{Fig:Plotgsnder} we show the  angles $\sigma$, solutions  of
the equation:
\bea
 \sigma: V_{eff}^2((r_\Ta(\ell_\star^+),\ell^-(r_\times^-))=K^2_\times
\eea
  versus the disk cusp $r_\times^-\in[r_{mbo}^-,r_{mso}^-]$,  for different counter--rotating angular momentum
$\ell_\star^+$  defining the inversion surfaces,   for different spins  signed on the panels.   Solutions are the angles $\sigma$ at  the crossing of the inversion surface with the  co--rotating  cusped tori $\cc_\times^-$ orbiting the central \textbf{BH}.  We note that different situations emerge for the faster and slower attractors.

Radii $\{r_{center}^-,r_{\times}^-,r_J^-,r^-_{\max}\}$, where $r^-_{\max}$ is the geometrical maximum projection on the equatorial plane,
are  related by the  curves   $\ell_\sigma^-=$constant,
 where, on the equatorial plane, there  is   $\left.\ell_\sigma^-\right|_{\sigma=1}=\ell^-$.
Fig.\il\ref{Fig:Plotgrimmunificatione} shows   the angles $\sigma(a)$, solutions of the equation
 $\ell_\sigma^-(r_\Ta(\ell^+_\bullet))=\ell_\star^-
$
  as function  of the dimensionless \textbf{BH}  spin $a$, where    the inversion surfaces with momentum $\ell^+_\bullet$   cross the curves  correspondent to    the geometrical maxima and the pressure critical  points  of the   toroids with angular momentum $\ell_\star^-$.  Solutions for   the  counter--rotating angular momentum $ \ell^+(\sigma)$ of the inversion surfaces,
  versus the angle $\sigma$ are shown  for  different spins in Fig.\il\ref{Fig:PlotMaxcrit099M}.
  The geometrical maximum represents a limiting value of the toroidal surface crossing the inversion surface in the inner or outer region.
\section{Conclusion}\label{Conclsusion}
Co--rotating toroids and proto-jets, orbiting a Kerr \textbf{BH},  can be   embedded, external  or partially contained  in the  spacetimes inversion surfaces, according to the \textbf{BHs} spins and the counter--rotating flows specific  angular momentum.
The inversion surfaces are located out of the outer ergoregion  of the Kerr \textbf{BH} spacetime, and they  are defined for
counter--rotating flows only, having specific angular momentum $\ell<0$.
We showed that a counter--rotating  toroid is always external to the inversion surfaces for  counter--rotating  flows and inversion surfaces specific angular momentum sufficiently large in magnitude, i.e.  $\ell^+<\ell_{mso}^+$.
The constraints on the inversion surfaces  with lower specific angular momentum in magnitude, crossing an outer counter--rotating toroid are provided in Sec.\il(\ref{Sec:Aspects}). The constraints  define the limiting angular momentum $\ell_G^+$ of Eq.\il(\ref{Eq:LG}) for the counter--rotating flows (see Fig.\il\ref{Fig:Plotsigi}).
Therefore in this analysis we concentrated on the counter--rotating  (and $u$counter--rotating)  flows inversion surfaces of the Kerr spacetimes in relation to the (inner) co--rotating toroids.

{Toroids are found as equi--potential surfaces from  the scalar function $V_{eff}$ of  Eq.\il(\ref{Eq:scond-d})  with constant  specific angular momentum.}

{In this analysis we have heavily relied on the fact that the inversion surface (defined as the loci of the vanishing toroidal flow velocity) and the surface separating the co-rotating and counter-rotating toroids,  do not coincide. This is in fact an essential aspect constituting the premise of this analysis, leading to the possibility that the fluid coming from an external counter-rotating disk can impact on a inner co-rotating disk with different conditions on the fluids toroidal velocity  components.
If the disks orbit  very far from the attractor,  for example having specific angular momentum parameter $\ell^\pm \in \mathbf{L}_3^\pm$,   there can be  no separation regions: the outer torus of  the pair could be for example a quiescent co--rotating torus.
On the other hand, focusing  on the  pair  composed by an  external counter-rotating cusped  disk and  an internal co-rotating disk, the tori separation region  is bounded above by $r_{mbo}^+$. (In the case of   quiescent tori  or proto-jets the situation is different.)
Therefore, this limit is  fixed exclusively by the spin of the attractor.
Then, the maximum radius of extension of the  co-rotating disk, is  bounded  by the maximum extension of the  disk outer edge, which is  evaluated in this analysis  accurately in Sec.\il(\ref{Sec:outer--edge-crossing}) and   illustrated, for example, in Figs\il(\ref{Fig:corso2}).
It general it  has to be
$r_{outer}^->r_{mso}^-$.
The case of inter--disks inversion sphere evidently occurs when
$r_\Ta\in]r_{outer}^-,r_{inner}^+[$, which is well clarified from Figs\il(\ref{Fig:corso}).
These considerations are reported for the equatorial plane, since the toroids are both orbiting  on the equatorial plane of the central attractor. However,  the inter--disks inversion sphere has a role also in the off--equatorial dynamics between the two toroids. This aspect is relevant,   since the tori considered in this analysis are geometrically  thick.   This situation  is  treated in Sec.\il(\ref{Sec:off-equatorial--crossing}).
 This difference  between the inversion surface and the tori separation regions leads to three different scenarios  for   tori embedded in an inversion surface,  toroids  located out of  the inversion surface or  crossing  the inversion surfaces  in particular points being partially contained in  the inversion surface.
}

All the co--rotating  toroids orbiting in the \textbf{BH} outer ergoregion   are embedded in all the  inversion surfaces and
will be ``shielded"  from contact with the $u$counter--rotating flows.
In general, however, the inversion surfaces are defined by a general, necessary but not sufficient,  condition on (the photons and) matter  which can be  ingoing into  the \textbf{BH}, or outgoing or  also moving along the \textbf{BH} spinning axis. On the other hand,  the $u$counter--rotating flows accreting into the central \textbf{BH},  have  inversion points, crossing the inversion surfaces  defined by the flow  (conserved) specific angular momentum $\ell<0$.
Thus, a co--rotating toroid, entirely contained in this   inversion surface,  is ``shielded" from   impact with the  $u$counter--rotating materials with constant specific angular momentum $\ell<0$,  accreting into the central \textbf{BH} (the flow being $u$co--rotating after crossing the flow inversion surfaces).
Although the inversion surfaces have a more general significance, in this case the effects of the flow inversion surfaces    emerge  as a consequence  of the spacetime frame--dragging. This is the case considered here   for    the inter--disks inversion surfaces, when the  $u$counter--rotating flows coming from the  outer accreting counter--rotating disk into  the central \textbf{BH},    impact on an inner co--rotating  toroid. We also studied  the case where one or two orbiting structures are proto--jets (inter proto-jets inversions surfaces), providing    the conditions when a  co--rotating disk or  co--rotating proto--jet can be  replenished with   $u$counter--rotating  matter and photons which is accreting into the central \textbf{BH} or, viceversa,  when  the counter--rotating matter and photons impacting on the surface   will be $u$co--rotating,  constrained according to
  the  flow parameters $\ell^\pm$ and the \textbf{BH} spin dimensionless $a$.

 In Table\il\ref{Table:tori}  the constraints on the flows and orbiting toroids, discussed in Sec.\il(\ref{Sec:lblu}), are  summarized.
We based part of our analysis  on the assumption  that an equatorial  accretion disk   inner edge is  located in the range $[r_{mbo}^\pm,r_{mso}^\pm]$ (for counter--rotating and co--rotating disks respectively).  The analysis of  the co--rotating  toroids in relation to the spacetime  inversion surfaces, developed in Sec.\il(\ref{Sec:Aspects}),   lead to distinguish sixteen classes of    \textbf{BH}   spacetimes according to the \textbf{BH} spins defined in Table\il\ref{Table:spins} and shown in Figs\il\ref{Fig:corsozoom1},\ref{Fig:corso},\ref{Fig:corso1}.
 (Boundary spins  in Table\il\ref{Table:spins}  result from  the intersection between the inversion surfaces radii $r_\Ta(\ell^+_\bullet)$  with the co--rotating  extended  geodesic structure --see Figs\il\ref{Fig:corso},\ref{Fig:corso1}.)
 The inversion surfaces properties in each  spacetime  class are summarized in Table\il\ref{Table:Regions-BHs}, where
each class is characterized by  the occurrence of a given inter--disks inversion surfaces,  or  constrained,  in terms of the parameters $(\ell^\pm,K^-,a)$, by the possibility of a given configuration.
We found that, in the spacetimes with spin  $a\in]0,a_{(s,c)}]$ the  inversion surfaces are always internal   to the outer co--rotating  toroids.
 The  centers $r_{center}^-$   are external to  the inversion surfaces,   in the spacetimes with  spins  $a\leq a_{(s,s)}$.
 For  $a\in]a_{(s,c)},a_{(c,s)}[$, tori  can be only partially included in an inversion surface.   Therefore,
for  $a<a_{(s,s)}$  co--rotating toroids could be only partially included  (with the inner region) in an inversion surface.  In this range a disk or proto-jet cusp could be  internal or external to  an inversion surface.
For  $a\geq a_{(s,s)}$  the toroid center can be contained in an inversion surface, and therefore  in these spacetimes a torus can be entirely embedded in the inversion surface,  for small co--rotating specific angular momentum $\ell^-$.
For  $a>a_{(s,s)}$  quiescent tori or, for the faster spinning attractors,
cusped tori,  can be or are internal to  the inversion surfaces  , according to the conditions in  Table\il\ref{Table:Regions-BHs}--some of these tori are  in the ergoregion.
When   $a\geq a_{((\bullet,\star))}$ (and from $a\geq a_{(s,s)}$ with more constraints), according to the specific angular momenta $\ell_\bullet^+\in\{\ell_{mco}^+,\ell_{mbo}^+,\ell_{mso}^+\}$,
a co--rotating  torus could be  entirely embedded in an inversion surface.

Hence, in  Sec.\il(\ref{Sec:inversion-crossing})    we investigated
 the  inversion surfaces crossing    the      co--rotating  toroids by  examining specifically    in Sec.\il(\ref{Sec:outer--edge-crossing})  the crossing with the        toroids outer edges. Results are illustrated   in Fig.\il\ref{Fig:corso2} and  Fig.\il\ref{Fig:PlotLibC}. In this case, the toroids with the outer edge crossing an inversion surface are  entirely embedded in the inversion surfaces and therefore can be totally shielded from impact from $u$counter--rotating  flows accreting into the \textbf{BH} with the constant specific angular momentum.
 The crossing can occur for large \textbf{BH} spin  and small momenta in magnitude of the inversion surface.
Tori with $\ell\in \mathbf{L_1^-}$ satisfy this condition in  the spacetimes with spins
$a>a_{(\bullet,s)}$ (for $a_{(\bullet,s)}=\{a_{(s,s)},a_{(b,s)},a_{(c,s)}\}$ according to the different angular momentum).

For  co--rotating toroids with specific angular momenta   $\ell^-\in \mathbf{L_2^-}$   we restricted the spacetimes to the spins
$a>a_{((b,\bullet))}$--for $a_{((b,\bullet))}=\{a_{((b,s))},a_{((b,b))},a_{((b,c))}\}$ (see Fig.\il\ref{Fig:corso2}--upper left panel).
 For the quiescent tori with  $\ell^-\in \mathbf{L_3^-}$,     these cases occur for
$a>a_{((c,\bullet))}$ for $a_{((c,\bullet))}=\{a_{((c,s))},a_{((c,b))},a_{((c,c))}\}$.

  Flows inversion surfaces  crossing the  toroids cusp, as shown in  Fig.\il\ref{Fig:PlotenIpno5finner}, has been  discussed in  Sec.\il(\ref{Sec:cusp-crossing}). In this case the toroids are external to the flow  inversion surfaces at fixed specific angular momentum.
 Results are illustrated in  Fig\il\ref{Fig:PlotenIpno5finner} and constrained in Table\il\ref{Table:Regions-BHs}.
We proved that, in general this condition  can occur  for (cusped) co--rotating tori   orbiting  \textbf{BHs} with
spin in the range   $a\in [a_{(\bullet,b)},a_{(\bullet,s)}[$ (for  the torus cusp)  and,  for the  co--rotating proto-jet cusps in the spacetimes with spin in the range $a\in [a_{(\bullet,c)},a_{(\bullet,b)}[$ (according to the  counter--rotating specific angular momenta $\ell_\bullet^+\in\{\ell_{mco}^+,\ell_{mbo}^+,\ell_{mso}^+\}$).

   When the flows  inversion surfaces cross  the toroids  centers,  the toroids are partially contained in the inversion surfaces.  This case  has been discussed  in  Sec.\il(\ref{Sec:torus-center-crossing}) and illustrated in   Fig.\il\ref{Fig:PlotgsnedibubbleC}.
We proved that it can be  $r_{center}^-(\mathbf{L_1^-})< r_\Ta(\mathbf{L_1^+})$ in the spacetimes with spin  $a\in]a_{(s,s)},a_{(b,s)}[$ (detailed constraints, for each spacetimes class, are summarized  in Table\il\ref{Table:Regions-BHs}). Disks and  proto--jets can be  partially embedded in the  inversion surfaces  for large \textbf{BH} spin ($a>a_{(s,s)}$) and for the faster  spinning \textbf{BHs} (as $a<a_{((c,c))}$), the toroids centers will be contained in the inversion surfaces.

  More in general we analyzed  the inversion surfaces crossing  the  toroids critical points of pressure and geometrical maxima
as   points of the
   curves $\ell^-_\sigma=$constant. This analysis is  discussed in  Sec.\il(\ref{Sec:geometric--crossing}). Results  are  illustrated  in  Fig.\il\ref{Fig:PlotMaxcrit099} and  Fig.\il\ref{Fig:PlotMaxcrit099sbcz}.

   Then,  the angles $\sigma$,  where    the inversion surfaces, with fixed specific angular momentum $\ell^+$,   cross the curves  of the   toroids extreme points are in Fig.\il\ref{Fig:Plotgrimmunificatione}, while the
 inversion surfaces  momentum  $\ell^+$  is shown in in  Fig.\il\ref{Fig:PlotMaxcrit099M}.
  We proved that the crossing with the inversion surface   $r_\Ta(\ell^+_{mso})$  occurs in the  spacetimes with spins   $a_{((\bullet,s))}$ and
  $a_{(s,s)}$ (according to the different co--rotating momenta $\ell_\bullet^-\in\{\ell_{mco}^-,\ell_{mbo}^-\}$).
  For the   inversion surface   $r_\Ta(\ell^+_{mbo})$, a crossing occurs in the spacetimes with spin   $a_{((\bullet,b))}$ and
  $a_{(b,s)}$.  For the inversion surface   $r_\Ta(\ell^+_{mco})$  there are the spins $a_{((\bullet,c))}$ and
  $a_{(c,s)}$.  Spins $a_{(\bullet,s)}$ (for  different counter--rotating momenta $\ell_\bullet^+\in\{\ell_{mco}^+,\ell_{mbo}^+,\ell_{mso}^+\}$) constrain the spacetimes characterized by  the  curves $\ell_\sigma^-=\ell_\star^-=\ell^-_{mso}$ crossings the inversion surfaces. Whereas, for the  crossings with   the extreme points curves
  $\ell_\star^-=\ell^-_{mbo}$, there are the  limiting spins $a_{((b,\bullet))}$, and for
  $\ell_\star^-=\ell^-_{mco}$,
  there are spins  $a_{((c,\bullet))}$. (All boundary spins distinguished   in this analysis are defined  in Table\il\ref{Table:spins}.)

On the other hand, the off--equatorial crossing is often the  case  for  the partially contained toroids.
The off--equatorial properties are determined by the characteristics of the specific accretion disk model and regulated  by the  inversion surfaces characteristics,   which  are fixed exclusively by the background geometry.  The inversion surfaces crossing  the toroids  on    planes different from the equatorial have been  considered in   Sec.\il(\ref{Sec:off-equatorial--crossing}) and
 the results of this analysis were illustrated Figs\il\ref{Fig:Plotgrimmunificatione},\ref{Fig:Plotgsnder} and Figs\il\ref{Fig:PlotMaxcrit099},\ref{Fig:PlotMaxcrit099sbcz},\ref{Fig:PlotMaxcrit099M}, summarizing the main aspects of the crossings at different parameters values:
 we showed the  angles $\sigma$ where  the inversion surface crosses   the  co--rotating  cusped tori $\cc_\times^-$ in  Fig.\il\ref{Fig:Plotgsnder};
the angles $\sigma(a)$,  where    the inversion surfaces with momentum $\ell^+_\bullet$   cross the curves  correspondent to    the geometrical maxima and the pressure critical  points  of the   toroids with angular momentum $\ell_\star^-$ were showed in Fig.\il\ref{Fig:Plotgrimmunificatione}.  Solutions for   the  counter--rotating angular momentum $ \ell^+(\sigma)$ of the inversion surfaces,
  versus the angle $\sigma$  were illustrated,  for  different spins, in Fig.\il\ref{Fig:PlotMaxcrit099M}.

Finally we expect that the  observational properties  on the inversion surfaces could   depend strongly on the  processes time--scales as related to  the  time flow reaches the inversion  points. However, the inversion  surfaces  could  be a      remarkably active  part of the accreting flux,    particularly in the region of the  \textbf{BH} poles (where the inversion  surfaces are very close) and the equatorial plane (when the accretion flows mostly leaves the equatorial  accretion disks inner edges) and,  eventually characterized  by an increase of the flow luminosity and temperature.

\section*{Data availability}
There are no new data associated with this article.
No new data were generated or analysed in support of this research.

\end{document}